\documentclass[preprint,12pt]{elsarticle}
\usepackage{booktabs}
\usepackage{hyperref}
\usepackage{float}

\makeatletter
\def\ps@pprintTitle{%
   \let\@oddhead\@empty
   \let\@evenhead\@empty
   \def\@oddfoot{}%
   \let\@evenfoot\@oddfoot}
\makeatother

\usepackage{amssymb}
\usepackage{comment}
\usepackage{amsmath}
\usepackage{bm}
\begin{document}

\begin{frontmatter}



%\includegraphics[width=\textwidth, keepaspectratio]{graphicalabstract.pdf}

%\item We investigate the robustness of PINNs to noise for inverse problems.
%\item PINNs are simple and scale with the computational complexity of the problem, but are %outperformed by a basic baseline.
%\item Sufficiently noisy data causes PINNs to disregard the physics loss.
%\item PINNs have biased parameter estimation.
%\item New early stopping strategies are needed for noisy inverse problems.

\title{Examining the robustness of
Physics-Informed Neural Networks to noise for Inverse Problems} %% Article title


\author[label1]{Aleksandra Jekic} %% Author name
\author[label1]{Afroditi Natsaridou} %% Author name
\author[label2]{Signe Riemer-Sørensen} %% Author name
\author[label1,label3]{Helge Langseth} %% Author name
\author[label1,label3]{Odd Erik Gundersen} %% Author name

\affiliation[label1]{organization={Department of Computer Science, Norwegian University of Science and Technology},%Department and Organization
            addressline={Høgskoleringen 1}, 
            city={Trondheim},
            postcode={7034}, 
            state={Trøndelag},
            country={Norway}}

\affiliation[label2]{organization={Department of Mathematics and Cybernetics, SINTEF Digital},%Department and Organization
            addressline={Forskningsveien 1}, 
            city={Oslo},
            postcode={0373}, 
            state={Oslo},
            country={Norway}}

\affiliation[label3]{organization={Aneo AI Research},%Department and Organization
            addressline={Klæbuveien 118}, 
            city={Trondheim},
            postcode={7031}, 
            state={Trøndelag},
            country={Norway}}

%% Abstract
\begin{abstract}Approximating solutions to partial differential equations (PDEs) is fundamental for the modeling of dynamical systems in science and engineering. Physics-informed neural networks (PINNs) are a recent machine learning-based approach, for which many properties and limitations remain unknown. 

PINNs are widely accepted as less computationally efficient and accurate than traditional methods for solving PDEs, such as the finite element method. However, PINNs are commonly claimed to show promise in solving inverse problems and handling noisy or incomplete data. We compare the performance of PINNs in solving inverse problems with that of a traditional approach using the finite element method combined with a numerical optimizer. The models are tested on viscosity identification in 1D Burgers' equation and in 2D/3D Taylor-Green Vortex, in all cases with additive Gaussian noise applied to training and validation data.

We find that while PINNs may require less human effort and specialized knowledge, they are outperformed by the traditional approach. For example, for 2D Taylor-Green Vortex with $\sigma=1$ noise, the baseline has a mean prediction RMSE of $0.0013$ compared to $0.01$ for the best PINN variation. However, PINNs scale better than the baseline with the computational complexity of the problem. We identify failures during training to be addressed if the PINN performance on noisy inverse problems is to become more competitive.

\end{abstract}


%% Keywords

\end{frontmatter}



\section{Introduction}
\label{sec1}

Partial differential equations (PDEs) are essential tools for modeling physical systems which are extensively used within engineering \cite{evans2010partial}. Solving a PDE is called the \textit{forward problem}. The most widely used methods are backed by a deep theoretical understanding of their limitations and their capabilities on various problems \cite{ryaben2006theoretical}. In many real-life cases, the PDE may not be fully defined. For \textit{inverse problems} \cite{vogel02}, the goal is to identify something missing in the physical model such as a parameter or an initial condition.

Physics-informed neural networks (PINN) leverage the flexibility of neural networks while combining it with physics knowledge to solve both the forward and inverse problems of PDEs within a unified framework \cite{raissi2019physics}. Note that PINNs were introduced to solve inverse problems with static parameters \cite{raissi2019physics}. There has been some limited work to extend PINNs to function-valued inverse problems \cite{MIAO2023133945, thakur}, but we consider this outside of our current scope.

Theoretical understanding of the capabilities and limitations of PINNs remains limited compared to standard methods \cite{cuomo2022scientificmachinelearningphysicsinformed}. Despite extensive literature on the subject \cite{mcgreivy2024weakbaselinesreportingbiases, karninat, grossmann2023physicsinformedneuralnetworksbeat, FERNANDEZDELAMATA2023128415, chuang2022experiencereportphysicsinformedneural}, PINNs are widely recognized to be inferior to traditional methods for solving the forward problem of PDEs. As an example, \citet{chuang2022experiencereportphysicsinformedneural} report that PINN takes 32 hours to achieve the same accuracy as a traditional solver running for less than 20 seconds. However, PINNs seem to be less vulnerable to the curse of dimensionality \cite{HU2024106369}. PINNs may also solve inverse problems many traditional solvers cannot, such as those without known boundary conditions \cite{JIN2021109951, doi:10.1126/science.aaw4741}. 

Despite lower accuracy in the forward problem, PINNs may solve the inverse problem satisfactorily \cite{FERNANDEZDELAMATA2023128415, mcgreivy2024weakbaselinesreportingbiases, karninat}, and they have been applied with reasonable success in various fields \cite{sharma2024inversiondcresistivitydata, yang2024twostageimagingframeworkcombining, yokota2024identificationphysicalpropertiesacoustic, berardi2024inversephysicsinformedneuralnetworks}. CDFbench tests various physics machine learning models on predicting velocity fields of fluid mechanics systems \cite{luo2024cfdbenchlargescalebenchmarkmachine} while PINNacle benchmarks variations of PINN on different types of physics problems, including some inverse problems \cite{hao2023pinnaclecomprehensivebenchmarkphysicsinformed}. However, as noted by \citet{Karnakov_2023}, to the best of our knowledge there exists no benchmark to test how PINNs compare with traditional solvers on inverse problems, in particular inverse Navier-Stokes problems with noise.

Numerous efforts have been made to enhance PINN performance \cite{wang2023expertsguidetrainingphysicsinformed}, such as by balancing the gradients of each training loss, domain decomposition, Fourier space embedding and learning earlier time steps of the physical system first \cite{wang2020understandingmitigatinggradientpathologies, Kharazmi_2021, tancik2020fourierfeaturesletnetworks, wang2022respectingcausalityneedtraining}. For the inverse problem, improvement has been demonstrated in parameter identification by using different optimizer for parameter identification and network parameters \cite{new2024equationidentificationfluidflows}. Using an ensemble of PINNs has also been suggested to achieve more accurate parameter identification \cite{aliakbari}. ODIL is a proposed drop-in replacement for PINNs which directly optimizes discrete formulations of PDEs, removing the neural network \cite{Karnakov_2023}.

\subsection{Contributions and outline}

We explore the claim that PINNs are effective for inverse problems. As our baseline, we use the general purpose method of combining the finite element method (FEM) with an optimizer. The most efficient solver-based approaches involve significant tailoring to the individual problem. Our purpose is to compare PINNs to a similarly easily accessible but reasonable choice of standard methods, without extensive tuning. We perform tests on parameter identification with varying levels of noise and data availability on a series of increasingly difficult fluid problems leading to the following contributions:

{\small
\begin{itemize}
    \item We show that while PINNs require less manual labor and specialized knowledge, a simple baseline with a FEM solver and SLSQP optimizer largely outperforms PINNs on noisy inverse problems.
    \item We find that the relative performance of PINNs appears to improve with the computational complexity of the problem, consistent with existing literature \cite{HU2024106369}.
    \item We identify two particular problems PINNs have on noisy inverse problems:
        \begin{itemize}
        \item The PINNs were consistently biased in their parameter estimation, eventually locking the model into a poor solution of the PDE.
        \item With adaptive weights \cite{wang2020understandingmitigatinggradientpathologies}, needed for harder problems, the physics loss grew very large for noisy training data. In these cases, the physics loss does not have the desired effect of correcting the noise.
        \end{itemize}
    \item We find that identifying the best model during training is not trivial. New early stopping strategies are needed for noisy inverse problems.
\end{itemize}}

The remainder of the paper is outlined as follows: In Section \ref{sec:background}, we provide an overview of PINNs, the finite element method, and the mathematical formulation of the test problems. In Section \ref{sec:general}, the shared setup of the experiments is described. Section \ref{sec:experiments} presents the experimental details and their results. The code can be found on GitHub \footnote{https://github.com/aleksjek9/pinnrobustness/}. Section \ref{sec:discussion} discusses the results of the experiments and in Section \ref{sec:conclusion} we conclude.

\section{Background}
\label{sec:background}

\subsection{Physics-Informed Neural Networks}
\label{sec:pinn}

In a standard neural network, an input vector $\bm x$ is passed through layers $a_1(\bm x;\bm \theta_1), \dots, a_n(\bm{\hat x}_{n-1};\bm \theta_n)$ of weighted sums and non-linear activation functions. $a_j$, $\bm{\hat x}_{j-1}$, $\bm \theta_{j}$, $j = 1 \dots n$ are respectively the output, the input and the parameters (biases and weights) of layer $j$. We may consider the full neural network as a parameterized single function \cite{Goodfellow-et-al-2016}:

\begin{equation}
a(\bm x;\bm \theta)=a_n(a_{n-1}(\cdots a_1(\bm x; \bm \theta_1))).
\end{equation}

\noindent
$a(\bm x;\bm \theta)$ can be both a scalar function and a vector function $\bm a(\bm x;\bm \theta)$. To evaluate the parameters, we define a loss function $\mathcal{L}_{\text{data}}$ which measures the discrepancy between the predicted output $\bm{\hat{y}}_i$ and the true value $\bm y_i$ for observations available in the training set, totaling $n_{\text{data}}$. For regression type problems, a common choice is the Mean Squared Error (MSE):

\begin{equation}
\mathcal{L}_{\text{data}} = \frac{1}{n_{\text{data}}} \sum_{i=1}^{n_{\text{data}}} \left\|\bm{\hat{y}}_i - \bm y_i\right\|_2^2.\label{mse}
\end{equation}

\noindent
In physics-informed neural networks (PINNs) \cite{raissi2019physics}, in addition to the above loss function, we include a second loss function:

\begin{equation}
\mathcal{L}_{\text{phys}} = \frac{1}{n_{\text{phys}}} \sum_{i=1}^{n_{\text{phys}}} \left\|\mathcal{R}_{\theta}(\bm x_i)\right\|_2^2.\label{phys}
\end{equation}

\noindent
$ \mathcal{R}_{\theta}(\bm x)$ is a function which calculates the residual of the differential equation we want to impose on the solution. Leveraging the automatic differentiation capabilities available in popular machine learning libraries, we can easily calculate the derivates of $a(\bm x; \bm \theta)$ to check adherence with the differential equation. Missing parameters in the PDE can be optimized along with the parameters of the neural network. The two losses in \autoref{mse} and \autoref{phys} are summed to get the total loss minimized by the optimizer: $\mathcal{L}_{\text{total}} = w\mathcal{L}_{\text{data}} + \mathcal{L}_{\text{phys}}$, where $w$ is used to balance the weights.

In our experiments, we used a popular algorithm which adjusts $w$ to ensure that the magnitudes from the two losses are similar \cite{wang2020understandingmitigatinggradientpathologies}. In early stopping \cite{Goodfellow-et-al-2016}, we exclude a part of the training set, called the validation set, and use it to approximate when the model has achieved maximum performance. 

When training PINNs, the optimizer Adam \cite{kingma2017adammethodstochasticoptimization} is often applied first to the loss function since it is considered effective for widely exploring search optimization landscapes. The quasi-Newtonian optimization method L-BFGS (Limited-Memory Broyden-Fletcher-Goldfarb-Shanno) is applied at the end to converge more thoroughly to the closest minimum \cite{URBAN2025113656}. To our knowledge, since neural network optimization landscapes are highly nonconvex, we cannot expect monotonous learning for L-BFGS with line search \cite{lewis2013nonsmoothness}.

\subsection{Finite Element Method}

The Finite Element Method (FEM) is a general purpose method for solving differential equations \cite{richard_courant_1943}. The procedure works by projecting the solution onto simpler functions, called basis functions, resulting in a system of equations which, when solved, give the closest approximate fit to the solution. In some ways, it is similar to how PINNs estimate the solution by representing it as a neural network and minimizing the losses.

The differential equation in its standard form is known as the \textit{strong formulation}. We project it by multiplying by a simpler function $v(x, t)$ on both sides, which is referred to as a test function in the literature, and by integrating. This gives us the projected \textit{weak form} of the problem:

\begin{equation}
\label{femeq}
\int_{\Omega} \left(\frac{\partial u(x,t)}{\partial t} +N[u(x,t)]\right) \ v(x, t) \ dx= 0.
\end{equation}

\noindent
Here $N[u(x, t)]$ is a linear differential operator which encodes the behavior of the system while $u(x, t)$ is the solution we are looking for, defined over a domain $\Omega$ on some specific time interval $[0, T]$ with respect to initial and boundary conditions. This description can be extended to any dimension, but we assume scalar variables here for simplicity.

There are many methods for how to proceed from \autoref{femeq}. One of the most well-established methods for deriving and solving the weak form is the Galerkin method \cite{seshu2003finiteelement}, in which the same basis functions $\{\phi_i(x)\}_{i \in \mathbb{N}}$, are used for both the test functions $v(x, t)$ and trial functions $u(x, t)$. We get a matrix equation with the following structure to solve:

\begin{equation}
\mathbf{M} \dot{\mathbf{a}}(t) + \mathbf{A} \mathbf{a}(t) + \mathbf{B} \mathbf{a}(t) = 0
\end{equation}

\noindent
The matrices are defined by the exact problem and the specific approach taken. There are many variations of FEM which can be applied to solve problems both in space and time. The steps to derive the weak form depend on the equation to be solved and its boundary conditions. Custom algorithms are often necessary for acceptable performance on different problems \cite{liu2022eighty}.

\subsubsection{Sequential Least Squares Programming}

To estimate an unknown parameter in a differential equation using observed data, we first make an initial guess. With this guess, we solve the differential equation using the finite element method. The data loss, measuring the discrepancy between the solution and the training data, is used to update the guess. This process is repeated until convergence \cite{vogel02}.

In the experiments, we use Sequential Least Squares Programming (SLSQP) \cite{kraft1988software}, a more robust variation of Sequential Quadratic Programming (SQP). Both algorithms are efficient for non-linear programming problems. Our goal is to find the minimum of the function $\mathcal{L}_{\text{data}}(\bm{x})$, which represents the data loss with regards to the unknown parameter $\bm{x}$. At the function minimum, the gradient $\nabla\mathcal{L}_{\text{data}}(\bm{x})$ will be zero. To seek this zero, the function (and any additional constraint functions) are linearized with Taylor expansions. In SQP, the problem is solved with Newton's method which uses the gradient to find which direction to move and by how much. However, SLSQP uses a quasi-Newtonian method such as BFGS to avoid calculating the Hessian. Constraints are handled with a nonlinear least squares method.

\subsection{Test problems}

We will perform our experiments on the Burgers' equation and the Navier-Stokes equations. The systems are related in the sense that Burgers' equation is a simplification of the Navier-Stokes equations. All the systems used for our experiments are commonly used benchmark problems \cite{raissi2019physics, mcgreivy2024weakbaselinesreportingbiases, luo2024cfdbenchlargescalebenchmarkmachine}.

The actual implementations of each system used in the experiments, such as boundary conditions and parameters, will be described in Chapter \ref{sec:experiments}. 

Although we will specifically work on Burgers' equation in 1D and Navier-Stokes in 2D and 3D, we will describe the systems in $d$-dimensional form.

\subsubsection{Burgers' equation}
Burgers' equation is a non-linear partial differential equation. It models the evolution of a velocity field $\bm u(\bm x, t)$ as a function of space and time:

\begin{align}
\frac{\partial \mathbf{u}(\bm x,t)}{\partial t} 
+ \mathbf{u}(\bm x,t) \cdot \nabla \mathbf{u}(\bm x,t) 
&= \nu \nabla^2 \mathbf{u}(\bm x,t), \notag \\
\mathbf{x} &\in \Omega \subset \mathbb{R}^n, 
\quad t \in [0, \infty), 
\quad \mathbf{u} \in \mathbb{R}^n.
\end{align}

\noindent
Here $\frac{\partial \mathbf{u}(\bm x,t)}{\partial t}$ describes how the velocity field changes in time. $\mathbf{u}(\bm x,t) \cdot \nabla \mathbf{u}(\bm x,t)$ is the non-linear advection term which determines how the velocity field changes as a function of its current velocity and its position. $\nu \nabla^2 \mathbf{u}(\bm x,t)$ 
is the diffusion term which describes how the velocity is evened out in the flow.

Depending on the initial conditions and viscosity, Burgers' equation can develop shocks when discontinuities arise in the solution as $\frac{\partial u(\bm x, t)}{\partial \bm x} \to -\infty$. Intuitively, shocks can be thought of as hard corners forming in the wave. Shocks pose a challenge for common approaches to solving differential equations since standard methods assume differentiable and smooth functions.

\subsubsection{Incompressible Navier-Stokes equations}

Like Burgers' equation, the Navier-Stokes equations are a set of non-linear PDEs which model the evolution of a velocity field $\bm u(\bm x, t)$ as a function of time and space. The incompressible form is used to describe liquids whose density does not change with pressure and temperature. The momentum equation in the incompressible Navier-Stokes equations has the general form:

\begin{align}
\frac{\partial \mathbf{u}(\bm x,t)}{\partial t} 
+ \mathbf{u}(\bm x,t) \cdot \nabla \mathbf{u}(\bm x,t) 
&= -\frac{1}{\rho} \nabla p(\bm x,t) 
+ \nu \nabla^2 \mathbf{u}(\bm x,t), \notag \\
\mathbf{x} &\in \Omega \subset \mathbb{R}^n, 
\quad t \in [0, \infty), 
\quad \mathbf{u} \in \mathbb{R}^n,
\end{align}

\noindent
with the requirement of continuity:

\begin{equation}
\nabla \cdot \mathbf{u}(\bm x,t) = 0.
\end{equation}

\noindent
As in Burgers' equation, $\frac{\partial \mathbf{u}(\bm x,t)}{\partial t}$, $\mathbf{u}(\bm x,t) \cdot \nabla \mathbf{u}(\bm x,t)$ and $\nu \nabla^2 \mathbf{u}(\bm x,t)$ describe the change in time, advection and diffusion, respectively. $-\frac{1}{\rho} \nabla p(\bm x,t)$ is the pressure term which evens out differences in pressure. $\nabla \cdot \mathbf{u}(\bm x,t) = 0$ simply states the incompressibility in mathematical terms: The mass of the fluid in any volume does not change. Burgers' equation can be seen as Navier-Stokes simplified, neglecting pressure and with no compressibility constraint.

Under some conditions, the Navier-Stokes equations exhibit turbulence. In turbulent flow, many small whirls form which interact in complex ways on small scales. We get a chaotic system where small changes in the initial conditions may lead to very large changes in the solution. The Reynolds number is a heuristic used to capture the turbulence of a system. Standard PINNs are not expected to work well with chaos \cite{wang2022respectingcausalityneedtraining}. For this reason, we only consider Navier-Stokes systems with low Reynolds numbers ($<2000$). However, for Taylor-Green Vortex this is sufficient for some turbulence.

\section{General experimental setup}
\label{sec:general}

The most efficient traditional solver-based approaches require significant tailoring to each individual problem. Our goal is to compare PINNs to a traditional general-purpose method which is feasible to run on limited hardware using existing libraries. We want to consider how reasonable implementations would practically measure up against each other in real life application. This is not a controlled scaling study, which is not common in machine learning where the effects of different parameters may be highly irregular. For example, a larger neural network may not give higher accuracy. The results must be understood in light of our methodological limitations. We chose to follow a paper which specifically considered how to compare standard methods to machine learning \cite{mcgreivy2024weakbaselinesreportingbiases}. To calibrate the models, we matched the models on accuracy when tested on noiseless data as well as we could.

As baseline, we employed a traditional Finite Element Method (FEM) solver combined with an SLSQP optimizer. For the optimizer, we used the SLSQP implementation from SciPy \cite{2020SciPy-NMeth}. For the 1D Burgers' equation, we used the optional FEniCS library pyadjoint \cite{mitusch2019dolfin} to differentiate the solver and calculate gradients for the optimizer. However, this approach required the data and solver grid structure to be identical, which was unfeasible for the Navier-Stokes test problems because of the more complex grids required to solve these problems. In these cases, we employed finite differences with a three-point stencil from SciPy. This produced more accurate gradients than alternative methods discussed in the FEniCS community \cite{dolfinadjoint-issue7}.

\begin{figure}[!ht!]
\centering
\includegraphics[width=0.49\textwidth]{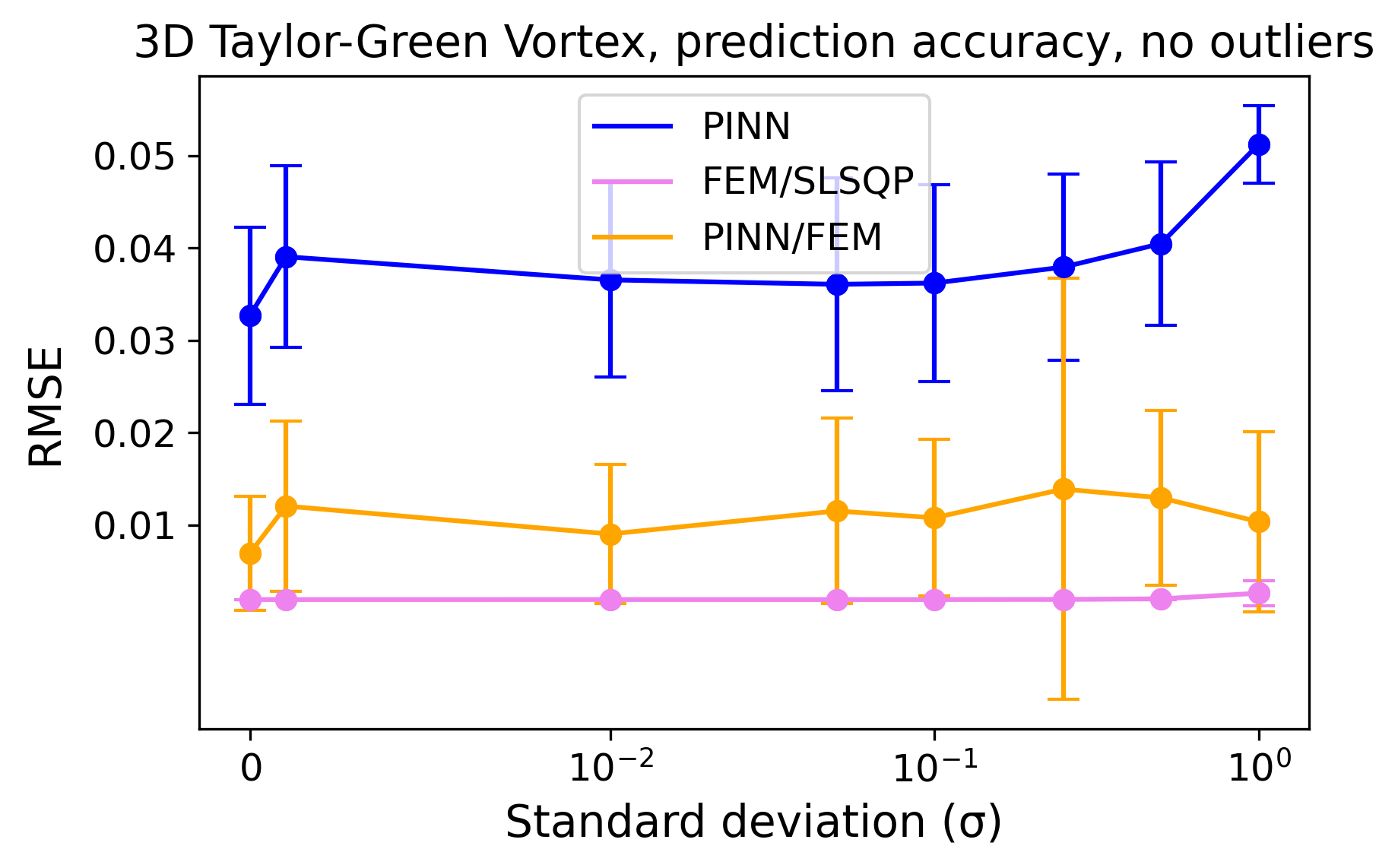}
\includegraphics[width=0.49\textwidth]{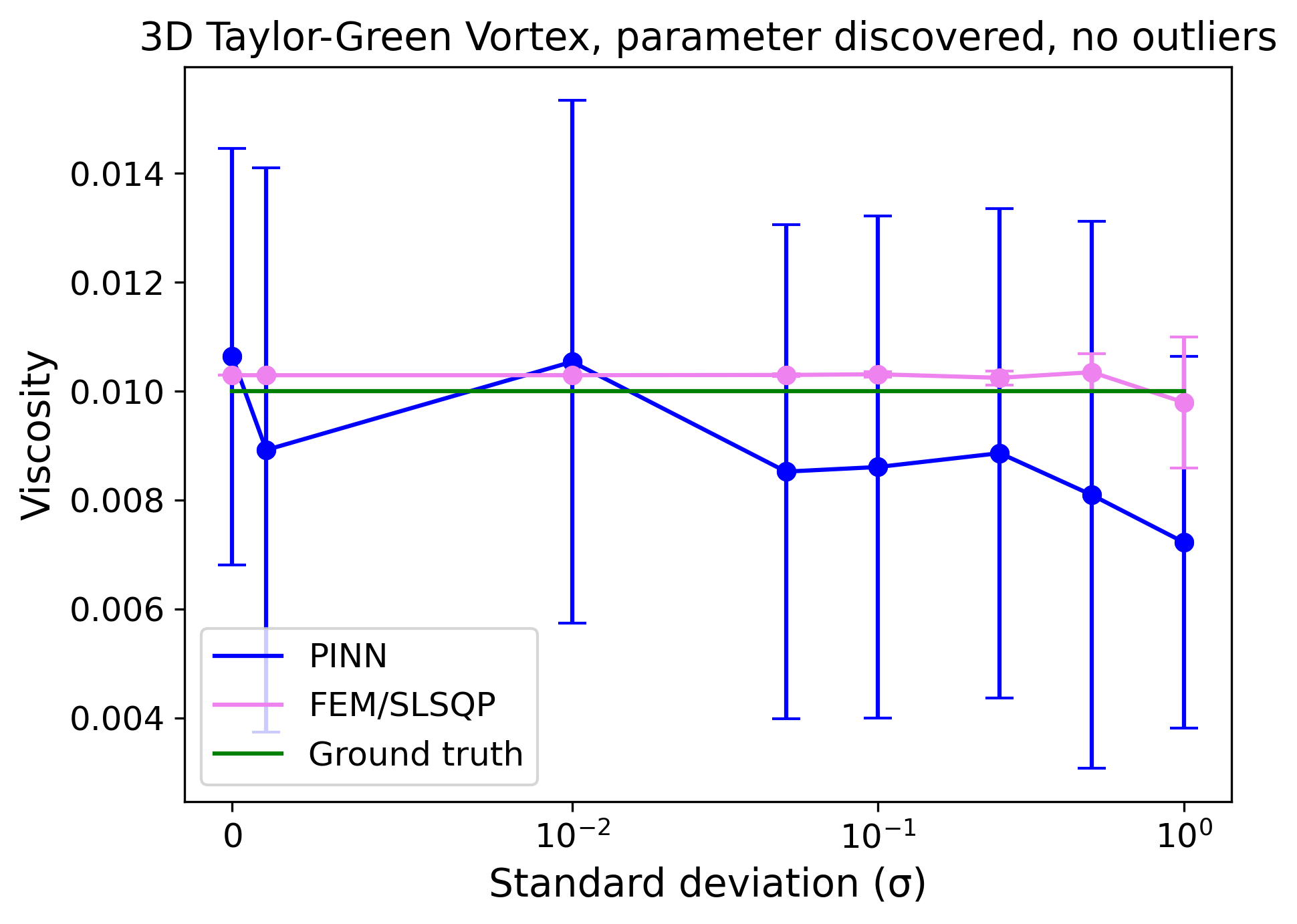}
\includegraphics[width=0.49\textwidth]{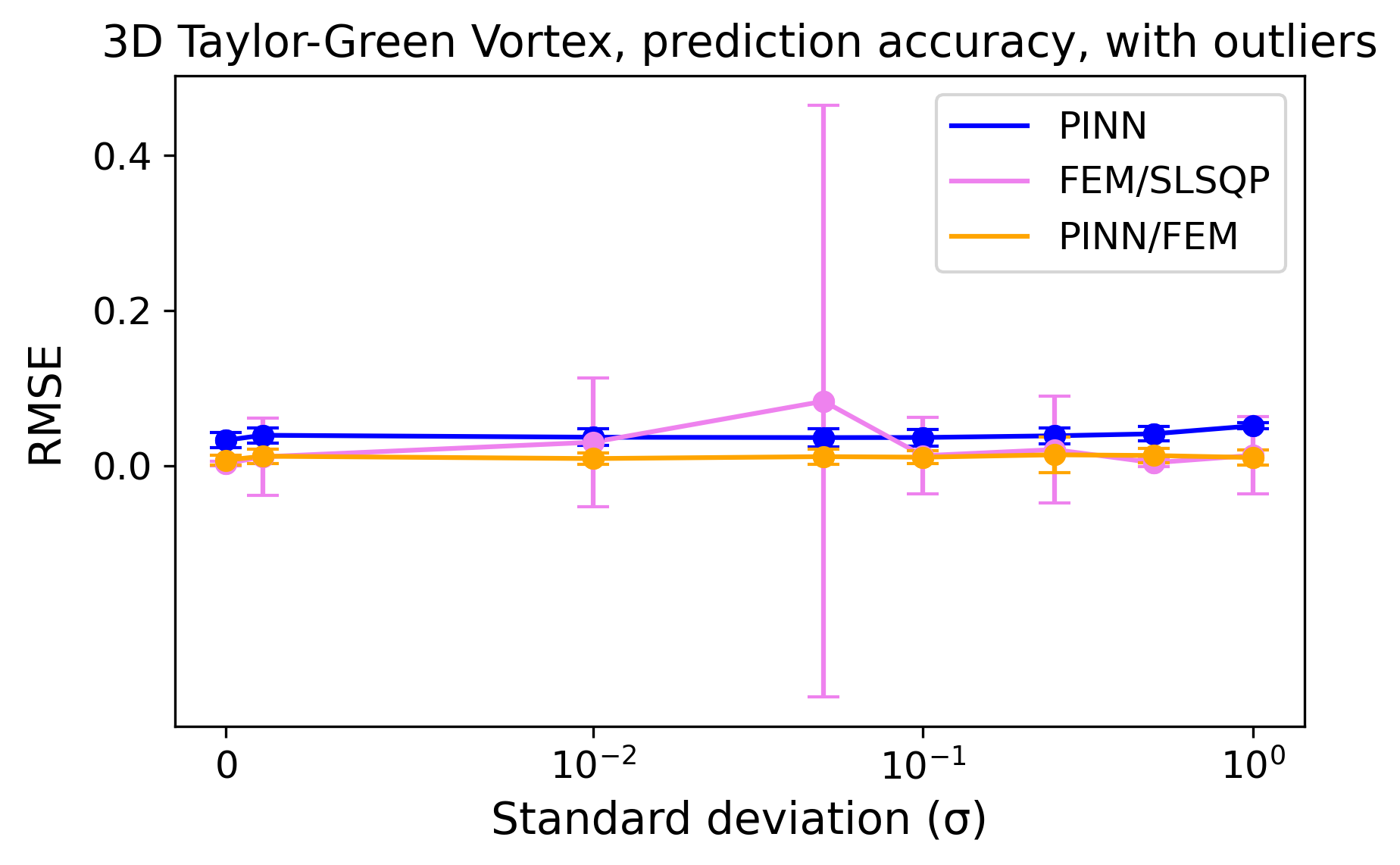}
\includegraphics[width=0.49\textwidth]{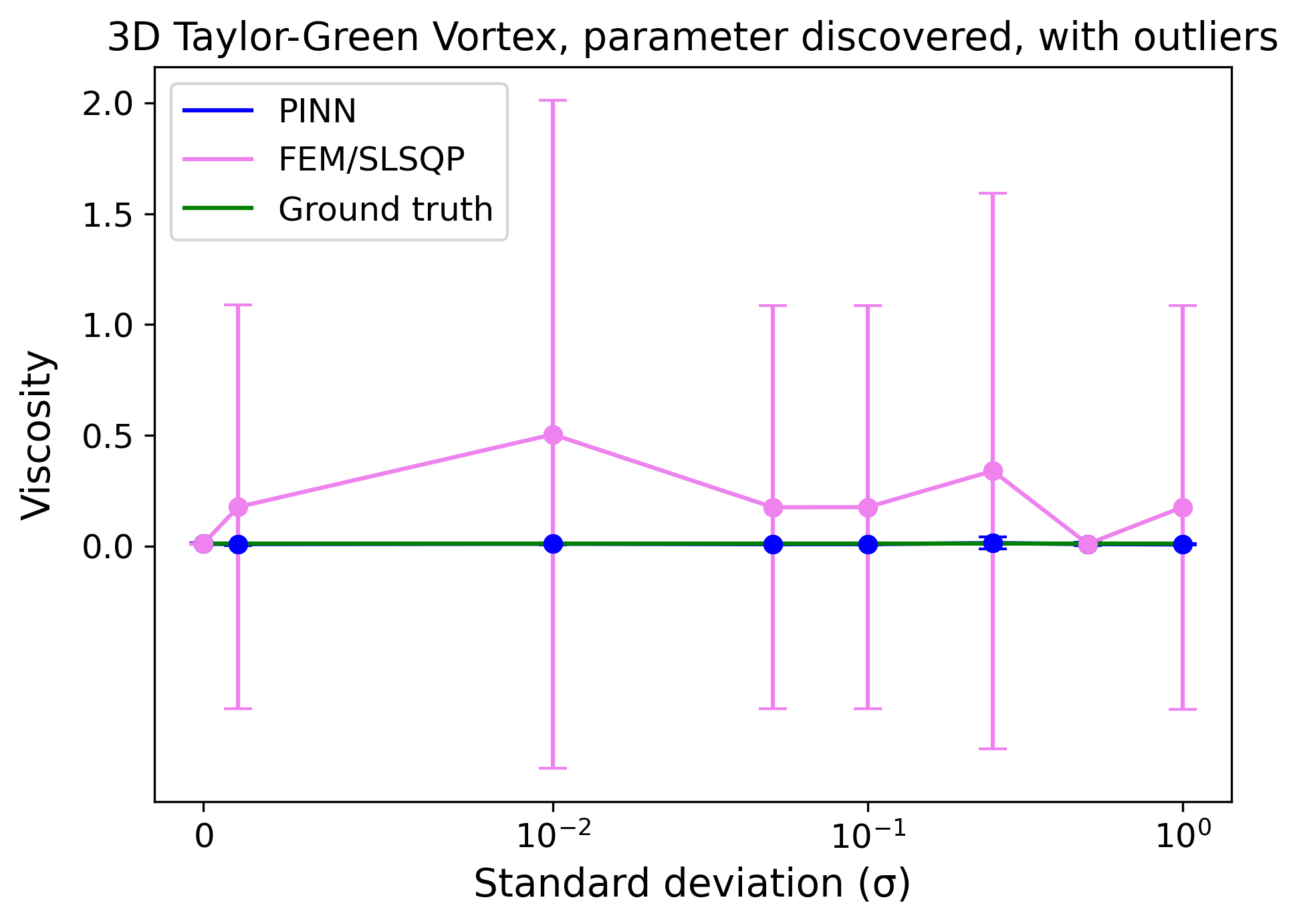}
\caption{Results for 3D Taylor-Green vortex on test set based on 30 runs of each model. For visibility, the x-axes are displayed in a symmetric logarithmic scale with a linear threshold of 0.01. Standard deviation ($\sigma$) is the noise added to the training data. Top left: Prediction accuracies with outliers removed. Top right: Parameter accuracies with outliers removed. Bottom left: Prediction accuracies with outliers included. Bottom right: Parameter accuracies with outliers included.} \label{fig:green3}
\end{figure}

The PINN models were implemented using PyTorch \cite{paszke2019pytorchimperativestylehighperformance}. Note that for PINN, we provide results for two types of models. One is a pure PINN model (marked "PINN" in the plots in Figures \ref{fig:burgers1}, \ref{fig:green1}, \ref{fig:green3}) where we get predictions with a forward pass on the trained neural network. In the "PINN/FEM" model, we collect the estimated parameter from the trained model and solve the differential equation using the same FEM solver as for the baseline. As is standard in machine learning, a small amount of the training data was separated and used as validation data in the PINN experiments. To get prediction RMSEs, we only compared predicted velocities and not the pressure. 

Note that we report the training losses in MSE, the quantity optimized during training. However, we report the predictions in RMSE. This is because the small variation in MSE makes the prediction graphs hard to read. For completeness, the same graphs with MSE are included in \ref{msegraphs}.

The experiments were conducted on the Idun high-performance computing cluster at the Norwegian University of Science and Technology (NTNU) \cite{ntnu_idun}. The compute nodes are mainly equipped with Intel Xeon processors and NVIDIA GPUs. For most of the PINN experiments, we used NVIDIA A100 GPUs. Training PINNs on Navier-Stokes problems may require days depending on the GPU and problem instance, making hyperparameter optimization computationally impractical \cite{chuang2022experiencereportphysicsinformedneural}. For this reason, hyperparameters are commonly selected by manual tuning, and for Navier-Stokes, we used hyperparameter settings consistent with related literature. For the 1D Burgers' equation, we performed extensive hyperparameter tuning using Optuna \cite{akiba2019optuna}. We also investigated the effect of individual hyperparameters. We observed a high degree of robustness with respect to hyperparameter selection. The final hyperparameters were chosen to favor performance on the inverse problem.

For 1D Burgers' equation, we used the same data as in the original PINN paper \footnote{\url{https://github.com/maziarraissi/PINNs/blob/master/appendix/Data/burgers_shock.mat}} \cite{raissi2019physics}. For 1D Burgers' equation with two parameters, we used an analytical solution based on the Cole-Hopf transform, which is described in additional detail in \ref{twoparamet}. For 2D Navier-Stokes, we used an analytical solution. For the specific 3D Navier-Stokes problem that we tested, neither an analytical solution nor pre-existing ground truth data were available. For this reason, we generated reference data using the spectral/hp element framework Nektar++ \cite{CANTWELL2015205}. We used a scheme similar to the quasi-3D scheme in the official Nektar++ 3D Taylor-Green Vortex tutorial \cite{nektar_tgv_tutorial}. 

The noise added to the data was Gaussian with zero mean, using the library NumPy \cite{harris2020array}. In all the experiments, we only added noise to the labels in the training data, but never to the initial condition and boundary condition data. Noise was also added to the validation set labels. This is a standard choice in machine learning, since if you were to have data without noise available, it would be much more valuable as training data. By $\sigma$, we denote the standard deviation of the noise. By SNR, we denote the variance of the training data without noise divided by the variance of the noise.

Due to the Central Limit Theorem \cite{walpole2012probability}, Gaussian noise can be used to approximate sensor noise. However, many of the noise levels we tested, in particular for the 1D Burgers' equation, are extreme. For 1D Burgers' equation, all the values stay within $[-1, 1]$. In other words, with $\sigma=1$ the magnitude of the noise should at least be as large as the original signal. With $\sigma=25$, noise dominates the original signal by a large amount. In real life, noise on scales such as $\sigma=0.1$ are more likely. However, we did not want to make any assumptions about the specific application of PINNs. Rather, we wanted to explore the practical properties and limits of PINNs in a general way.

For 1D Burgers' equation, we chose not to use early stopping based on empirical experience. For the Navier-Stokes problems, we used the validation set to determine which model to use. However, for this to work we had to exclude early models from considerations. Especially with noise, PINNs may do well on the validation set before they have started learning the physics meaningfully. This means the optimizer does not have time to adjust the unknown physics parameter in the physics loss and the inverse problem is not solved. Additionally, both PINN and baseline model were bounded to output minimum $\nu = \pi/1000$. This is because the FEM solver is unable to handle large Reynolds numbers and may break with low viscosity values.

\section{Experiments}
\label{sec:experiments}

\subsection{1D Burgers' equation}

The ground truth viscosity term to solve for was set to $\nu=0.01/\pi$ following the original PINN paper \cite{raissi2019physics}. Likewise, $t \in [0,0.99]$ and $x \in [-1, 1]$. The following initial conditions and Dirichlet boundary conditions were used:

\begin{equation}
u(x, 0) = -\sin{(x\pi)}, \hspace{30pt} u(-1, t) = u(1, t) = 0 \, .
\end{equation}

\noindent
Under these conditions, the system develops shock-like formations around $t=0.5$, which are known to be difficult to solve \cite{raissi2019physics}. Analytical solutions are possible for $\nu > 0$ using the Cole-Hopf transformation \cite{evans2010partial}. We also tested 1D Burgers' equation with two unknown parameters, adding a second physical parameter $a$ to the equation, described more extensively in \ref{twoparamet}.

\subsubsection{Specific experimental setup}

For the training and test data for Burgers' equation, the same data was used as in the original PINN paper \cite{raissi2019physics}. The total dataset consisted of 25600 observations of which 10\% was used as training data and 2\% as validation data. The training and validation data were randomly sampled from the full dataset. For each model, we trained 30 random initializations for which we present the mean and standard deviation of the performance. The noise levels tested for the inverse problem, both in $\sigma$ and SNR, are given in \ref{noiselevels1d}. Remaining implementation details are in \ref{1dapp}.

\subsubsection{Results, 1D Burgers' equation}

\begin{figure}[!ht!]
\centering
\includegraphics[width=0.49\textwidth]{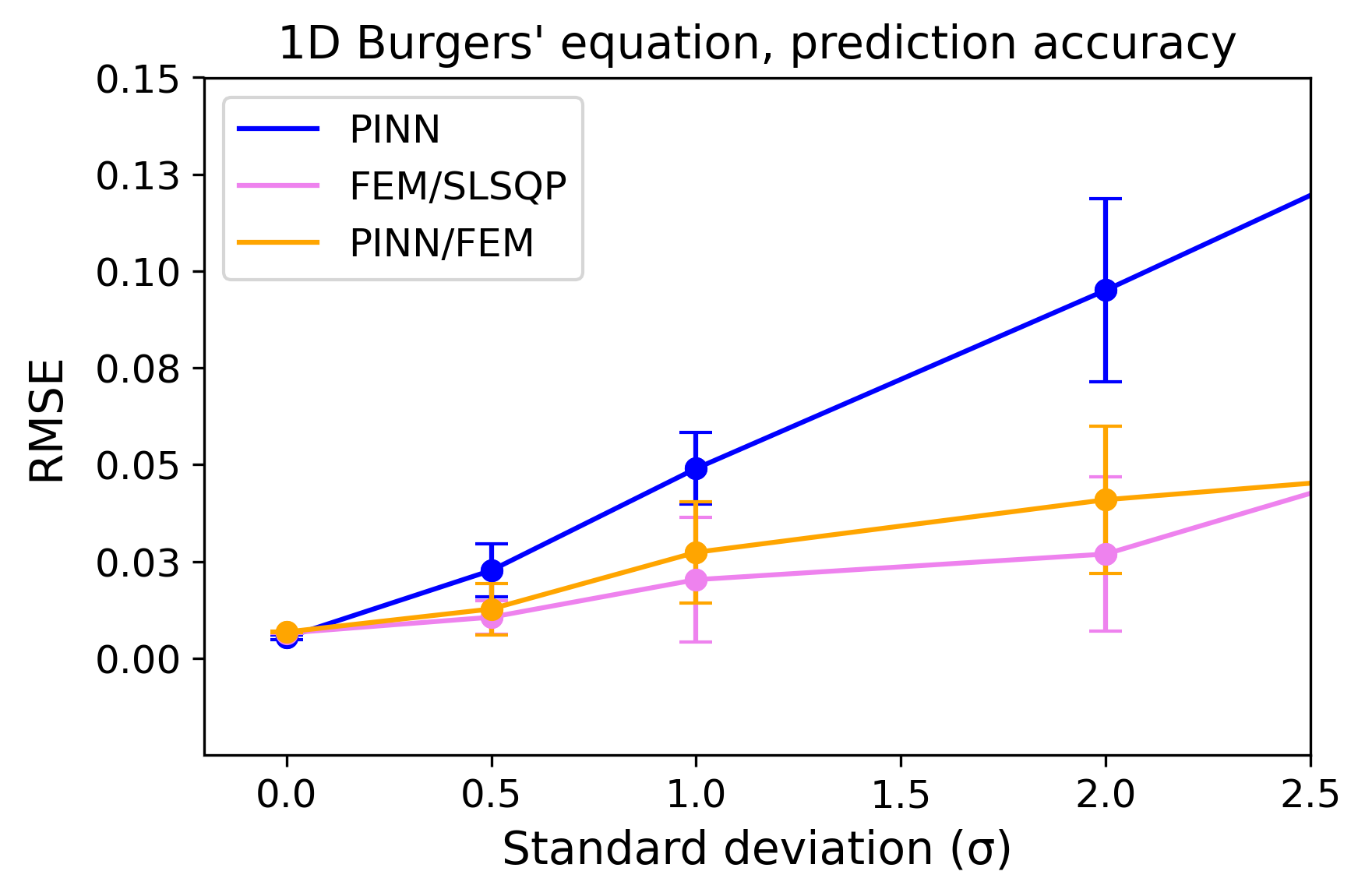}
\includegraphics[width=0.49\textwidth]{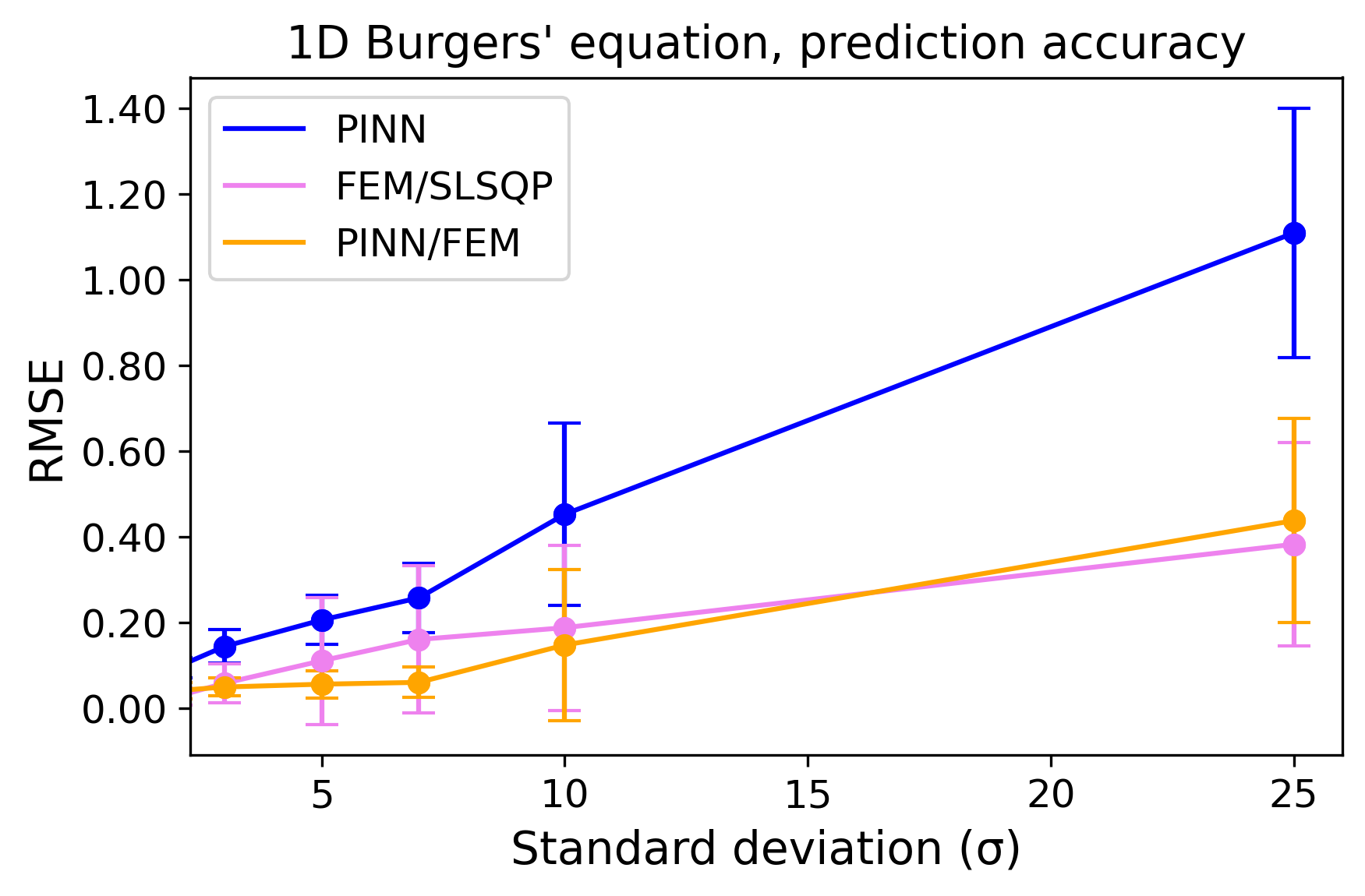}
\includegraphics[width=0.49\textwidth]{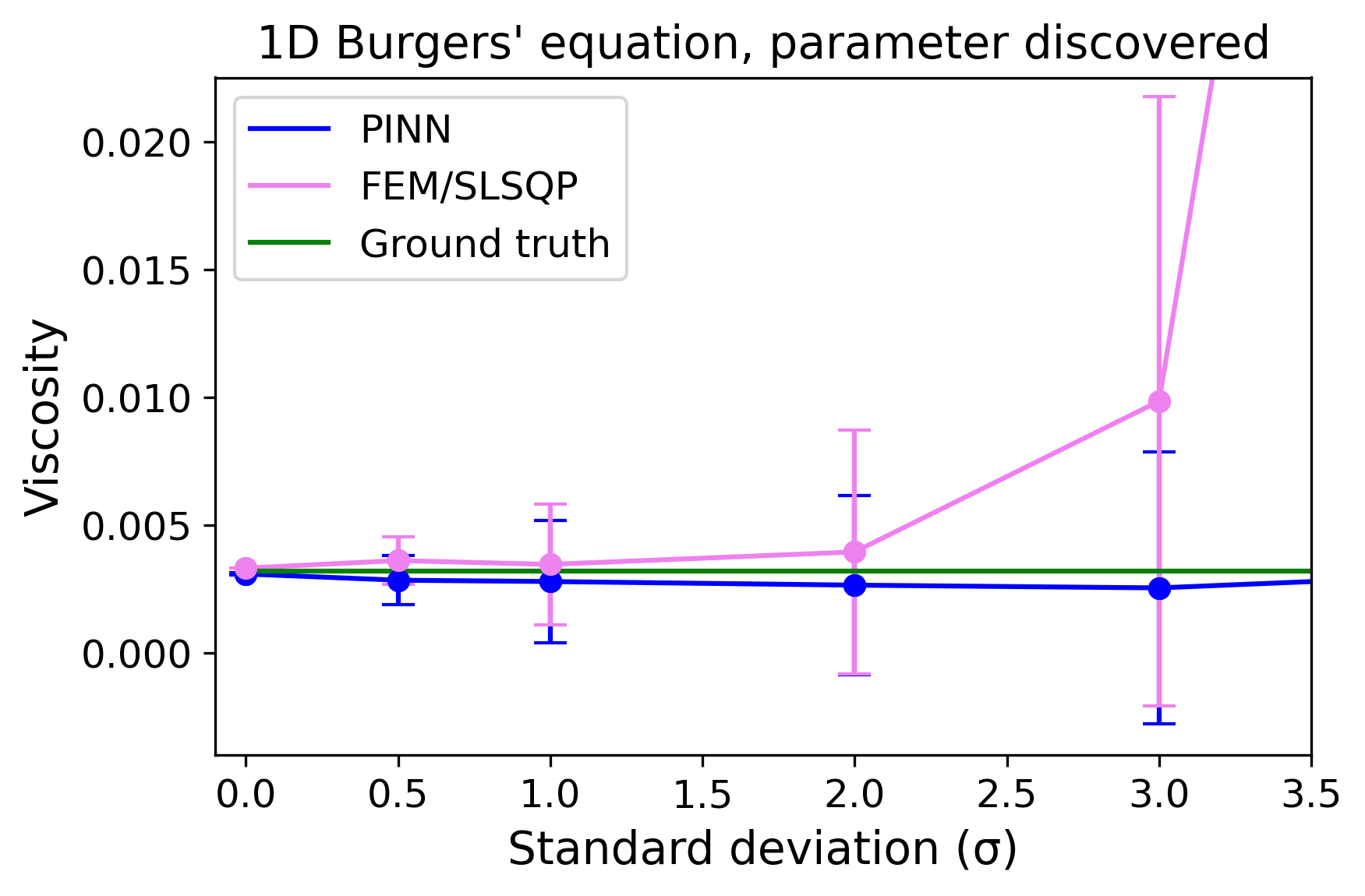}
\includegraphics[width=0.49\textwidth]{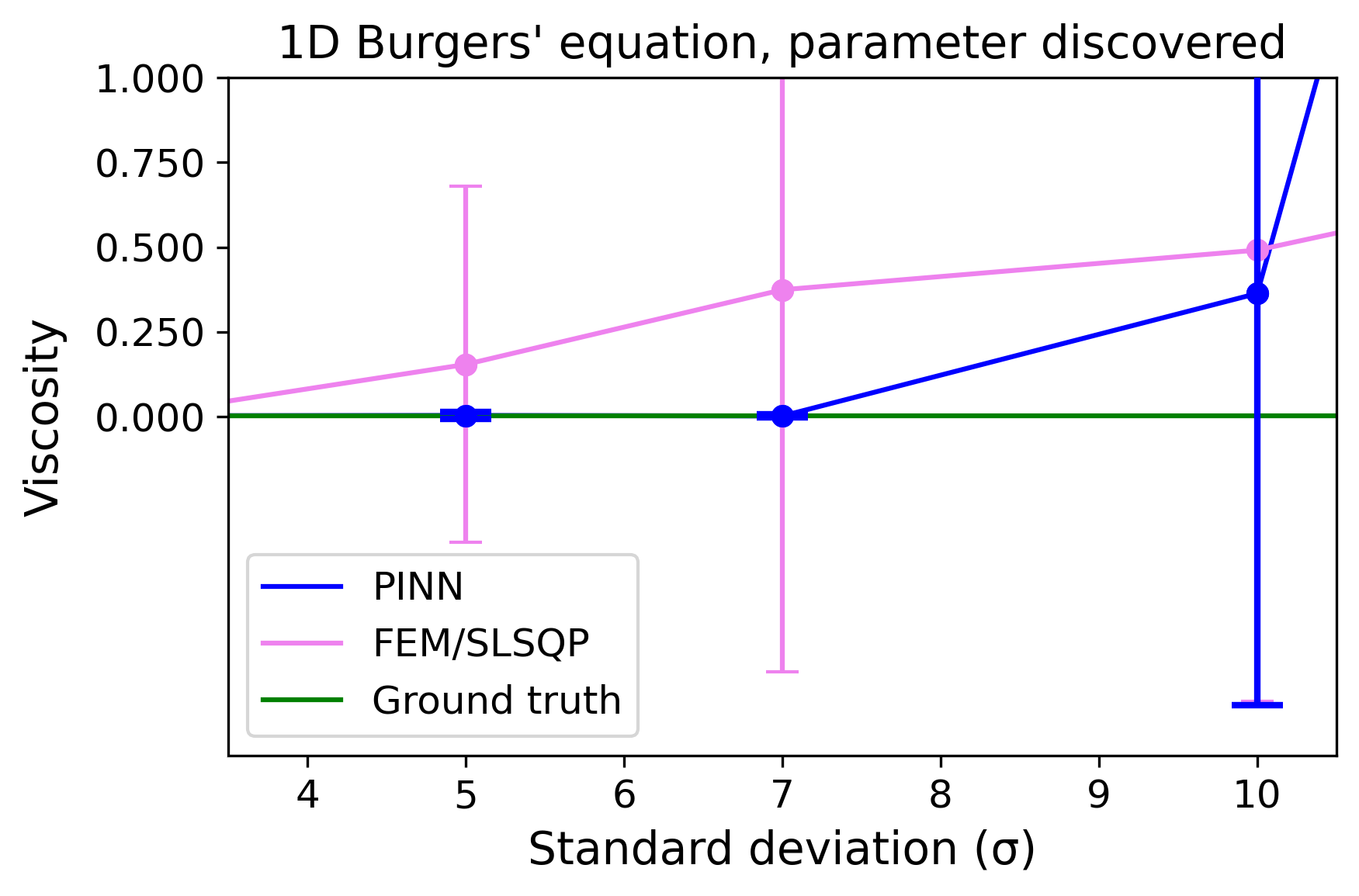}
\caption{Results for 1D Burgers' equation on test set based on 30 runs of each model. Standard deviation ($\sigma$) is the noise added to the training data. Top left: Prediction accuracies, $\sigma \in [0, 2]$. Top right: Prediction accuracies, $\sigma \in [3, 25]$. Bottom left: Parameter predictions, $\sigma \in [0, 3]$. Bottom right: Parameter predictions, $\sigma \in [5, 10]$.}\label{fig:burgers1}
\end{figure}

\begin{figure}[!ht]
\centering
\includegraphics[width=0.49\textwidth]{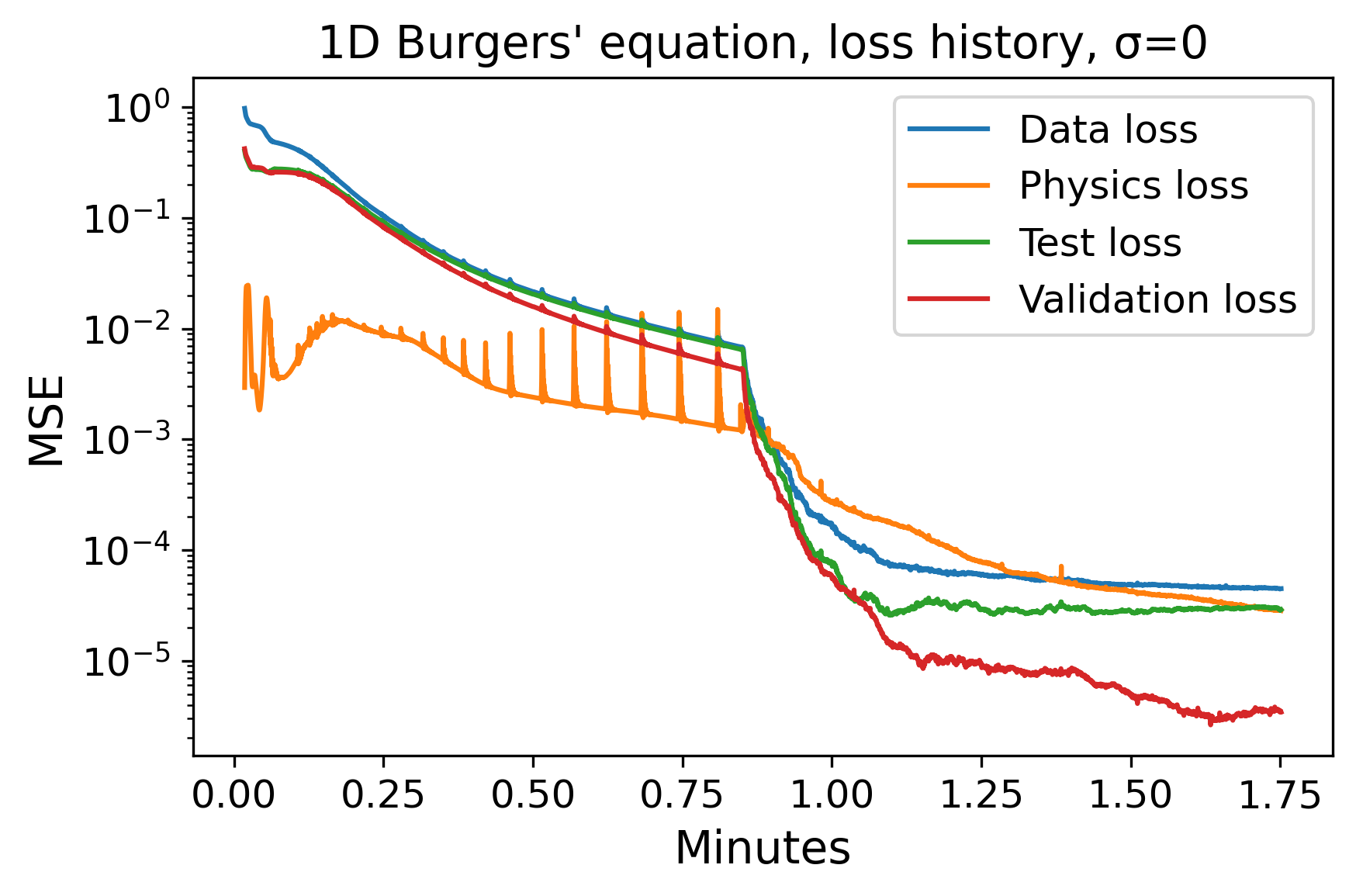}
\includegraphics[width=0.49\textwidth]{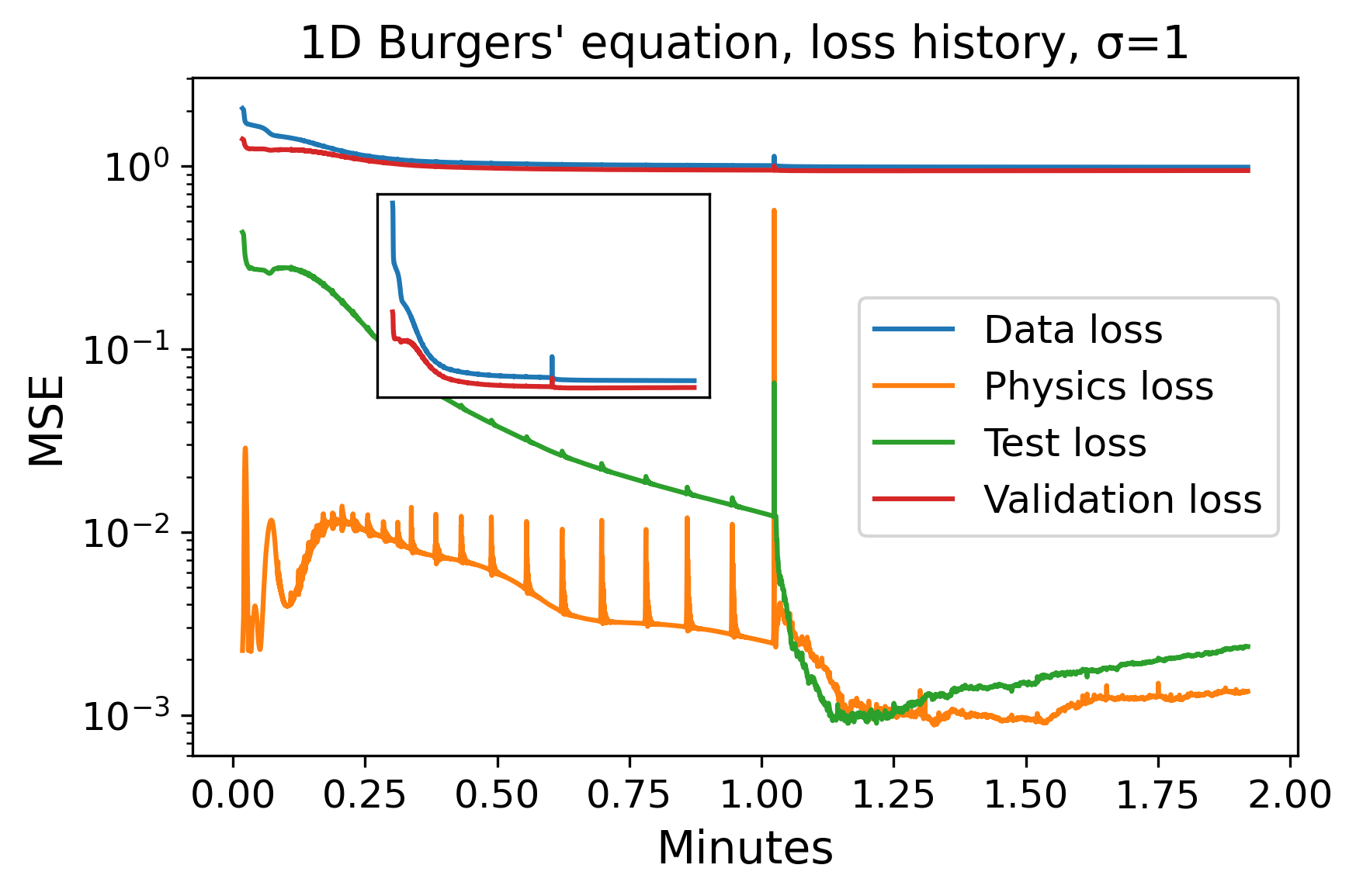}
\includegraphics[width=0.49\textwidth]{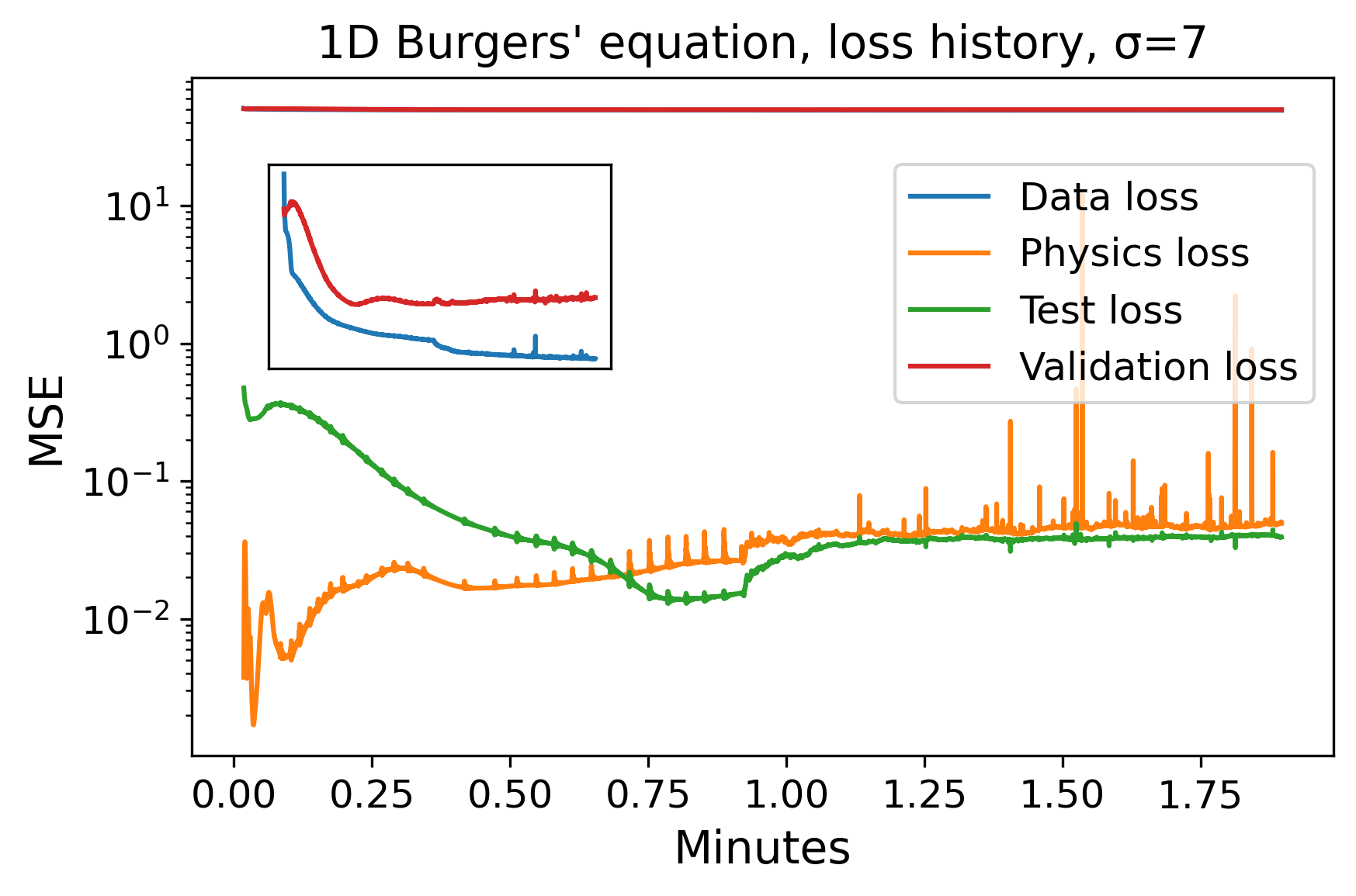}
\includegraphics[width=0.49\textwidth]{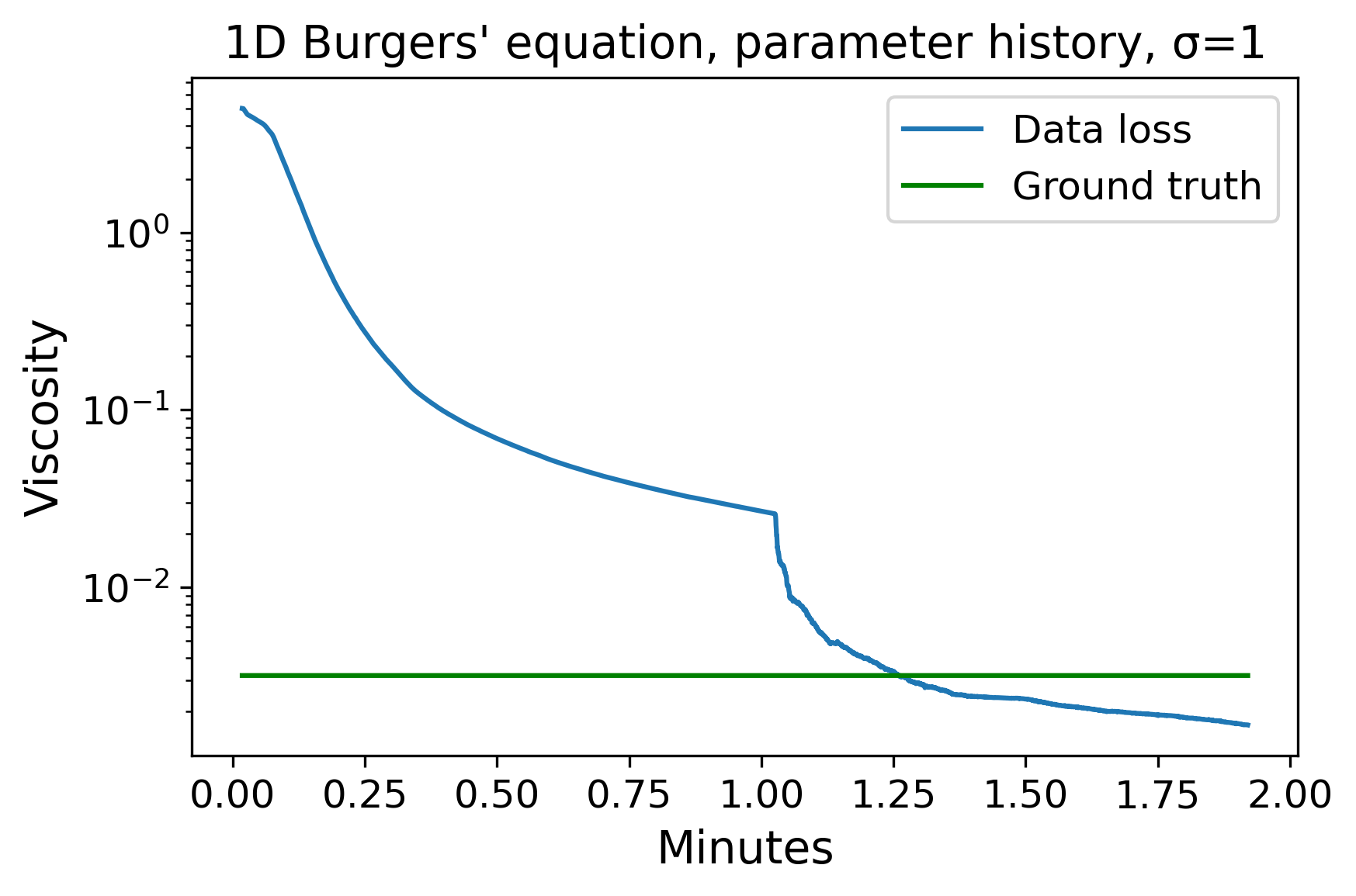}
\caption{1D Burgers' equation loss history from training and parameter history from training. Note that the validation data, like the training data, has noise added to it which can explain the similar performance of their losses on high noise levels. The test loss only evaluates predictive accuracy, not agreement with differential equation. Top left: Loss history during PINN training, $\sigma=0$. Top right: Loss history for $\sigma=1$. Bottom left: Loss history for $\sigma=7$. Bottom right: Parameter learning history for $\sigma=1$. All the results of the losses exclude the weights used for loss balancing.}\label{fig:burgers2}
\end{figure}

Figure \ref{fig:burgers1} shows the results of the experiments. By inspecting the tables in \ref{1dtests}, we see that for $\sigma=0$, the PINN model has the lowest RMSE accuracy of the three models tested. However, the PINN model takes 110 seconds on average to run using GPUs while our not optimized FEM/SLSQP model takes 25 seconds on a single core CPU. The peak memory use of both models are similar, around 500 MB. As the noise increases, the variation in prediction accuracy for both models begins to increase as well. In particular, PINN tends to predict very high parameter values at higher frequency when the noise increases. As can be seen in \ref{1dparametertest}, for PINN the skewness of the prediction distribution increase in a right-leaning direction. Without noise, all models perform well with RMSE of the order of $10^{-3}$.

Despite many of our tests being on extreme levels of noise, we observe that FEM/SLSQP significantly outperforms the classical PINN at all noise levels. However, the results of FEM/SLSQP and the PINN/FEM model cannot be distinguished with statistical significance, aside from at $\sigma=0.05$ where FEM/SLSQP is highly likely to have lower error than PINN/FEM.

PINN exhibits a small negative bias for all noise levels, including $\sigma=0$, as seen in \ref{1dparametertest}. However, the bias is only statistically significant for noise levels 0, 3, 5, 7, 25. FEM/SLSQP has a positive bias, but these results are only significant for the highest noise levels. With $\sigma=25$, which can be seen in \ref{addgraph}, PINNs have the highest mean estimates. (However, PINN also has a very large variation.)

Figure \ref{fig:burgers2} shows the loss history during training. The observed history for $\sigma = 0$ is typical for the 1D Burgers' equation tests without noise. While running the Adam optimizer, we see characteristic spikes in both the data loss and physics loss as the model tries to balance them. When L-BFGS is turned on (after slightly less than 1 minute), we get a mostly smooth convergence towards a local minimum as would be expected from this optimizer.

With both $\sigma = 1$ and $\sigma = 7$, the data loss is largely monotonic. The physics loss develops in a more complicated way with more erratic spikes. For $\sigma=7$, the physics loss makes a jump around 0.9 minutes. After the test loss has reached its lowest point, the model ends up converging to a worse state. However, there is no way to tell from the validation loss when this lowest point happened, which supports our choice to not use early stopping for this problem. The remaining loss histories from our experiments not included here show similar behavior in the presence of noise.

The minimum test loss approximately occurs when the parameter estimates are closest to the ground truth. After, PINN tends to estimate a lower viscosity on average and converge on a worse total loss. This is also supported by the previously mentioned statistical tests in \ref{1dparametertest}, which show that PINN tends to have a slight negative bias. This behavior is prominent in the example loss history with $\sigma = 1$ and $\sigma = 7$. For $\sigma=0$, this can only be weakly seen. However, as more noise is added to the training data, the model becomes increasingly likely to estimate the viscosity too high.

As described in \ref{twoparamet}, we also tested 1D Burgers' equation with two unknown parameters. As we can see in \ref{1drmsetwoparameter}, the models had similar performance as with one unknown parameter with slightly worse overall performance. Some of the issues that we shall later see in the other experiments with more complex systems are more pronounced in the problem with two parameters, such as large physics losses. Solving for two parameters produces effects similar to increasing system complexity in other experiments.

\subsection{2D Taylor-Green Vortex}

For the 2D Navier-Stokes experiment, the Taylor-Green vortex problem was selected. This problem is defined by the initial conditions:

\begin{equation}
\begin{split}
u(x, y, 0) = \sin{x}\cos{y},\\
v(x, y, 0) = -\cos{x}\sin{y},\\
p(x, y, 0) = \frac{\rho}{4}\left( \cos{2x} + \cos{2y} \right).
\end{split}
\end{equation}

\noindent
The ground truth viscosity term to be solved for was set to $\nu = 0.1$ while $\rho = 1$ is known. The variables were restricted to $t \in [0, 2.5]$, $x \in [0, 2\pi]$ and $y \in [0, 2\pi]$. This problem roughly corresponds to $Re=100$ which is well within laminar flow.  We assume $\bm q_{2D}(x, y,t)$ is a state vector containing velocities $u$, $v$ and the pressure $p$ at any given position and time in the 2D domain. The following periodic boundary conditions were used:

\begin{equation}
\begin{split}
\bm q_{2D}(2\pi, y, t) = \bm q_{2D}(0, y,t ), \\
\bm q_{2D}(x, 2\pi, t) = \bm q_{2D}(x, 0, t).
\end{split}
\end{equation}

\noindent
The analytical solution for this system is as follows:

\begin{equation}
\begin{split}
u(x, y, t) &= e^{-2 \nu t} \sin{x}\cos{y},\\
v(x, y, t) &= -e^{-2 \nu t} \cos{x}\sin{y},\\
p(x, y, t) &= e^{-4 \nu t}\frac{\rho}{4}\left( \cos{2x} + \cos{2y} \right).
\end{split}
\end{equation}

\noindent
Intuitively, the system can be described as a set of four evenly spaced whirls with velocity dissipating according to the value of viscosity.

\subsubsection{Specific experimental setup}
\label{subsec:2dspec}

Following \citet{raissi2019physics}, we trained the PINN model $\bm a(\bm x;\bm \theta)$ to learn the stream function, $\bm \psi(x, y, t;\bm \theta)$, along with the pressure field $p(x, y, t;\bm \theta)$:

\begin{equation}
    \bm a(\bm x; \bm \theta) =
\begin{bmatrix}
\bm \psi(x, y, t; \bm \theta) \\
p(x, y, t; \bm \theta)
\end{bmatrix}.
\end{equation}

\noindent
Instead of having to learn the velocity for each dimension $x$ and $y$, the network only needs to learn a single stream output for all dimensions. This is achieved using the identities $\frac{\partial \bm \psi(x, y, t)}{dy} = u(x, y, t)$ and $\frac{\partial \bm \psi(x, y, t)}{dx} = -v(x,y, t)$. The $\bm \psi(\bm x)$ formulation assumes incompressibility, so we only need to impose the momentum equation to get the correct physics with this approach.

\begin{figure}[ht!]
\centering
\includegraphics[width=0.49\textwidth]{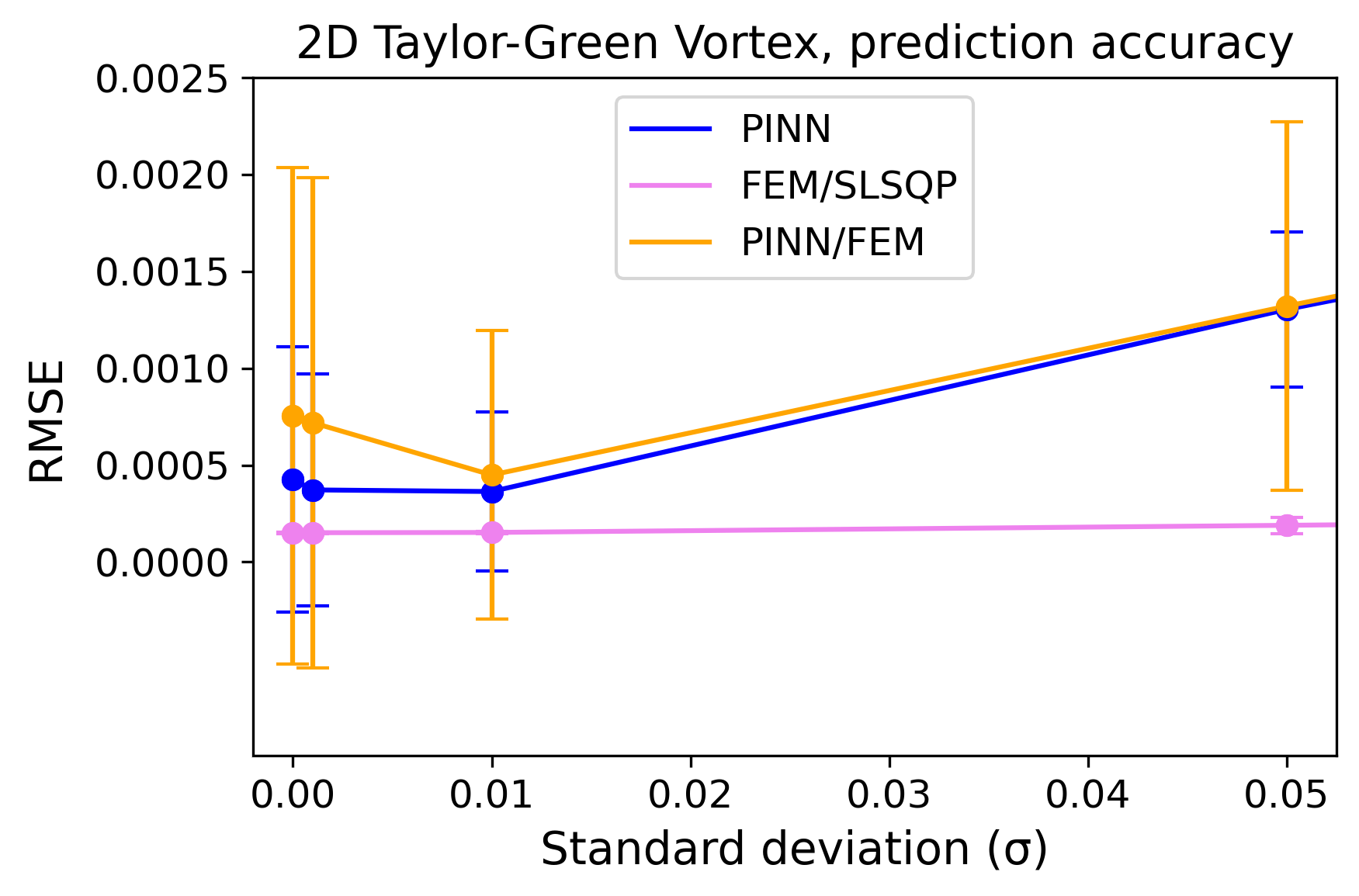}
\includegraphics[width=0.49\textwidth]{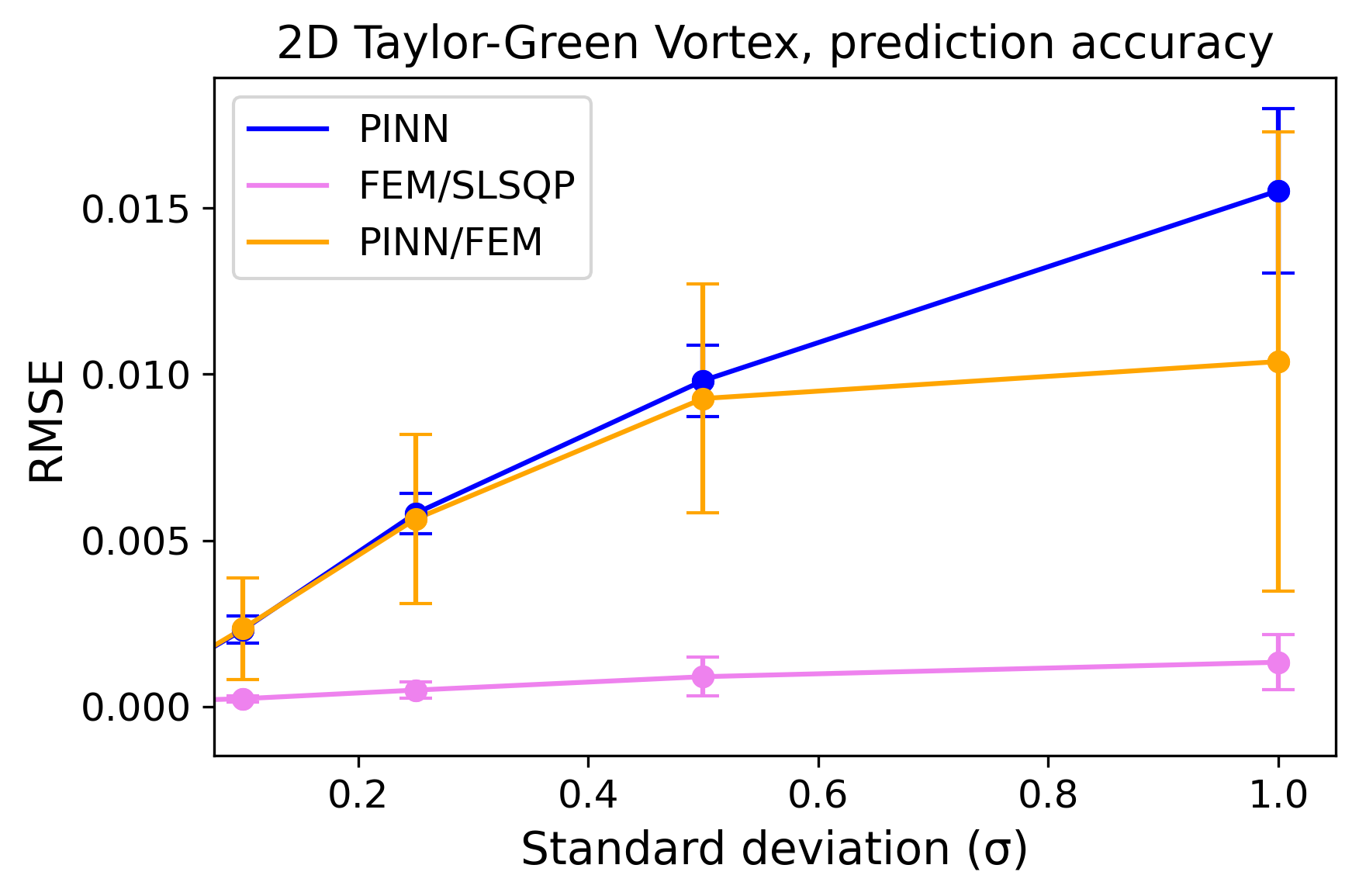}
\includegraphics[width=0.49\textwidth]{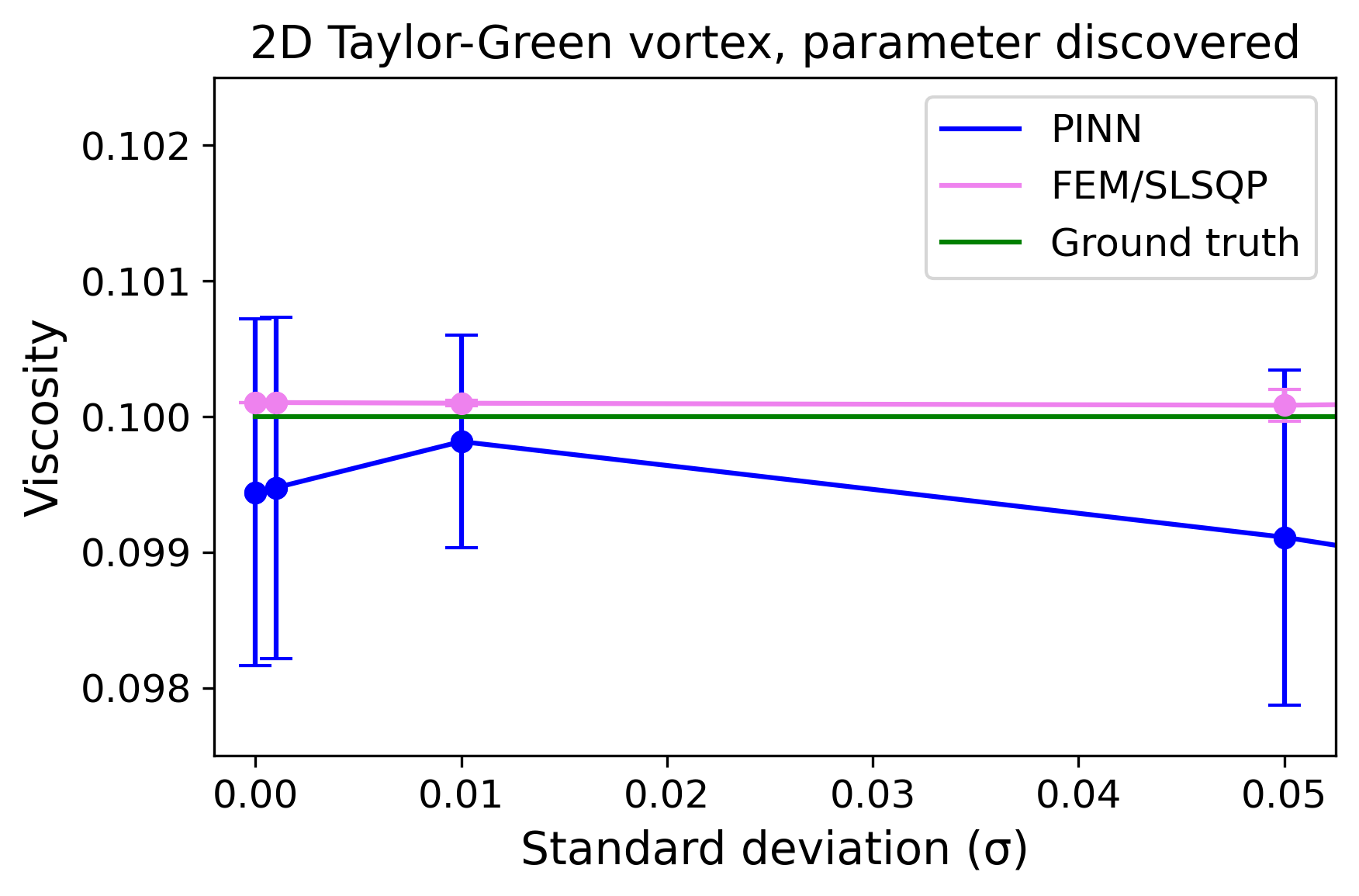}
\includegraphics[width=0.49\textwidth]{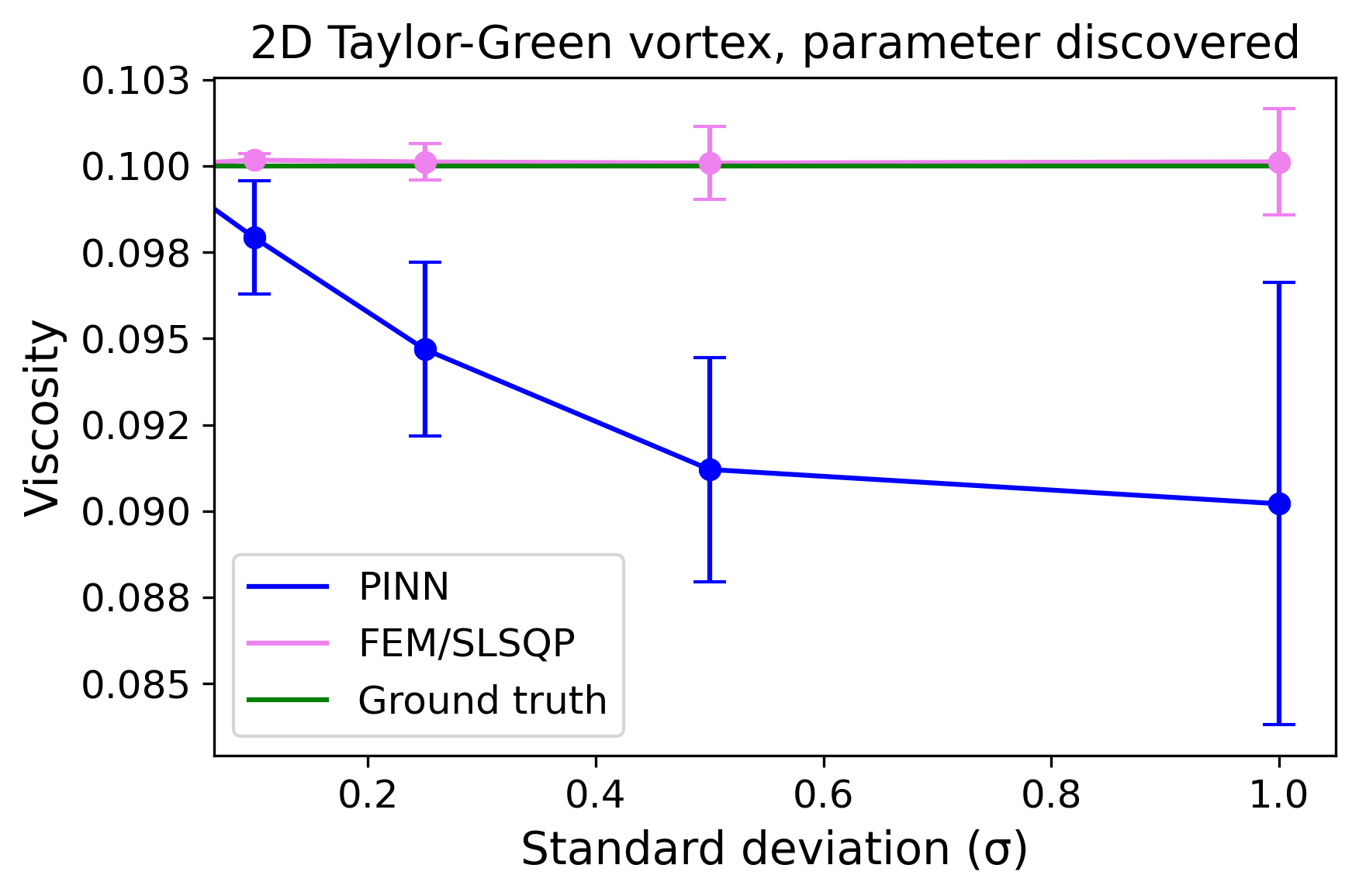}
\caption{Results for 2D Taylor-Green vortex on test set based on 30 runs of each model. Standard deviation ($\sigma$) is the noise added to the training data. Top left: Prediction accuracies, $\sigma \in [0, 0.05]$. Top right: Prediction accuracies, $\sigma \in [0.1, 1]$. Bottom left: Parameter accuracies, $\sigma \in [0, 0.05]$. Bottom right: Parameter accuracies, $\sigma \in [0.1, 1]$.} \label{fig:green1}
\end{figure}

The total dataset consisted of $412 \ 776$ observations uniformly sampled in space and time from the analytical solution. We used 24\% for training, 4.8\% for validation and imitated hyperparameters used for successful training on 2D Navier-Stokes in the literature \cite{raissi2019physics}. To consistently get acceptable performance, we had to use a gradient balancing algorithm on the weights of the losses \cite{wang2020understandingmitigatinggradientpathologies}. Our implementation is standard, but more information can be found in \ref{adaptinfo}. We trained 30 random initializations and present the mean and standard deviations of the results. The noise levels tested for the inverse problem, both in $\sigma$ and SNR, are given in \ref{noiselevels2d}. Remaining exact implementation details can be found in \ref{2dapp}. 

\subsubsection{Results, 2D Taylor-Green Vortex}

\begin{figure}[ht!]
\centering
\includegraphics[width=0.49\textwidth]{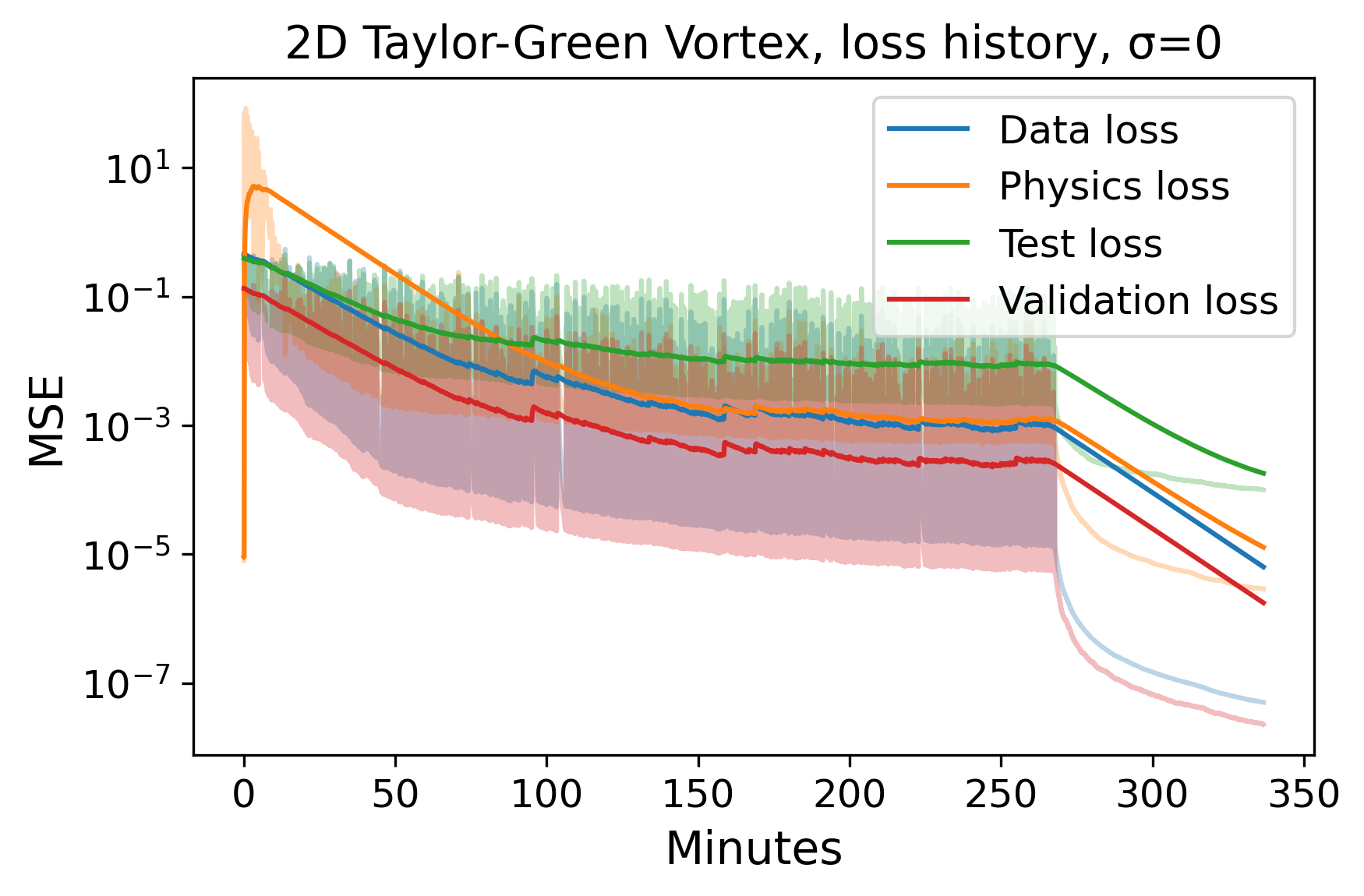}
\includegraphics[width=0.49\textwidth]{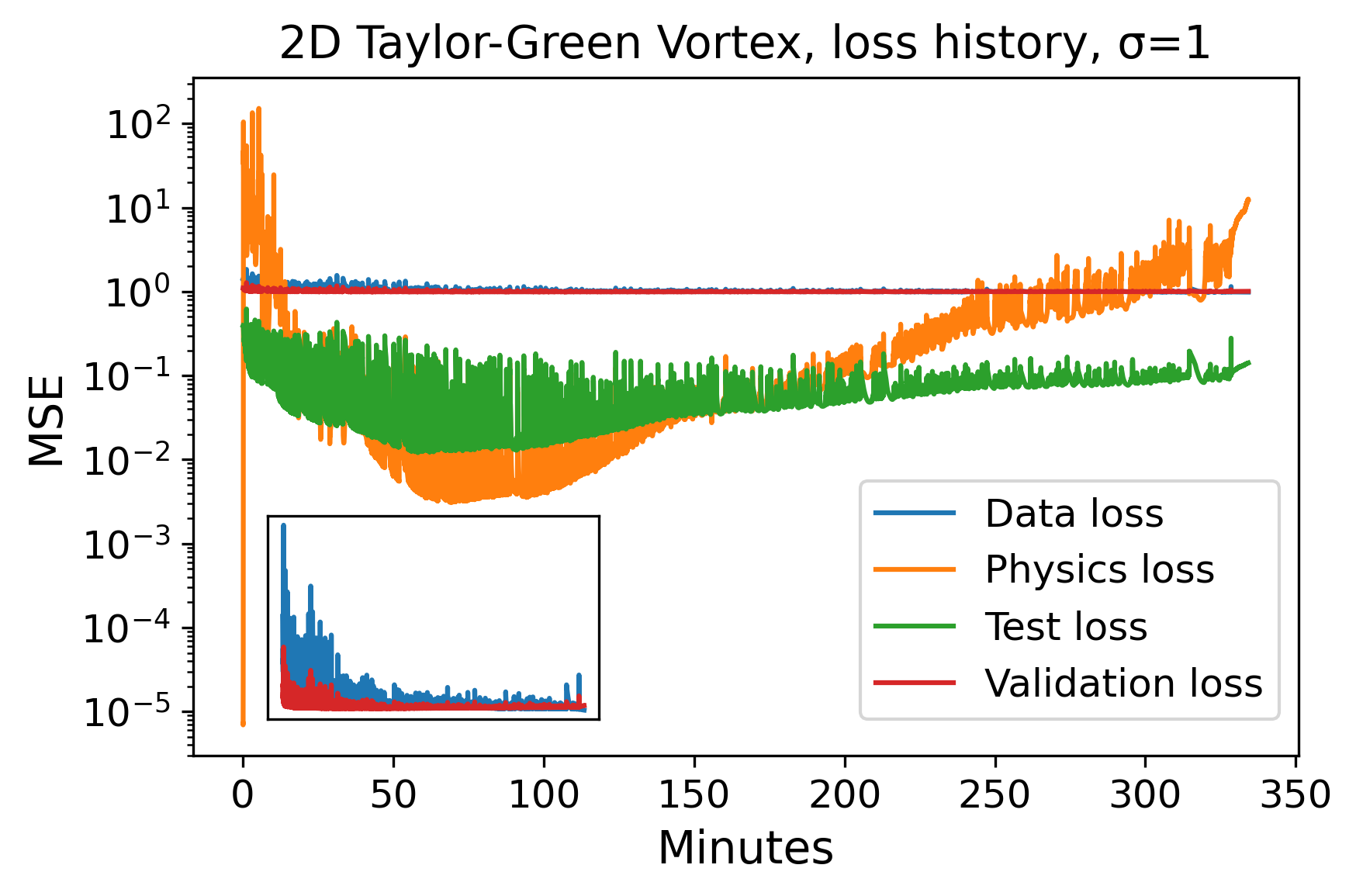}
\includegraphics[width=0.49\textwidth]{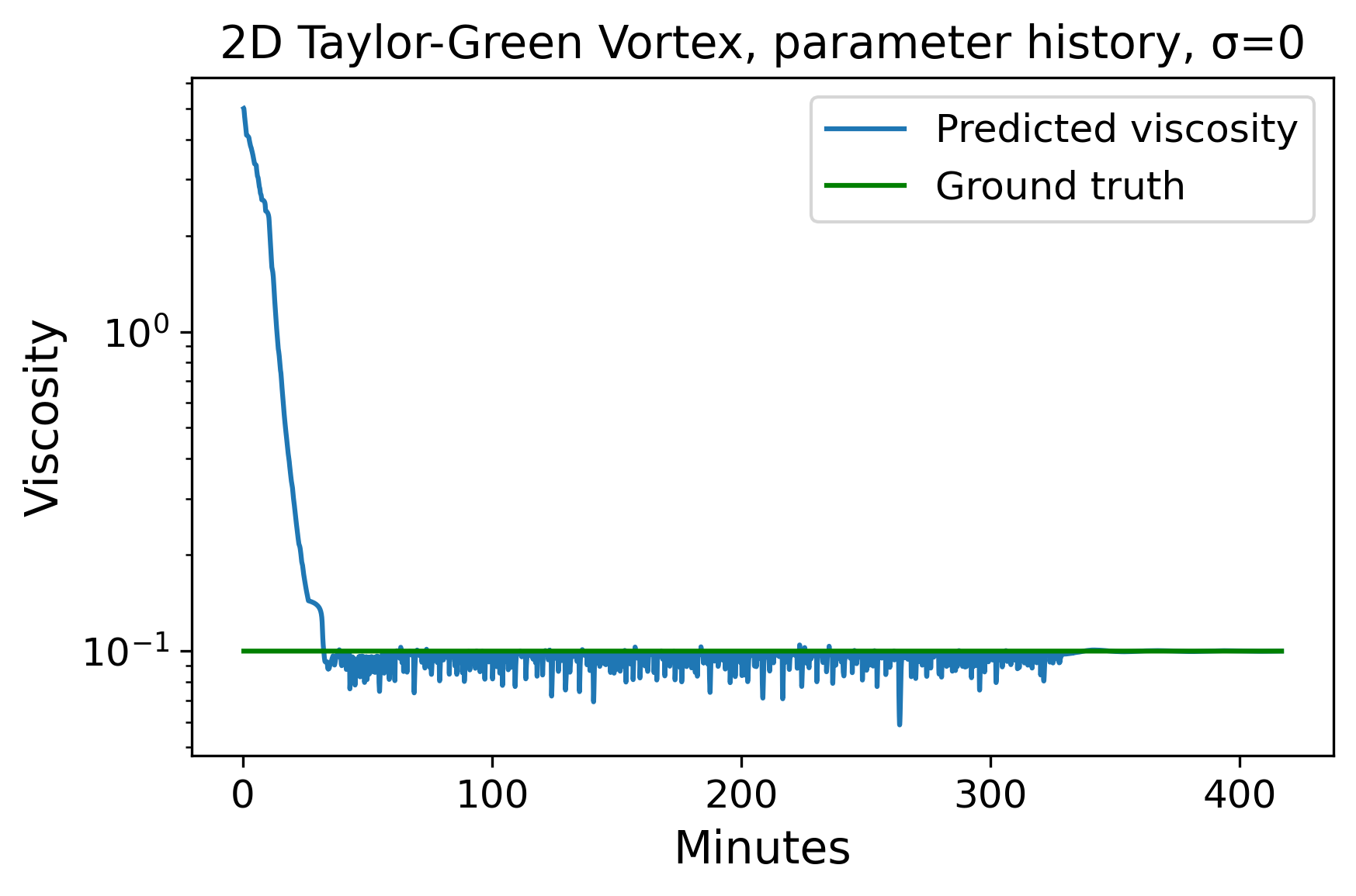}
\includegraphics[width=0.49\textwidth]{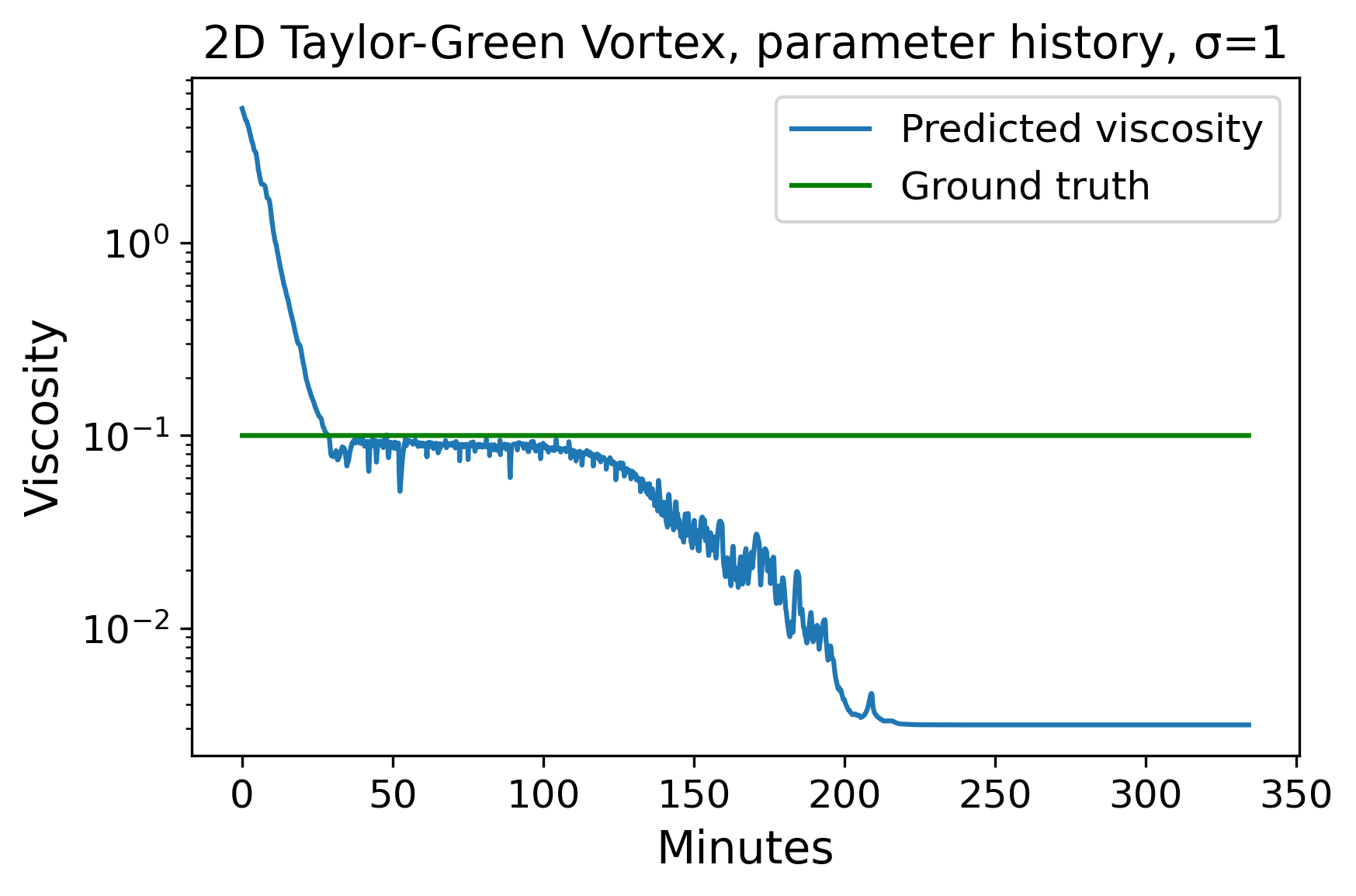}
\caption{2D Taylor-Green vortex training history. Note that noise is added to the validation data which may explain poor validation loss performance. The test loss only evaluates predictive accuracy, not agreement with differential equation. Top left: Loss history during PINN training, $\sigma=0$. Top right: Loss history for $\sigma=1$. Bottom left: Parameter learning history, $\sigma=0$. Bottom right: Parameter learning history, $\sigma=1$. All the results of the losses exclude the weights used for loss balancing.}\label{fig:green2}
\end{figure}

\autoref{fig:green1} shows that FEM/SLSQP on average outperforms the other models for 2D Taylor-Green Vortex on all tested noise levels. Where the distributions overlap, we have confirmed that FEM/SLSQP is superior with statistical significance in \ref{2dstat_test}. On the other hand, PINN takes 6.6 hours on average while FEM takes 11.2 hours on average. PINN also had lower peak memory use than the baseline. We used 10 cores running in parallel for FEM/SLSQP and each individual core required 966 MBs at most. For PINN we only used one core, with a peak memory use of around 1088 MB.

We can see that the parameter estimation for 2D Taylor-Green Vortex has larger variation than for 1D Burgers' equation on the noise levels for which we did experiments on both systems. As with RMSE predictions, FEM/SLSQP outperforms PINN at predicting the correct unknown parameter. As opposed to 1D Burgers' equation, the skewness of the estimated parameter distribution leans left rather than right for PINN, and does not grow as evenly with the noise.

As with 1D Burgers' equation, PINN tends to have a small negative bias on parameter estimation for 2D Taylor-Green Vortex. The bias is statistically significant from noise level $\sigma=0.05$ to $\sigma=1$. In contrast with the results for 1D Burgers' equation, the mean of the predicted parameter continues to decrease with noise rather than to eventually blow up. FEM/SLSQP has a small positive bias on lower noise levels. Interestingly, PINN has the best performance at $\sigma=0.01$. However, the advantage is slight and, for this noise level, PINN also has a larger range in accuracy than for lower noise levels.

In \autoref{fig:green2}, the loss histories fluctuate even more than they did for 1D Burgers' equation. For visibility at $\sigma = 0$, we have applied weighted average exponential smoothing with $\alpha=0.001$ \cite{hyndman2021chapter8_1}. On lower noise levels, the loss history resembles the behavior of 1D Burgers' equation: After switching to L-BFGS, we often see a smooth convergence towards a better model. At lower noise levels, PINNs rapidly converge near ground truth but show little subsequent improvement.

As noise increases, a different phenomenon becomes increasingly prevalent. While the PINN model first tends to balance physics and data training well, at some point it starts to disregard the physics loss. We can see for $\sigma = 1$ in \autoref{fig:green2} that the physics loss, which is the mean square average over the data, approaches $10^1$. In the parameter estimation histories for higher numbered $\sigma$, it often stays near the ground truth for a while before training starts to fail. The best models are always from before this transition.

\subsection{3D Taylor-Green Vortex}

For the implementation of 3D Navier-Stokes, the Taylor-Green vortex problem was used. In 3D, it is defined by the initial conditions:

\begin{equation}
\begin{split}
u(x, y, z, 0) &= \sin{x}\cos{y}\cos{z},\\
v(x, y, z, 0) &= -\cos{x}\sin{y}\cos{z},\\
w(x, y, z, 0) &= 0,\\
p(x, y, z, 0) &= 0.
\end{split}
\end{equation}

\noindent
The ground truth viscosity term was chosen to be $\nu = 0.01$. This roughly corresponds to the Reynolds number 1000, which for the Taylor-Green Vortex is starting to become turbulent. We wanted to increase the complexity of the problem as well as add a dimension to get more variety in the test systems. The variables were restricted to $t \in [0, 2.5]$, $x \in [-\pi, \pi]$, $y \in [-\pi, \pi]$ and $z \in [0, 2\pi]$. We assume $\bm q_{3D}(x, y, z, t)$ is a state vector containing velocities $u$, $v$, $w$ and the pressure $p$ at any given position and time in the 3D domain. The following periodic boundary conditions were used:

\begin{equation}
\begin{split}
\bm q_{3D}(-\pi, y, z, t) = \bm q_{3D}(\pi, y, z, t), \\
\bm q_{3D}(x, -\pi, z, t) = \bm q_{3D}(x, \pi, z, t), \\
\bm q_{3D}(x, y, 0, t) = \bm q_{3D}(x, y, 2\pi, t).
\end{split}
\end{equation}

\noindent
Intuitively, 3D Navier-Stokes consists of layers of the 2D version of the problem where the direction of the flow is opposite on regular intervals depending on the z-coordinate. Due to small turbulence, velocity starts to build and creates some velocity in the z direction making the layers of whirls interact with each other chaotically. There is no analytical solution for the chosen parameters.

\subsubsection{Specific experimental setup}

3D Navier-Stokes cannot be reduced to a single stream function, and alternative formulations that decrease the number of required functions are non-trivial. For this reason, we trained the PINN model to learn the individual velocities as outputs. The total dataset consisted of $1\,100\,000$ observations randomly sampled from the data we generated using Nektar++ as described in Section \ref{sec:general}. We used 25\% as training data and 5\% as validation data which is similar as what we used for the 2D problem. The proportion of training and validation data was selected through manual experimentation to find what was required to reliably solve the problem with PINN at $\sigma=0$. Here as well, we used a gradient balancing algorithm for the loss weights \cite{wang2020understandingmitigatinggradientpathologies}. 30 random initializations were trained of each model and we present the mean and standard deviations of the results. The noise levels tested for the inverse problem, both in $\sigma$ and SNR, are given in \ref{noiselevels3d}. Other implementation details can be found in \ref{3dapp}.

\subsubsection{Results, 3D Taylor-Green Vortex}
\label{sec:3dresult}

\begin{figure}[!ht!]
\centering
\includegraphics[width=0.49\textwidth]{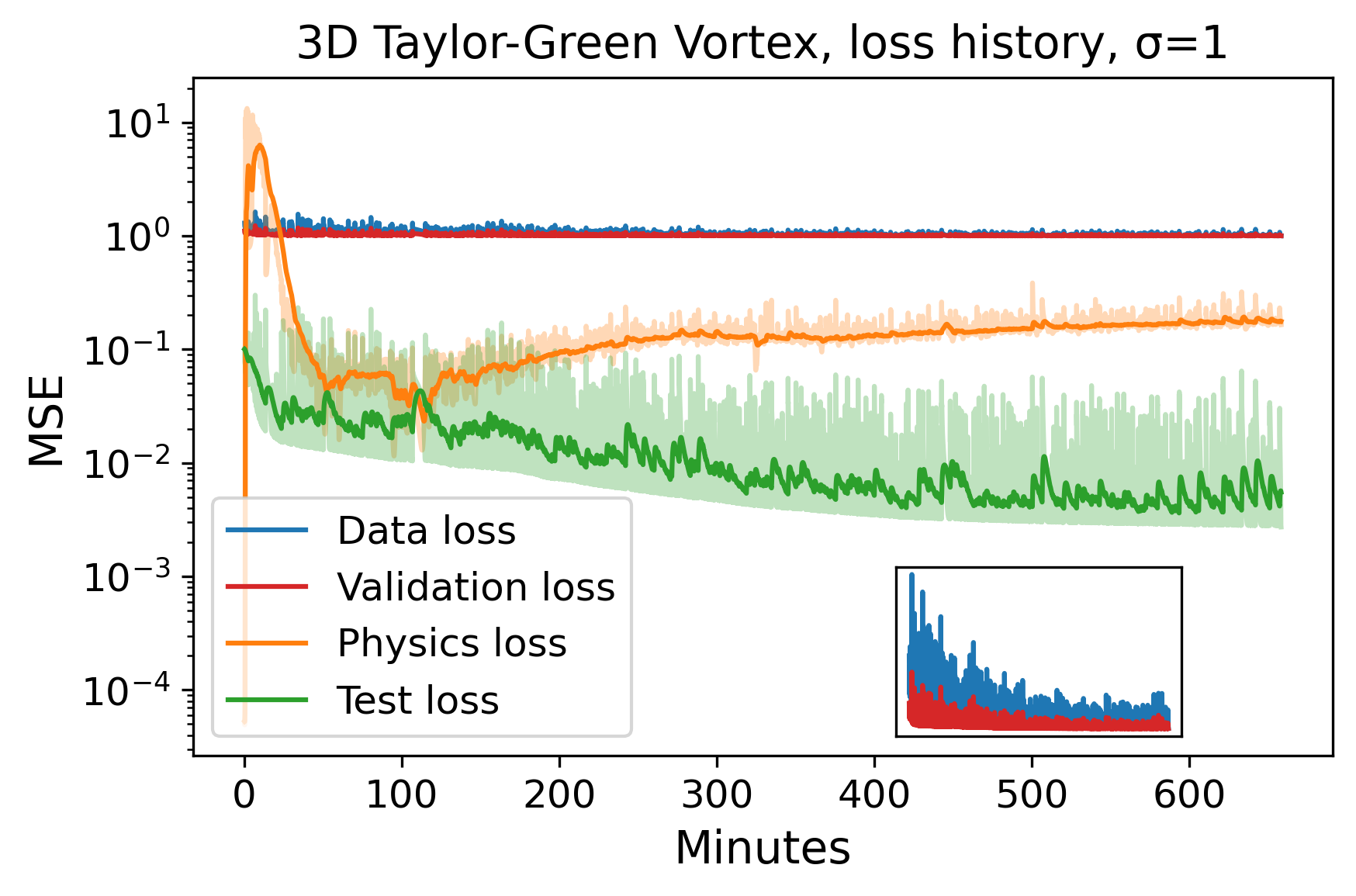}
\includegraphics[width=0.49\textwidth]{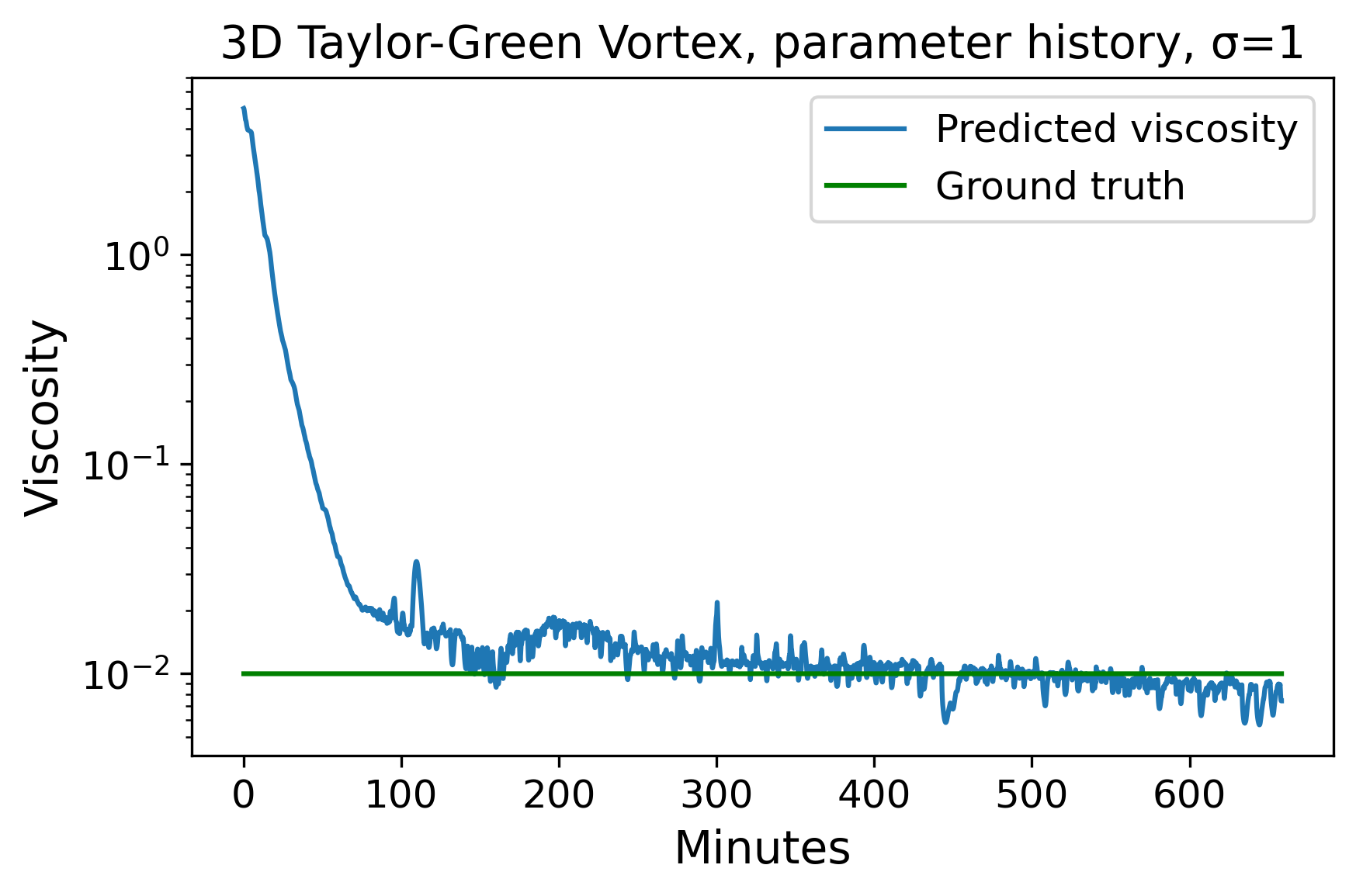}
\caption{3D Taylor-Green vortex training histories. Note that the validation data may contain noise which can explain poor validation loss performance. The test loss only evaluates predictive accuracy, not agreement with differential equation. Left: Loss history for PINN, $\sigma=1$. Weighted average exponential smoothing with $\alpha=0.001$ was used \cite{hyndman2021chapter8_1}. Right: Parameter learning history, $\sigma=1$. All the results of the losses exclude the weights used for loss balancing.}\label{fig:green4}
\end{figure}

For the 3D experiments, both PINN and FEM/SLSQP sometimes catastrophically fail. Both PINN and FEM/SLSQP occasionally get stuck on or near the initial guess. Note that these failures, in a practical setting, could easily be identified and excluded. We define outliers using interquartile range (IQR) \cite{walpole2012probability}. Values below $Q_1 - 1.5 \cdot IQR$ and values above $Q_3 + 1.5 \cdot IQR$ were identified as outliers. For FEM/SLSQP these failures account for about 20\% of the runs. For PINN, it is less than 1\%. However, in more than 25\% of runs, PINN is stuck on the lowest parameter allowed. This is not captured by our outlier definition and relative to the task it is a decent estimation, so we have chosen not to consider these results outliers.

No negative bias can be proven for PINN while FEM has a small positive bias on several noise levels. When we include the outlier runs, the standard deviations of FEM are also very large. For instance, it is 0.38 for $\sigma=0.05$. According to our statistical tests in \ref{3dwithoutlierstests}, FEM still outperforms both PINN and PINN/FEM with statistical significance. When we exclude the outliers for FEM, the same effect becomes larger. With outliers excluded, FEM/SLSQP's parameter identification accuracy also becomes similar as in 2D. However, it must be noted that the baseline's failure rate is very high. Even though FEM is very likely to outperform PINN in most cases, it is not straightforward to claim that the baseline is better than PINN.

For 3D Taylor-Green Vortex, FEM/SLSQP is a lot more expensive to run than PINN. On average, FEM/SLSQP takes 19.9 hours to solve the inverse problem. PINN takes 10.8 hours on average. The memory costs of FEM/SLSQP are also far greater. PINN's peak memory use is similar to as for 2D Taylor-Green Vortex, about 1100 MB. FEM/SLSQP has an average peak memory of almost 2000 MB for each of the 55 cores we used. The baseline does not scale well and requires HPC resources for 3D.

Even with outliers removed, all the models generally perform worse than on the 2D problem. This result is expected since by adding a dimension with equal size as the others, the domain becomes much larger. The lower ground truth viscosity problem also increases the Reynolds number, making the system harder, potentially leading to small amounts of chaos as well.

Comparing the loss and parameter estimation histories in \autoref{fig:green4}, we observe the same phenomena as in 2D Taylor-Green Vortex. In the most successful runs, we may get a smooth improvement when L-BFGS is applied in the end. In the less successful runs, the physics loss tends to blow up. In the 2D case, these phenomena seem tied to the level of noise with the physics loss exploding becoming increasingly likely with higher levels of noise. For 3D, the phenomena seem to be independent of the noise.

\section{Discussion}
\label{sec:discussion}

\subsection{1D Burgers' equation}
\label{sec:burgdisc}

The results from the 1D Burgers' equation experiments raise an interesting question. As seen in \autoref{fig:burgers2}, for $\sigma = 1$, the PINN test loss exhibits typical overfitting behavior after around 1.25 minutes. Since we have much of the differential equation and noisy data, we could naively expect that minimizing both losses, would always approximate the solution better. The physics equation should prohibit the model from overfitting on the noise.

One contribution to this failure may be the tendency of PINNs to estimate the viscosity as lower than the ground truth. As the physics loss gets sufficiently small, PINNs may eventually decrease the loss significantly by making the unknown viscosity smaller. When the differential equation thus becomes incorrect, it is less compatible with the data. The model may not be able to improve anymore, stuck in a local minima, and it will converge on a worse solution than earlier in training. This is similar to the "reward hacking" problem common in reinforcement learning where the model finds a way to optimize its loss which does not serve the purpose of training \cite{rlblogpost}. As seen in \ref{1dparametertest}, PINN often has a negative bias.

While test loss identifies the optimal model, using only training and validation losses is insufficient. Developing a method to detect optimal stopping points from these losses alone would enable early stopping tailored to PINNs in noisy inverse problems. Since the validation loss only tells us how the model is doing on the forward problem, it is not always indicative of the inverse problem. In many cases already, understanding and accommodating the difference of PINNs from ordinary machine learning has helped develop the competitiveness of PINNs in other aspects \cite{wang2020understandingmitigatinggradientpathologies, taylor2022optimizingoptimizerdatadriven}.

Even if the FEM/SLSQP baseline is not significantly more accurate than PINN/FEM on the 1D Burgers' equation, it is certainly faster. For some applications this difference may not make a practical difference. At the same time FEM/SLSQP is considerably more difficult to set up and requires more specialized knowledge. We need a weak form of our PDE, the grid must be specified and the matrix solver chosen. In many real-life settings, PINNs may be more appropriate given the existence of user-friendly ML packages.

\subsection{2D Taylor-Green Vortex}

We noted that for this problem PINN was faster than the FEM/SLSQP baseline. As discussed in Section \ref{sec:general}, a careful comparison of computational cost is outside of our scope. A significant part of the cost of FEM/SLSQP consists in calculating the points of the training set to compare against the solution. The large amount of training data may be biased towards PINN in that FEM/SLSQP may not need as much training data as a neural network to achieve equal results. This is a typical challenge in comparing traditional methods with machine learning methods: Both training on different data sets and training on the same data set may be considered unfair.

Shorter training times may also be achievable for PINN. With $\sigma=0$, the model finds parameter values close to the ground truth in less than an hour, if we properly know when to stop training. With the applied gradient weighting algorithm, PINN may require less epochs than presented in the literature.  This supports our claim in Section \ref{sec:burgdisc} that we may be able to improve PINNs if we develop appropriate early stopping strategies.

For 2D Navier-Stokes, PINN has the same tendency to estimate the viscosity as lower than the ground truth as we observed with 1D Burgers' equation. An example of this can be seen in the loss histories in \autoref{fig:green2}. This bias is also confirmed by statistical tests found in \ref{2dstat_test}. However, with higher noise a different behavior begins to dominate. Increasingly with noise, PINNs will at some point during training disregard the physics loss and let it grow by a few magnitudes. For higher noise levels, such as $\sigma = 7$ for Burger's equation in Section \ref{fig:burgers2}, we can already see some of this tendency although less prominent. We also see this more frequently on lower noise levels in the 1D Burgers' equation tests with two unknown parameters. However, we hypothesize this phenomena is exacerbated by the adaptive weight algorithm we used for the Navier-Stokes problems \cite{wang2020understandingmitigatinggradientpathologies}. The weight on the data loss, calculated by dividing the maximum gradient of the physics loss by the average gradient of the data loss, may make it more profitable to overfit on the noise than to consider the physics. As the physics become incompatible with the model of the noisy data, the optimizer will eventually disregard the physics loss. Contrary to expectations, the physics loss does not effectively constrain the adverse effects of noise in PINNs, in particular when the adaptive weight gradient balancing algorithm is used.

\subsection{3D Taylor-Green Vortex}

Interestingly, many of the problems which we only saw for higher noise levels with 2D Taylor-Green Vortex are already present without noise with 3D Taylor-Green Vortex. A possible explanation is that increasing the difficulty of the problem has a similar effect as adding noise to the training data. This is also supported by the 1D Burgers' equations experiments with two parameters instead of only one, where we see more of the problems that appear with 2D and 3D Taylor-Green Vortex. We may hypothesize that overfitting on a simplified model of a difficult physical system can become more profitable than learning the physics, just as when overfitting on noise.

PINN does not show a consistent negative bias for 3D Taylor-Green Vortex as it did with 1D Burgers' equation and 2D Taylor-Green Vortex. The mean biases go both directions, but are only statistically significant at $\sigma = 1$ (where the bias is negative). A possible explanation is that learning issues due to the higher complexity in 3D occlude this simpler learning problem.

We observe that PINN seems to scale better with the computational complexity of the problem compared to FEM/SLSQP. The cost for 3D Taylor-Green Vortex is not much higher than it was for the 2D version of the problem, in particular the memory use. The baseline, FEM/SLSQP, scales much worse. Running it on the 3D problem is significantly more expensive and requires HPC resources. FEM/SLSQP is also highly likely to catastrophically fail compared to PINN. On the other hand, as with FEM, PINNs accuracy degrades with the 3D problem compared to the 2D problem. Our observation that PINNs may scale well supports previous claims in the literature \cite{HU2024106369}.

\section{Conclusion}
\label{sec:conclusion}

Through three different experiments, we have observed that a simple baseline using FEM and an optimizer, mostly outperforms PINNs in accuracy on noisy inverse problems. However, this observation is complicated by the baseline's memory costs for Navier-Stokes and its frequent catastrophic failure rate (about 20\%) for the 3D test problem. The limits of our experiments also make discussion about time cost difficult. To the best of our knowledge, this is the first benchmark comparing PINNs with traditional solvers on inverse problems, in particular inverse Navier-Stokes problems with noise. However, PINNs do require less manual labor and specialized knowledge compared to the FEM-based baseline. We have also found that relative to the baseline, PINNs appear to scale better with the computational complexity of the problem as memory costs change little. This is consistent with findings in existing literature where PINNs excel in extremely high dimensions \cite{HU2024106369}.

We have identified two difficulties which appear to limit the performance of PINNs on noisy inverse problems. There is a "reward hacking" phenomenon, the parameter learning is biased by the optimization process. This causes training to converge to a suboptimal solution. Adaptive weighting also tends to break down with noisy training data. Contrary to our expectations, the physics loss fails to constrain the adverse effects of noise.

Furthermore, we have found that it is not trivial to identify the best model during training. Neither the validation loss nor physics loss tends to give a good indication according to the test loss. Developing new early stopping strategies for PINNs may be crucial to make the models more competitive on noisy inverse problems. The future of PINN research appears to lie in better tailoring neural networks to the physics domain in particular \cite{wang2023expertsguidetrainingphysicsinformed}.

\bibliographystyle{elsarticle-num-names} 
\bibliography{references.bib}

\appendix

\section{Implementation details}

\subsection{1D Burgers' equation}
\label{1dapp}

In the FEM model, we used the following weak formulation:

\begin{equation}
\begin{split}
\int_\Omega \frac{\partial u(x,t)}{\partial t} v(x,t) \, dx &+ \int_\Omega u(x,t) \frac{\partial u(x,t)}{\partial x} v(x,t) \, dx \\ &+ \nu \int_\Omega \frac{\partial u(x,t)}{\partial x} \frac{\partial v(x,t)}{\partial x} \, dx = 0.
\end{split}
\end{equation}

\noindent
To represent the solution vector field $u(x, t)$, we assume the PINN model $a(\bm x_i;\bm \theta)$ where $\bm x_i$ contains a single space and time coordinate from the physics training data $x$ and $t$. Substituting this into the 1D Burgers' partial differential equation, we get the following residual to measure:

\begin{equation}
\mathcal{R}_{\theta}(\bm x_i) = \frac{\partial a(\bm x_i;\bm \theta)}{\partial x_i(t)} + a(\bm x_i;\bm \theta) \frac{\partial a(\bm x_i;\bm \theta)}{\partial x_i(x)} - \nu \frac{\partial^2 a(\bm x_i;\bm \theta)}{\partial x_i(x)^2}.
\end{equation}

\begin{table}[H]
\centering
\caption{1D Burgers' equation, shared setup.}
\footnotesize
\renewcommand{\arraystretch}{1.3} 
\begin{tabular}{@{}ll@{}}
\toprule
\textbf{Aspect} & \textbf{Details} \\
\midrule
Repetitions & 30 runs per experiment. \\
Noise levels ($\sigma$) & 0, 0.5, 1, 2, 3, 5, 7, 10, 25. \\
Spatial resolution & $dx = \frac{2}{255} \approx 0.00784$. \\
Temporal resolution & $dt = 0.01$. \\
Dataset & From PINN paper \cite{raissi2019physics}. \\
Total data points & 100 time steps $\cdot$ 256 positions = 25600. \\
Training data & 10 time steps $\cdot$ 256 = 2560 (10\%). \\
Validation data & 2 time steps $\cdot$ 256 = 512. \\
Test data & 88 time steps $\cdot$ 256 = 22528 (88 \%).\\
Initial condition data & 160 randomly generated points. \\
Boundary condition data & 80 randomly generated points. \\
Physics data & 2540 random $(x,t)$ points. \\
\bottomrule
\end{tabular}
\label{tab:data_setup}
\end{table}

\begin{table}[H]
\centering
\caption{1D Burgers' model setups, FEM/SLSQP setup.}
\footnotesize
\renewcommand{\arraystretch}{1.3} 
\begin{tabular}{@{}ll@{}}
\toprule
\textbf{Aspect} & \textbf{Details} \\
\midrule
Basis functions & 1st degree polynomials. \\
Training data & Training and validation sets. \\
Discretization & Same as training data. \\
Solver & Newton, max 50 iter, no preconditioner. \\
Initial guess & $\nu = 5$. \\
SLSQP settings & Ftol=$1e-16$, max iter 100, bounds $[-5, 5]$. \\
Gradient calc. & Adjoint method (pyadjoint). \\
\bottomrule
\end{tabular}
\label{tab:fem_setup}
\end{table}

\begin{table}[H]
\centering
\caption{1D Burgers' model setups, PINN setup.}
\footnotesize
\renewcommand{\arraystretch}{1.3} 
\begin{tabular}{@{}ll@{}}
\toprule
\textbf{Aspect} & \textbf{Details} \\
\midrule
Architecture & 4 layers, 20 neurons/layer. \\
Activation & Tanh. \\
Batch size & Full-batch.\\
Training data & Training data, BC, IC, physics points. \\
Training epochs & 4000 (ADAM) + 4000 (L-BFGS). \\
Early stopping & Best results without. \\
Adaptive weights & Best results without.\\
Parameter search & Logarithmic scale.\\
Initial guess & $\nu = 5$. \\
Loss weight & $w = 1$. \\
Adam, learning rate & 0.001 (default).\\
L-BFGS, learning rate & 1 (default).\\
L-BFGS, history size & 50. \\
L-BFGS, line search & Strong Wolfe.\\
L-BFGS, rtol and atol & 1e-16, 1e-16.\\
\bottomrule
\end{tabular}
\label{tab:pinn_setup}
\end{table}

\subsection{2D Navier-Stokes equations}
\label{2dapp}

For FEM, we used the algorithm Chorin's projection \cite{CHORIN196712}. Each time step requires solving three systems. First we solve for $u^{temp}$, ignoring pressure:

\begin{equation}
\int_\Omega \frac{\mathbf{u}^{temp} - \mathbf{u}^{n}}{\Delta t} \cdot \mathbf{v} \, dx
+ \int_\Omega (\mathbf{u}^{n} \cdot \nabla \mathbf{u}^{n}) \cdot \mathbf{v} \, dx
+ \nu \int_\Omega \nabla \mathbf{u}^{temp} : \nabla \mathbf{v} \, dx = 0.
\end{equation}

\noindent
Next, we solve for the pressure $p$ using the velocity $u^{temp}$ from last step:

\begin{equation}
\int_{\Omega} \nabla p \cdot \nabla q \, dx
+ \frac{1}{\Delta t} \int_{\Omega} \left( \nabla \cdot \mathbf{u}^{temp} \right) q \, dx = 0,
\end{equation}

\noindent
where $q$ is a test function. With $p$ we can get the corrected, real $u^{n+1}$:

\begin{equation}
\int_{\Omega} \mathbf{u}^{n+1} \cdot \mathbf{v} \, dx
- \int_{\Omega} \mathbf{u}^{temp} \cdot \mathbf{v} \, dx
+ \Delta t \int_{\Omega} \nabla p^{n+1} \cdot \mathbf{v} \, dx = 0.
\end{equation}

\noindent
For incompressible Navier-Stokes systems with periodic boundary conditions, the pressure is only unique up to an additive constant. In our implementation, this gauge condition is not enforced during the discrete solve. We obtain a pressure field with a fixed gauge by subtracting the spatial average of the pressure, thus enforcing a zero-mean pressure condition. Since we only compare velocity predictions in our study, this choice does not affect the results.

By using the identities from \ref{subsec:2dspec}, by substitution we have $u_i=\frac{\partial \bm a(\bm x_i;\bm \theta)}{dx_i(y)}$, $v_i=-\frac{\partial \bm a(\bm x_i;\bm \theta)}{dx_i(x)}$ and the residual functions for the physics loss:

\begin{equation}
\begin{split}
\mathcal{R}_{1,\theta}(\bm x_i) &= \frac{\partial u_i}{\partial x_i(t)} + u_i \frac{\partial u_i}{\partial x_i(x)} + v_i \frac{\partial u_i}{\partial x_i(y)} + \frac{\partial p(\bm x_i; \bm \theta)}{\partial x_i(x)} - \nu \left( \frac{\partial^2 u_i}{\partial x_i(x)^2} + \frac{\partial^2 u_i}{\partial x_i(y)^2} \right),\\
\mathcal{R}_{2,\theta}(\bm x_i) &= \frac{\partial v_i}{\partial x_i(t)} + u_i \frac{\partial v_i}{\partial x_i(x)} + v_i \frac{\partial v_i}{\partial x_i(y)} + \frac{\partial p(\bm x_i; \bm \theta)}{\partial x_i(y)} - \nu \left( \frac{\partial^2 v_i}{\partial x_i(x)^2} + \frac{\partial^2 v_i}{\partial x_i(y)^2} \right),\\
\mathcal{R}_{\theta}(\mathbf{x}_i) &= \left(\mathcal{R}_{1,\theta}(\mathbf{x}_i),\,\mathcal{R}_{2,\theta}(\mathbf{x}_i)\right)^T.
\end{split}
\end{equation}

Here we assumed that each $\bm x_i$ contains the coordinates $x, y$ and time $t$.

\begin{table}[H]
\centering
\caption{2D Navier-Stokes equation, shared setup.}
\footnotesize
\renewcommand{\arraystretch}{1.3} 
\begin{tabular}{@{}ll@{}}
\toprule
\textbf{Aspect} & \textbf{Details shared} \\
\midrule
Repetitions & 30 runs per experiment. \\
Noise levels ($\sigma$) & 0, 0.001, 0.01, 0.05, 0.1, 0.25, 0.5, 1. \\
Spatial resolution & $dx = dy = 0.05$. \\
Temporal resolution & $dt = 0.1$. \\
Dataset & Generated from analytical solution. \\
Total data points & 26 time steps $\cdot$ 126 $\cdot$ 126 = 412776 observations. \\
Training data & 99225 random samples, $\approx 24\%$ of total. \\
Validation data & 19845 random samples, $\approx 4.8\%$ of total. \\
Test data & 293706 random samples, $\approx 71.12\%$ of total. \\
Initial condition data & 4000 randomly generated positions. \\
Boundary condition data & 4000 randomly generated positions. \\
Physics data & The $\bm x$ inputs from the training data. \\
\bottomrule
\end{tabular}
\label{tab:data_setup1}
\end{table}

\begin{table}[H]
\centering
\caption{2D Navier-Stokes, FEM/SLSQP setup.}
\footnotesize
\renewcommand{\arraystretch}{1.3} 
\begin{tabular}{@{}ll@{}}
\toprule
\textbf{Aspect} & \textbf{Details for FEM/SLSQP} \\
\midrule
Basis functions & Cubic polynomial for velocity, \\ & quadratic polynomial for pressure. \\
Training data & Training and validation sets. \\
Discretization & $64^2 = 4096, \quad dx = dy = \frac{2\pi}{64} \approx$ \\
& $0.098175, dt=0.01$. \\
Solver & Gmres solver with amg preconditioner.\\ & Abstol = 1e-12, rtol=1e-10, \\ & maximum iterations 5000.\\
Initial guess & $\nu = 5$. \\
SLSQP settings & Ftol=$1e-16$, max iter 100, \\ & bounds $[0.003141, 5]$. \\
Gradient calc. & 3-point finite difference (SciPy). \\
\bottomrule
\end{tabular}
\label{tab:fem_setup1}
\end{table}

\begin{table}[H]
\centering
\caption{2D Navier-Stokes, PINN setup.}
\footnotesize
\renewcommand{\arraystretch}{1.3} 
\begin{tabular}{@{}ll@{}}
\toprule
\textbf{Aspect} & \textbf{Details for PINN} \\
\midrule
Architecture & 9 layers, 20 neurons/layer. \\
Activation & Tanh. \\
Batch size & Full-batch.\\
Training data & Training data, BC, IC, physics points. \\
Training epochs & 200000 (ADAM) + 50000 (L-BFGS). \\
Early stopping & Best model, based on the validation set, is loaded at end of training. \\
                & Models from the first 10,000 epochs are ignored.\\
Adaptive weights & Yes, was needed \cite{wang2020understandingmitigatinggradientpathologies}.\\
Parameter search & Softplus used to bound to $[0.003141, \infty)$.\\
Initial guess & $\nu = 5$. \\
Adam, learning rate & 0.001 (default).\\
L-BFGS, learning rate & 1 (default).\\
L-BFGS, history size & 50.\\
L-BFGS, line search & Strong Wolfe.\\
L-BFGS, rtol and atol & 1e-16, 1e-16.\\
\bottomrule
\end{tabular}
\label{tab:pinn_setup1}
\end{table}

\subsection{3D Navier-Stokes equations}
\label{3dapp}

As for 2D Navier-Stokes, we used Chorin's projection described in $d$-dimensional form. For the physics loss, the residual function was implemented as follows, assuming each $\bm x_i$ has coordinates $x, y, z$ and time $t$:

\begin{align}
\mathcal{R}_{1, \theta}(\bm x_i) &\notag = \frac{\partial u(\bm x_i; \bm \theta)}{\partial x_i(t)} + u(\bm x_i; \bm \theta) \frac{\partial u(\bm x_i; \bm \theta)}{\partial x_i(x)} + v(\bm x_i; \bm \theta) \frac{\partial u(\bm x_i; \bm \theta)}{\partial x_i(y)} + w(\bm x_i; \bm \theta) \frac{\partial u(\bm x_i; \bm \theta)}{\partial x_i(z)} +\\ \frac{1}{\rho} &\frac{\partial p(\bm x_i; \bm \theta)}{\partial x_i(x)} - \nu \left( \frac{\partial^2 u(\bm x_i; \bm \theta)}{\partial x_i(x)^2} + \frac{\partial^2 u(\bm x_i; \bm \theta)}{\partial x_i(y)^2} + \frac{\partial^2 u(\bm x_i; \bm \theta)}{\partial x_i(z)^2} \right),\\
\mathcal{R}_{2,\theta}(\bm x_i) &\notag= \frac{\partial v(\bm x_i; \bm \theta)}{\partial x_i(t)} + u(\bm x_i; \bm \theta) \frac{\partial v(\bm x_i; \bm \theta)}{\partial x_i(x)} + v(\bm x_i; \bm \theta) \frac{\partial v(\bm x_i; \bm \theta)}{\partial x_i(y)} + w(\bm x_i; \bm \theta) \frac{\partial v(\bm x_i; \bm \theta)}{\partial x_i(z)} +\\ \frac{1}{\rho} &\frac{\partial p(\bm x_i; \bm \theta)}{\partial x_i(y)} - \nu \left( \frac{\partial^2 v(\bm x_i; \bm \theta)}{\partial x_i(x)^2} + \frac{\partial^2 v(\bm x_i; \bm \theta)}{\partial x_i(y)^2} + \frac{\partial^2 v(\bm x_i; \bm \theta)}{\partial x_i(z)^2} \right),\\
\mathcal{R}_{3,\theta}(\bm x_i) &\notag = \frac{\partial w(\bm x_i; \bm \theta)}{\partial x_i(t)} + u(\bm x_i; \bm \theta) \frac{\partial w(\bm x_i; \bm \theta)}{\partial x_i(x)} + v(\bm x_i; \bm \theta) \frac{\partial w(\bm x_i; \bm \theta)}{\partial x_i(y)} + w(\bm x_i; \bm \theta) \frac{\partial w(\bm x_i; \bm \theta)}{\partial x_i(z)} +\\ &\frac{1}{\rho} \frac{\partial p(\bm x_i; \bm \theta)}{\partial x_i(z)} - \nu \left( \frac{\partial^2 w(\bm x_i; \bm \theta)}{\partial x_i(x)^2} + \frac{\partial^2 w(\bm x_i; \bm \theta)}{\partial x_i(y)^2} + \frac{\partial^2 w(\bm x_i; \bm \theta)}{\partial x_i(z)^2} \right),\\
\mathcal{R}_{4, \theta}(\bm x_i) &= \frac{\partial u(\bm x_i; \bm \theta)}{\partial x_i(x)} + \frac{\partial v(\bm x_i; \bm \theta)}{\partial x_i(y)} + \frac{\partial w(\bm x_i; \bm \theta)}{\partial x_i(z)},\\
\mathcal{R}_{\theta}(\mathbf{x}_i) &=\left(\mathcal{R}_{1,\theta}(\mathbf{x}_i),\mathcal{R}_{2\theta}(\mathbf{x}_i),\mathcal{R}_{3,\theta}(\mathbf{x}_i),\mathcal{R}_{4,\theta}(\mathbf{x}_i)\right)^T.
\end{align}

\begin{table}[H]
\centering
\caption{3D Navier-Stokes equation, shared setup.}
\footnotesize
\renewcommand{\arraystretch}{1.3} 
\begin{tabular}{@{}ll@{}}
\toprule
\textbf{Aspect} & \textbf{Details} \\
\midrule
Repetitions & 30 runs per experiment. \\
Noise levels ($\sigma$) & 0, 0.001, 0.01, 0.05, 0.1, 0.25, 0.5, 1. \\
Spatial resolution & $dx = dy = \frac{2\pi}{128}\approx 0.0490876$. \\
Temporal resolution & Solved with $dt=0.005$, saved with $dt=0.25$. \\
Dataset & Generated using Nektar++ \cite{CANTWELL2015205}.\\
Total data points & 1100000 random samples from generated solution. \\
Training data & 275000 random samples, $=25\%$ of total.\\
Test data & 770000 random samples, $=75\%$ of total.\\
Validation data & 55000 random samples, $=5\%$ of total. \\
Initial condition data & 16000 randomly generated positions. \\
Boundary condition data & 8000 random positions from generated solution. \\ & None shared with training data.  \\
Physics data & The $\bm x$ inputs from the training data. \\
\bottomrule
\end{tabular}
\label{tab:data_setup2}
\end{table}

\begin{table}[H]
\centering
\caption{3D Navier-Stokes equation, FEM/SLSQP setup.}
\footnotesize
\renewcommand{\arraystretch}{1.3} 
\begin{tabular}{@{}ll@{}}
\toprule
\textbf{Aspect} & \textbf{Details} \\
\midrule
Basis functions & Quadratic polynomial for velocity, \\ & 1st degree polynomial for pressure. \\
Training data & Training and validation sets. \\
Discretization & $32^3 = 32768, dx = dy = dz = \frac{2\pi}{32} \approx$\\
& $0.1963495, dt=0.05$.\\
Solver & Gmres solver with amg preconditioner.\\ & Abstol = 1e-12, rtol=1e-10, \\ & maximum iterations 50000.\\
Initial guess & $\nu = 5$. \\
SLSQP settings & Ftol=1e-16, max iter 100, \\ & bounds $[0.003141, 5]$. \\
Gradient calc. & 3-point finite difference (SciPy). \\
\bottomrule
\end{tabular}
\label{tab:fem_setup2}
\end{table}

\begin{table}[H]
\centering
\caption{3D Navier-Stokes equation, PINN setup.}
\footnotesize
\renewcommand{\arraystretch}{1.3} 
\begin{tabular}{@{}ll@{}}
\toprule
\textbf{Aspect} & \textbf{Details} \\
\midrule
Architecture & 9 layers, 20 neurons/layer. \\
Activation & Tanh. \\
Batch size & Full-batch. \\
Training data & Training data, BC, IC, physics points. \\
Training epochs & 200000 (ADAM) + 50000 (L-BFGS). \\
Early stopping & Best model, based on the validation set, is loaded at end of training. \\
                & Models from the first 10,000 epochs are ignored.\\
Adaptive weights & Yes, was needed. \cite{wang2020understandingmitigatinggradientpathologies}.\\
Parameter search & Softplus used to bound to $[0.003141, \infty)$.\\
Initial guess & $\nu = 5$. \\
Adam, learning rate & 0.001 (default). \\
L-BFGS, learning rate & 1 (default).\\
L-BFGS, history size & 50. \\
L-BFGS, line search & Strong Wolfe. \\
L-BFGS, rtol and atol & 1e-16, 1e-16.\\
\bottomrule
\end{tabular}
\label{tab:pinn_setup2}
\end{table}

\subsection{Adaptive weights}
\label{adaptinfo}

For the Navier-Stokes problems, we used a gradient balancing algorithm \cite{wang2020understandingmitigatinggradientpathologies}. At regular intervals, the algorithm calculates the difference in magnitude of the maximum physics loss gradients and the average data loss gradients:

\begin{equation}
    \hat{w} = \frac{\max(|\nabla \mathcal{L}_{\text{phys}}(\bm x)|)}{\overline{|\nabla \mathcal{L}_{\text{data}}(\bm x)|}}.
\end{equation}

\noindent
The data weight $w$ is updated using $\hat{w}$, with $\alpha$ commonly chosen as $0.9$:

\begin{equation}
    w \leftarrow (1 - \alpha)w + \alpha\hat{w}.
\end{equation}

The weights are updated every 10 epochs, which is the frequency discussed in the paper \cite{wang2020understandingmitigatinggradientpathologies}. We tried lower update frequencies, but this did not give satisfactory performance. Note that the weights are detached from the computational graph. Boundary condition and initial condition data is combined with the rest of the training data and they are all weighted together as one. The physics loss is not weighted. Below are some sample graphs demonstrating how the weights evolve during training.

\begin{figure}[H]
    \centering
    \includegraphics[width=1\linewidth]{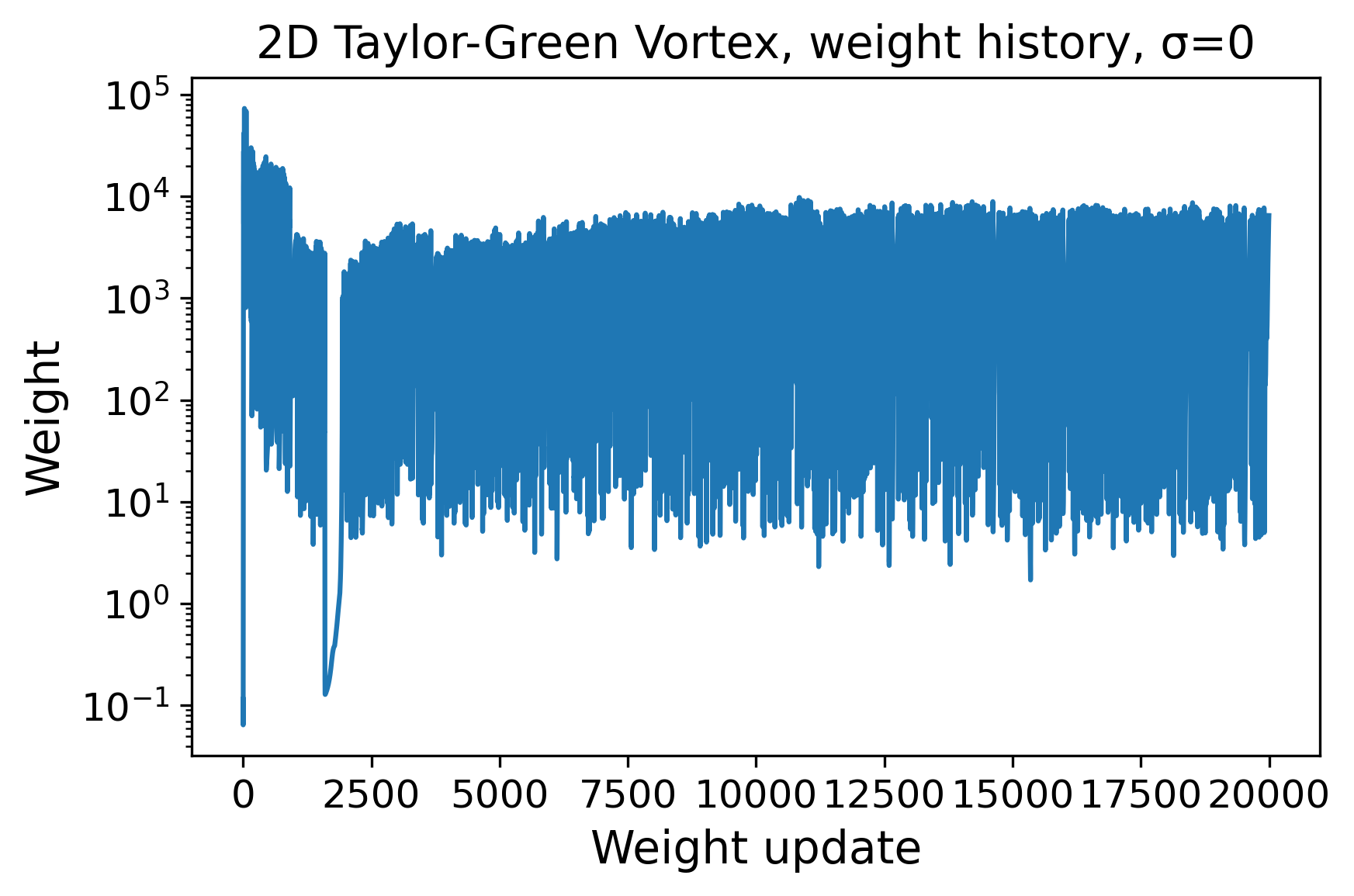}
\end{figure}

\begin{figure}[H]
    \centering
    \includegraphics[width=1\linewidth]{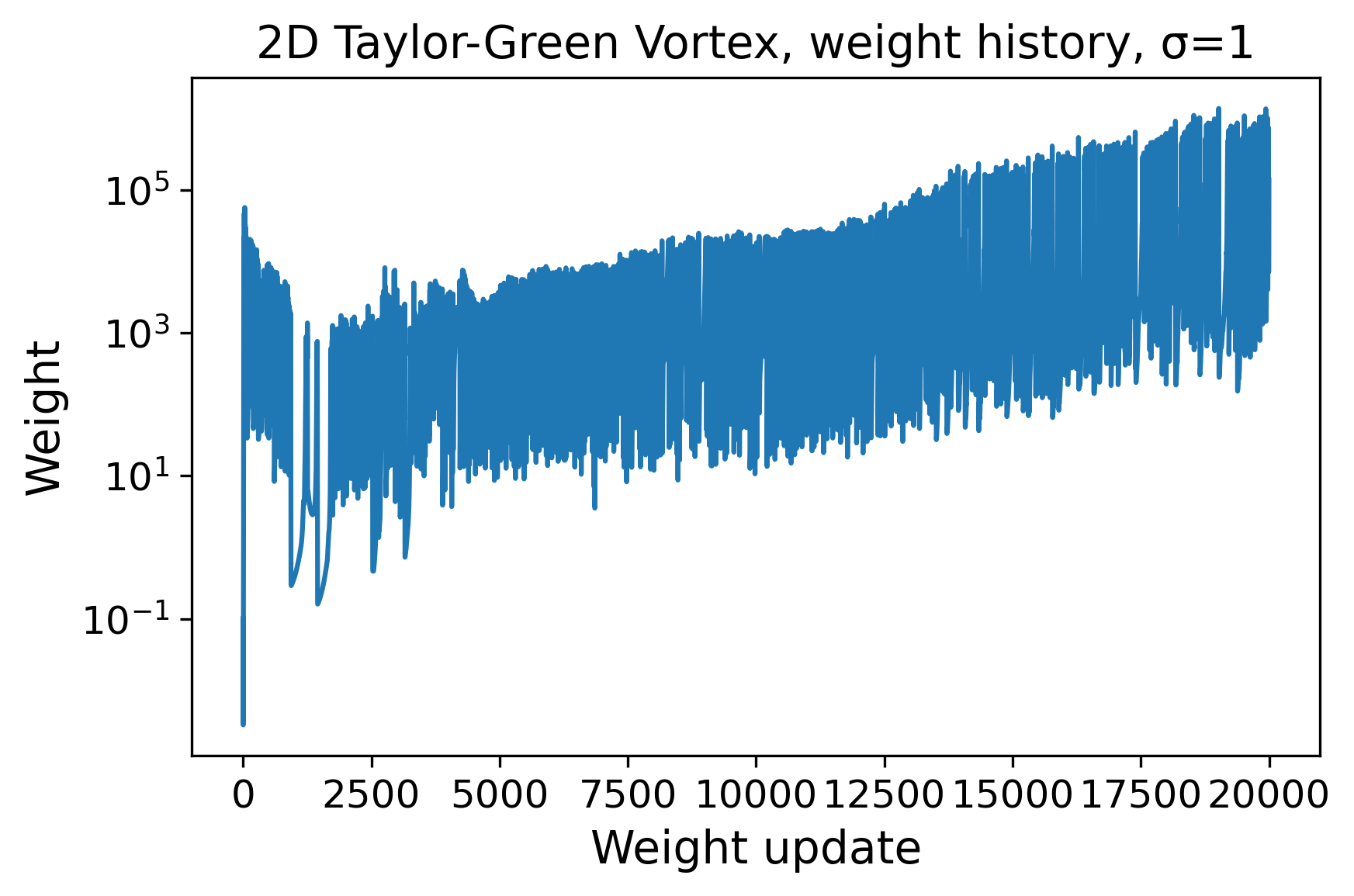}
    \includegraphics[width=1\linewidth]{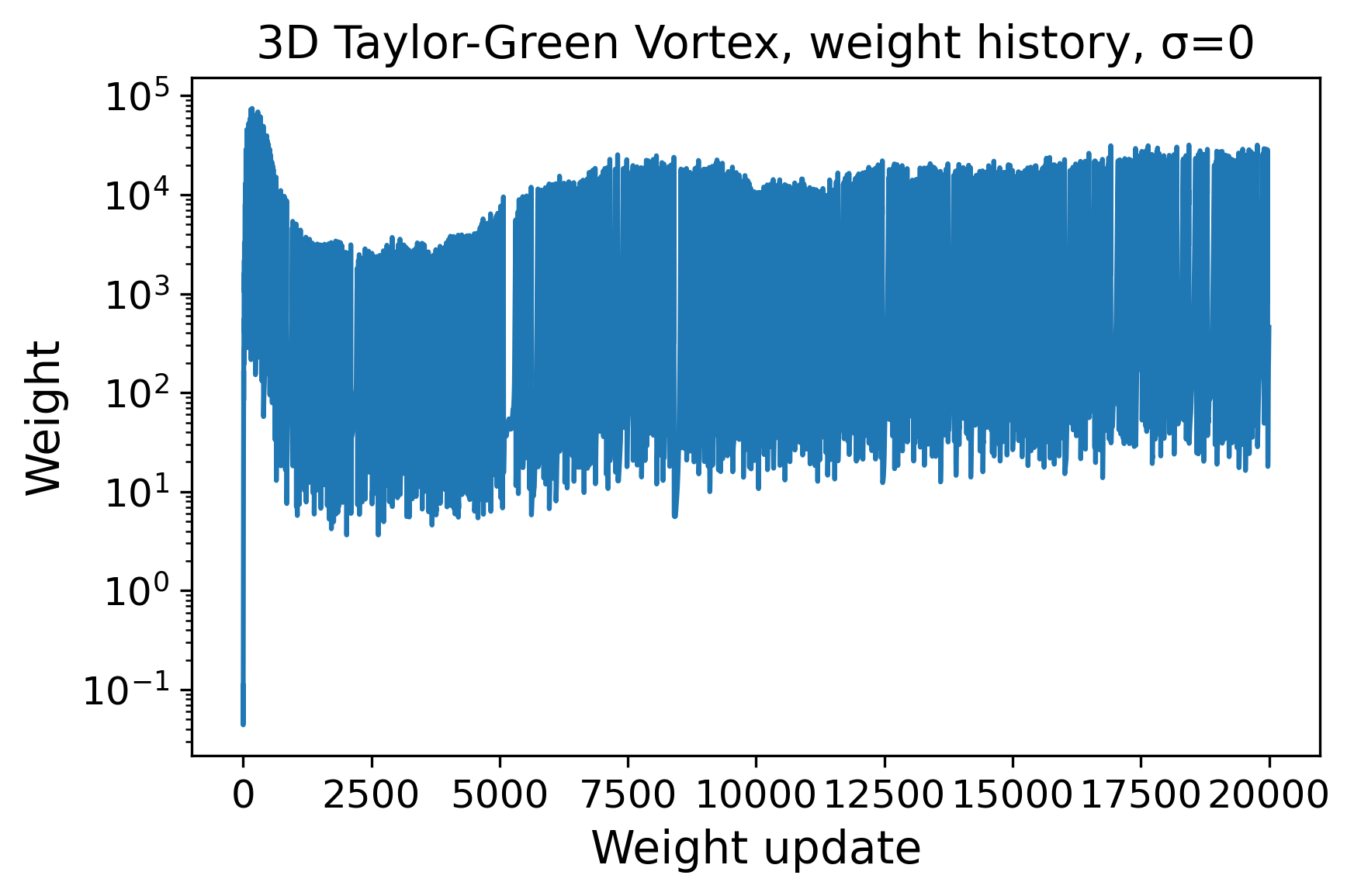}
\end{figure}

\begin{figure}[H]
    \centering
    \includegraphics[width=1\linewidth]{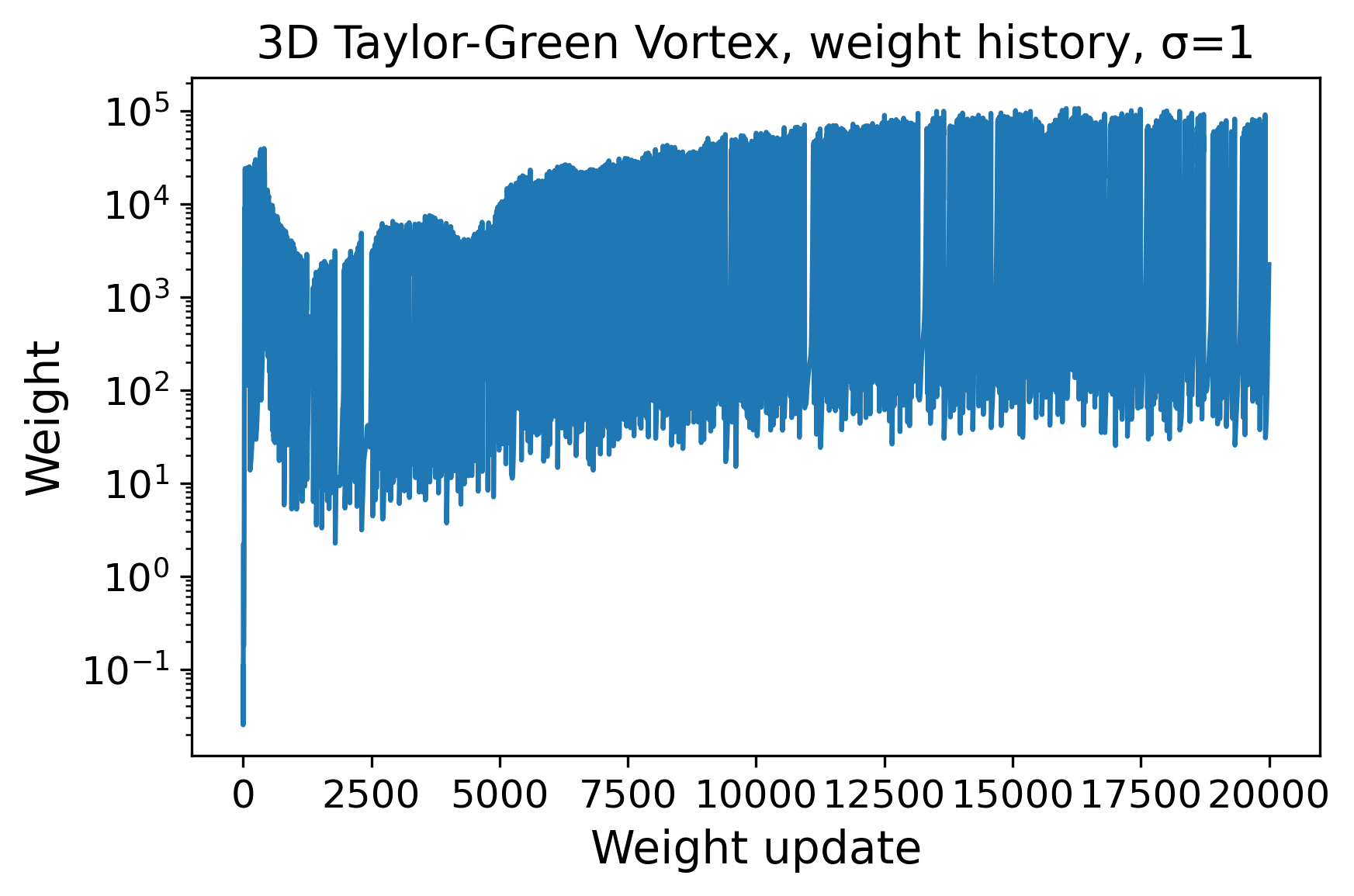}
\end{figure}

Visually the graphs are similar to what is shown in the original paper \cite{wang2020understandingmitigatinggradientpathologies}, but the weight values are much larger. Note that for 2D Taylor-Green Vortex with $\sigma=1$ and 3D Taylor-Green Vortex, where there was struggle with apparent overfitting, the weights tend to grow during training.

\section{Prediction results}

In this section we report the mean and standard deviation for each model's RMSE performance on the test set for all the problems studied. The difference between each pair of models is quantified using the Vargha-Delaney $A$ statistic \cite{Vargha2000ACA}. Its value indicates a probability between 0 and 1 that the reference model (the first model in each set of tables) has larger RMSE than the other model. To quantify whether the distribution of the results of each pair of models are different to a statistically significant degree, we used the Brunner-Munzel test \cite{brunnermunzel}. This is justified by the results from the tests generally not being normally distributed. Note that Brunner-Munzel is undefined when the distributions of each model do not overlap at all. For the 1D Burgers' problems with $\sigma=0$, FEM is deterministic, and variance is 0 which also means Brunner-Munzel becomes undefined. For both these cases, tests are simply reported as $\mathrm{N/A}$. We provide the original $p$ values, as well as the $p$ values adjusted using the Holm family-wise error rate correction \cite{holmsture} and the Benjamini-Hochberg false discovery rate correction \cite{benjamini}. For the Holm-derived $p$ values, all the values below 0.05\% have been marked in bold.

\subsection{1D Burgers' equation}
\label{1dtests}

\begin{table}[H]
\centering
\label{tab:pinn_fem_descriptive}
\caption{PINN and FEM results with Vargha-Delaney A effect size.}
\begin{tabular}{c c c c}
\toprule
Noise & PINN & FEM & Vargha--Delaney $A$ \\
\midrule
0 & $0.005366 \pm 0.000595$ & $0.006515 \pm 0$ & $0.067$ \\
0.001 & $0.022701 \pm 0.006844$ & $0.010623 \pm 0.004350$ & $0.956$ \\
0.01 & $0.049035 \pm 0.009264$ & $0.020314 \pm 0.016021$ & $0.914$ \\
0.05 & $0.095079 \pm 0.023605$ & $0.026936 \pm 0.019883$ & $0.974$ \\
0.1 & $0.144118 \pm 0.039172$ & $0.058326 \pm 0.045087$ & $0.914$ \\
0.25 & $0.205870 \pm 0.057595$ & $0.110683 \pm 0.148179$ & $0.832$ \\
0.5 & $0.257481 \pm 0.080885$ & $0.160393 \pm 0.171301$ & $0.791$ \\
1 & $0.452368 \pm 0.212503$ & $0.187883 \pm 0.192528$ & $0.822$ \\

\bottomrule
\end{tabular}
\end{table}

\begin{table}[H]
\centering
\caption{Brunner-Munzel tests and adjusted p-values for PINN and FEM results.}
\label{tab:pinn_fem_tests}
\begin{tabular}{c c c c c}
\toprule
Noise &
Brunner--Munzel $W$ &
$p$ &
$p_{\mathrm{Holm}}$ &
$p_{\mathrm{FDR}}$ \\
\midrule
0 & $\mathrm{N/A}$ & $\mathrm{N/A}$ & $\mathrm{N/A}$ & $\mathrm{N/A}$ \\
0.001 & $-15.398$ & $2.537e-16$ & $\mathbf{1.522e-15}$ & $8.879e-16$ \\
0.01 & $-9.032$ & $4.370e-10$ & $\mathbf{1.748e-09}$ & $7.647e-10$ \\
0.05 & $-20.920$ & $1.023e-19$ & $\mathbf{7.160e-19}$ & $7.160e-19$ \\
0.1 & $-10.280$ & $3.039e-12$ & $\mathbf{1.520e-11}$ & $7.091e-12$ \\
0.25 & $-5.207$ & $1.280e-05$ & $\mathbf{2.560e-05}$ & $1.493e-05$ \\
0.5 & $-4.324$ & $1.427e-04$ & $\mathbf{1.427e-04}$ & $1.427e-04$ \\
1 & $-5.712$ & $1.243e-06$ & $\mathbf{3.729e-06}$ & $1.740e-06$ \\

\bottomrule
\end{tabular}
\end{table}

\begin{table}[H]
\centering
\label{tab:pinn_fem_descriptive}
\caption{FEM and PINN/FEM results with Vargha-Delaney A effect size.}
\begin{tabular}{c c c c}
\toprule
Noise & FEM & PINN/FEM & Vargha--Delaney $A$ \\
\midrule
0 & $0.006515 \pm 0$ & $0.006799 \pm 0.000140$ & $0.033$ \\
0.001 & $0.010623 \pm 0.004350$ & $0.012746 \pm 0.006623$ & $0.462$ \\
0.01 & $0.020314 \pm 0.016021$ & $0.027380 \pm 0.013051$ & $0.311$ \\
0.05 & $0.026936 \pm 0.019883$ & $0.040960 \pm 0.019025$ & $0.263$ \\
0.1 & $0.058326 \pm 0.045087$ & $0.049486 \pm 0.021213$ & $0.507$ \\
0.25 & $0.110683 \pm 0.148179$ & $0.056091 \pm 0.031897$ & $0.504$ \\
0.5 & $0.160393 \pm 0.171301$ & $0.060345 \pm 0.035793$ & $0.677$ \\
1 & $0.187883 \pm 0.192528$ & $0.147435 \pm 0.175961$ & $0.529$ \\

\bottomrule
\end{tabular}
\end{table}

\begin{table}[H]
\centering
\label{tab:pinn_fem_tests}

\caption{Brunner-Munzel tests and adjusted p-values for FEM and PINN/FEM results.}
\begin{tabular}{c c c c c}
\toprule
Noise &
Brunner--Munzel $W$ &
$p$ &
$p_{\mathrm{Holm}}$ &
$p_{\mathrm{FDR}}$ \\
\midrule
0 & $\mathrm{N/A}$ & $\mathrm{N/A}$ & $\mathrm{N/A}$ & $\mathrm{N/A}$ \\
0.001 & $0.482$ & $0.6321$ & $1.0000$ & $0.9565$ \\
0.01 & $2.573$ & $0.0127$ & $0.0763$ & $0.0445$ \\
0.05 & $3.405$ & $0.0013$ & $\mathbf{8.800e-03}$ & $0.0088$ \\
0.1 & $-0.083$ & $0.9342$ & $1.0000$ & $0.9565$ \\
0.25 & $-0.055$ & $0.9565$ & $1.0000$ & $0.9565$ \\
0.5 & $-2.347$ & $0.0245$ & $0.1223$ & $0.0571$ \\
1 & $-0.373$ & $0.7110$ & $1.0000$ & $0.9565$ \\

\bottomrule
\end{tabular}
\end{table}

\begin{table}[H]
\centering
\label{tab:pinn_fem_descriptive}
\caption{PINN and PINN/FEM results with Vargha-Delaney A effect size.}
\begin{tabular}{c c c c}
\toprule
Noise & PINN & PINN/FEM & Vargha--Delaney $A$ \\
\midrule
0 & $0.005366 \pm 0.000595$ & $0.006799 \pm 0.000140$ & $0.013$ \\
0.001 & $0.022701 \pm 0.006844$ & $0.012746 \pm 0.006623$ & $0.847$ \\
0.01 & $0.049035 \pm 0.009264$ & $0.027380 \pm 0.013051$ & $0.927$ \\
0.05 & $0.095079 \pm 0.023605$ & $0.040960 \pm 0.019025$ & $0.969$ \\
0.1 & $0.144118 \pm 0.039172$ & $0.049486 \pm 0.021213$ & $0.989$ \\
0.25 & $0.205870 \pm 0.057595$ & $0.056091 \pm 0.031897$ & $0.988$ \\
0.5 & $0.257481 \pm 0.080885$ & $0.060345 \pm 0.035793$ & $0.998$ \\
1 & $0.452368 \pm 0.212503$ & $0.147435 \pm 0.175961$ & $0.878$ \\

\bottomrule
\end{tabular}
\end{table}

\begin{table}[H]
\centering
\caption{Brunner-Munzel tests and adjusted p-values for PINN and PINN/FEM results.}
\label{tab:pinn_fem_tests}
\begin{tabular}{c c c c c}
\toprule
Noise &
Brunner--Munzel $W$ &
$p$ &
$p_{\mathrm{Holm}}$ &
$p_{\mathrm{FDR}}$ \\
\midrule
0 & $38.304$ & $2.931e-29$ & $\mathbf{1.466e-28}$ & $5.863e-29$ \\
0.001 & $-6.798$ & $2.676e-08$ & $\mathbf{2.676e-08}$ & $2.676e-08$ \\
0.01 & $-12.038$ & $2.228e-15$ & $\mathbf{6.685e-15}$ & $2.971e-15$ \\
0.05 & $-26.468$ & $7.718e-34$ & $\mathbf{4.631e-33}$ & $2.058e-33$ \\
0.1 & $-52.720$ & $2.058e-44$ & $\mathbf{1.441e-43}$ & $8.233e-44$ \\
0.25 & $-38.771$ & $9.420e-29$ & $\mathbf{3.768e-28}$ & $1.507e-28$ \\
0.5 & $-183.956$ & $1.679e-74$ & $\mathbf{1.343e-73}$ & $1.343e-73$ \\
1 & $-7.514$ & $1.142e-08$ & $\mathbf{2.284e-08}$ & $1.305e-08$ \\

\bottomrule
\end{tabular}
\end{table}

\subsection{1D Burgers' equation with two parameters}
\label{1drmsetwoparameter}

\begin{table}[H]
\centering
\label{tab:pinn_fem_descriptive}
\caption{PINN and FEM results with Vargha-Delaney A effect size.}
\begin{tabular}{c c c c}
\toprule
Noise & PINN & FEM & Vargha--Delaney $A$ \\
\midrule
0 & $0.006525 \pm 0.000048$ & $0.004709 \pm 0$ & $1.000$ \\
0.001 & $0.028642 \pm 0.007888$ & $0.013563 \pm 0.006890$ & $0.922$ \\
0.01 & $0.063422 \pm 0.013970$ & $0.022783 \pm 0.009990$ & $0.988$ \\
0.05 & $0.123037 \pm 0.030658$ & $0.043907 \pm 0.019644$ & $0.991$ \\
0.1 & $0.194091 \pm 0.081561$ & $0.060095 \pm 0.036979$ & $0.976$ \\
0.25 & $0.334672 \pm 0.208818$ & $0.128104 \pm 0.117891$ & $0.938$ \\
0.5 & $0.484270 \pm 0.267189$ & $0.183947 \pm 0.209537$ & $0.870$ \\
1 & $0.733201 \pm 0.405181$ & $0.287932 \pm 0.246003$ & $0.849$ \\

\bottomrule
\end{tabular}
\end{table}

\begin{table}[H]
\centering
\label{tab:pinn_fem_tests}
\caption{Brunner-Munzel tests and adjusted p-values for PINN and FEM results.}
\begin{tabular}{c c c c c}
\toprule
Noise &
Brunner--Munzel $W$ &
$p$ &
$p_{\mathrm{Holm}}$ &
$p_{\mathrm{FDR}}$ \\
\midrule
0 & $\mathrm{N/A}$ & $\mathrm{N/A}$ & $\mathrm{N/A}$ & $\mathrm{N/A}$ \\
0.001 & $-12.126$ & $3.650e-15$ & $\mathbf{1.460e-14}$ & $6.387e-15$ \\
0.01 & $-52.367$ & $2.276e-44$ & $\mathbf{1.594e-43}$ & $1.594e-43$ \\
0.05 & $-52.803$ & $1.488e-34$ & $\mathbf{8.928e-34}$ & $5.208e-34$ \\
0.1 & $-23.988$ & $2.868e-22$ & $\mathbf{1.434e-21}$ & $6.692e-22$ \\
0.25 & $-12.536$ & $3.325e-14$ & $\mathbf{9.975e-14}$ & $4.655e-14$ \\
0.5 & $-6.778$ & $1.138e-07$ & $\mathbf{1.138e-07}$ & $1.138e-07$ \\
1 & $-6.632$ & $3.979e-08$ & $\mathbf{7.957e-08}$ & $4.642e-08$ \\

\bottomrule
\end{tabular}
\end{table}

\begin{table}[H]
\centering
\label{tab:pinn_fem_descriptive}
\caption{FEM and PINN/FEM results with Vargha-Delaney A effect size.}
\begin{tabular}{c c c c}
\toprule
Noise & FEM & PINN/FEM & Vargha--Delaney $A$ \\
\midrule
0 & $0.004709 \pm 0$ & $0.004786 \pm 0.000033$ & $0.000$ \\
0.001 & $0.013563 \pm 0.006890$ & $0.017567 \pm 0.010422$ & $0.382$ \\
0.01 & $0.022783 \pm 0.009990$ & $0.044224 \pm 0.022587$ & $0.180$ \\
0.05 & $0.043907 \pm 0.019644$ & $0.093578 \pm 0.043949$ & $0.169$ \\
0.1 & $0.060095 \pm 0.036979$ & $0.106218 \pm 0.056491$ & $0.206$ \\
0.25 & $0.128104 \pm 0.117891$ & $0.135507 \pm 0.077178$ & $0.428$ \\
0.5 & $0.183947 \pm 0.209537$ & $0.176300 \pm 0.118212$ & $0.372$ \\
1 & $0.287932 \pm 0.246003$ & $0.236276 \pm 0.162077$ & $0.517$ \\

\bottomrule
\end{tabular}
\end{table}

\begin{table}[H]
\centering
\caption{Brunner-Munzel tests and adjusted p-values for FEM and PINN/FEM results.}
\label{tab:pinn_fem_tests}
\begin{tabular}{c c c c c}
\toprule
Noise &
Brunner--Munzel $W$ &
$p$ &
$p_{\mathrm{Holm}}$ &
$p_{\mathrm{FDR}}$ \\
\midrule
0 & $\mathrm{N/A}$ & $\mathrm{N/A}$ & $\mathrm{N/A}$ & $\mathrm{N/A}$ \\
0.001 & $1.607$ & $0.1135$ & $0.3974$ & $0.1589$ \\
0.01 & $5.467$ & $3.183e-06$ & $\mathbf{1.910e-05}$ & $1.114e-05$ \\
0.05 & $6.147$ & $2.347e-07$ & $\mathbf{1.643e-06}$ & $1.643e-06$ \\
0.1 & $5.080$ & $5.725e-06$ & $\mathbf{2.862e-05}$ & $1.336e-05$ \\
0.25 & $0.959$ & $0.3416$ & $0.6831$ & $0.3985$ \\
0.5 & $1.686$ & $0.0993$ & $0.3974$ & $0.1589$ \\
1 & $-0.212$ & $0.8329$ & $0.8329$ & $0.8329$ \\

\bottomrule
\end{tabular}
\end{table}

\begin{table}[H]
\centering
\label{tab:pinn_fem_descriptive}
\caption{PINN and PINN/FEM results with Vargha-Delaney A effect size.}
\begin{tabular}{c c c c}
\toprule
Noise & PINN & PINN/FEM & Vargha--Delaney $A$ \\
\midrule
0 & $0.006525 \pm 0.000048$ & $0.004786 \pm 0.000033$ & $1.000$ \\
0.001 & $0.028642 \pm 0.007888$ & $0.017567 \pm 0.010422$ & $0.832$ \\
0.01 & $0.063422 \pm 0.013970$ & $0.044224 \pm 0.022587$ & $0.784$ \\
0.05 & $0.123037 \pm 0.030658$ & $0.093578 \pm 0.043949$ & $0.706$ \\
0.1 & $0.194091 \pm 0.081561$ & $0.106218 \pm 0.056491$ & $0.866$ \\
0.25 & $0.334672 \pm 0.208818$ & $0.135507 \pm 0.077178$ & $0.896$ \\
0.5 & $0.484270 \pm 0.267189$ & $0.176300 \pm 0.118212$ & $0.921$ \\
1 & $0.733201 \pm 0.405181$ & $0.236276 \pm 0.162077$ & $0.944$ \\

\bottomrule
\end{tabular}
\end{table}

\begin{table}[H]
\centering
\label{tab:pinn_fem_tests}
\caption{Brunner-Munzel tests and adjusted p-values for PINN and PINN/FEM results.}
\begin{tabular}{c c c c c}
\toprule
Noise &
Brunner--Munzel $W$ &
$p$ &
$p_{\mathrm{Holm}}$ &
$p_{\mathrm{FDR}}$ \\
\midrule
0 & $\mathrm{N/A}$ & $\mathrm{N/A}$ & $\mathrm{N/A}$ & $\mathrm{N/A}$ \\
0.001 & $-5.694$ & $2.003e-06$ & $\mathbf{6.008e-06}$ & $2.804e-06$ \\
0.01 & $-4.561$ & $3.622e-05$ & $\mathbf{7.243e-05}$ & $4.225e-05$ \\
0.05 & $-3.024$ & $0.0039$ & $\mathbf{3.900e-03}$ & $0.0039$ \\
0.1 & $-7.310$ & $8.469e-09$ & $\mathbf{3.388e-08}$ & $1.482e-08$ \\
0.25 & $-9.102$ & $5.139e-11$ & $\mathbf{2.569e-10}$ & $1.199e-10$ \\
0.5 & $-11.928$ & $1.006e-15$ & $\mathbf{7.040e-15}$ & $6.471e-15$ \\
1 & $-13.773$ & $1.849e-15$ & $\mathbf{1.109e-14}$ & $6.471e-15$ \\

\bottomrule
\end{tabular}
\end{table}

\subsection{2D Taylor-Green Vortex}
\label{2dstat_test}

\begin{table}[H]
\centering
\label{tab:pinn_fem_descriptive}
\caption{PINN and FEM results with Vargha-Delaney A effect size.}
\begin{tabular}{c c c c}
\toprule
Noise & PINN & FEM & Vargha--Delaney $A$ \\
\midrule
0 & $0.000426 \pm 0.000686$ & $0.000151 \pm 0.000005$ & $0.257$ \\
0.001 & $0.000372 \pm 0.000597$ & $0.000151 \pm 0.000005$ & $0.268$ \\
0.01 & $0.000364 \pm 0.000410$ & $0.000152 \pm 0.000006$ & $1.000$ \\
0.05 & $0.001304 \pm 0.000400$ & $0.000189 \pm 0.000042$ & $1.000$ \\
0.1 & $0.002315 \pm 0.000403$ & $0.000240 \pm 0.000093$ & $1.000$ \\
0.25 & $0.005814 \pm 0.000603$ & $0.000499 \pm 0.000251$ & $1.000$ \\
0.5 & $0.009809 \pm 0.001076$ & $0.000902 \pm 0.000588$ & $1.000$ \\
1 & $0.015522 \pm 0.002475$ & $0.001336 \pm 0.000822$ & $1.000$ \\

\bottomrule
\end{tabular}
\end{table}

\begin{table}[H]
\centering
\caption{Brunner-Munzel tests and adjusted p-values for PINN and FEM results.}
\label{tab:pinn_fem_tests}
\begin{tabular}{c c c c c}
\toprule
Noise &
Brunner--Munzel $W$ &
$p$ &
$p_{\mathrm{Holm}}$ &
$p_{\mathrm{FDR}}$ \\
\midrule
0 & $3.055$ & $0.0048$ & $\mathbf{9.576e-03}$ & $0.0083$ \\
0.001 & $2.832$ & $0.0083$ & $\mathbf{9.576e-03}$ & $0.0083$ \\
0.01 & $\mathrm{N/A}$ & $\mathrm{N/A}$ & $\mathrm{N/A}$ & $\mathrm{N/A}$ \\
0.05 & $\mathrm{N/A}$ & $\mathrm{N/A}$ & $\mathrm{N/A}$ & $\mathrm{N/A}$ \\
0.1 & $\mathrm{N/A}$ & $\mathrm{N/A}$ & $\mathrm{N/A}$ & $\mathrm{N/A}$ \\
0.25 & $\mathrm{N/A}$ & $\mathrm{N/A}$ & $\mathrm{N/A}$ & $\mathrm{N/A}$ \\
0.5 & $\mathrm{N/A}$ & $\mathrm{N/A}$ & $\mathrm{N/A}$ & $\mathrm{N/A}$ \\
1 & $\mathrm{N/A}$ & $\mathrm{N/A}$ & $\mathrm{N/A}$ & $\mathrm{N/A}$ \\

\bottomrule
\end{tabular}
\end{table}

\begin{table}[H]
\centering
\caption{FEM and PINN/FEM results with Vargha-Delaney A effect size.}
\label{tab:pinn_fem_descriptive}
\begin{tabular}{c c c c}
\toprule
Noise & FEM & PINN/FEM & Vargha--Delaney $A$ \\
\midrule
0 & $0.000151 \pm 0.000005$ & $0.000755 \pm 0.001281$ & $0.059$ \\
0.001 & $0.000151 \pm 0.000005$ & $0.000719 \pm 0.001267$ & $0.077$ \\
0.01 & $0.000152 \pm 0.000006$ & $0.000449 \pm 0.000745$ & $0.061$ \\
0.05 & $0.000189 \pm 0.000042$ & $0.001320 \pm 0.000950$ & $0.079$ \\
0.1 & $0.000240 \pm 0.000093$ & $0.002353 \pm 0.001531$ & $0.064$ \\
0.25 & $0.000499 \pm 0.000251$ & $0.005645 \pm 0.002547$ & $0.020$ \\
0.5 & $0.000902 \pm 0.000588$ & $0.009270 \pm 0.003437$ & $0.000$ \\
1 & $0.001336 \pm 0.000822$ & $0.010382 \pm 0.006904$ & $0.043$ \\

\bottomrule
\end{tabular}
\end{table}

\begin{table}[H]
\centering
\caption{Brunner-Munzel tests and adjusted p-values for PINN and PINN/FEM results.}
\label{tab:pinn_fem_tests}
\begin{tabular}{c c c c c}
\toprule
Noise &
Brunner--Munzel $W$ &
$p$ &
$p_{\mathrm{Holm}}$ &
$p_{\mathrm{FDR}}$ \\
\midrule
0 & $14.208$ & $1.936e-16$ & $\mathbf{1.162e-15}$ & $6.252e-16$ \\
0.001 & $10.193$ & $1.985e-11$ & $\mathbf{3.971e-11}$ & $2.316e-11$ \\
0.01 & $12.760$ & $1.606e-14$ & $\mathbf{6.423e-14}$ & $2.810e-14$ \\
0.05 & $9.351$ & $2.130e-10$ & $\mathbf{2.130e-10}$ & $2.130e-10$ \\
0.1 & $11.308$ & $1.842e-12$ & $\mathbf{5.525e-12}$ & $2.578e-12$ \\
0.25 & $28.873$ & $3.530e-25$ & $\mathbf{2.471e-24}$ & $2.471e-24$ \\
0.5 & $\mathrm{N/A}$ & $\mathrm{N/A}$ & $\mathrm{N/A}$ & $\mathrm{N/A}$ \\
1 & $15.444$ & $2.679e-16$ & $\mathbf{1.340e-15}$ & $6.252e-16$ \\

\bottomrule
\end{tabular}
\end{table}

\begin{table}[H]
\centering
\caption{PINN and PINN/FEM results with Vargha-Delaney A effect size.}
\label{tab:pinn_fem_descriptive}
\begin{tabular}{c c c c}
\toprule
Noise & PINN & PINN/FEM & Vargha--Delaney $A$ \\
\midrule
0 & $0.000426 \pm 0.000686$ & $0.000755 \pm 0.001281$ & $0.197$ \\
0.001 & $0.000372 \pm 0.000597$ & $0.000719 \pm 0.001267$ & $0.196$ \\
0.01 & $0.000364 \pm 0.000410$ & $0.000449 \pm 0.000745$ & $0.531$ \\
0.05 & $0.001304 \pm 0.000400$ & $0.001320 \pm 0.000950$ & $0.567$ \\
0.1 & $0.002315 \pm 0.000403$ & $0.002353 \pm 0.001531$ & $0.518$ \\
0.25 & $0.005814 \pm 0.000603$ & $0.005645 \pm 0.002547$ & $0.527$ \\
0.5 & $0.009809 \pm 0.001076$ & $0.009270 \pm 0.003437$ & $0.582$ \\
1 & $0.015522 \pm 0.002475$ & $0.010382 \pm 0.006904$ & $0.876$ \\

\bottomrule
\end{tabular}
\end{table}

\begin{table}[H]
\centering
\caption{Brunner-Munzel tests and adjusted p-values for PINN and PINN/FEM results.}
\label{tab:pinn_fem_tests}
\begin{tabular}{c c c c c}
\toprule
Noise &
Brunner--Munzel $W$ &
$p$ &
$p_{\mathrm{Holm}}$ &
$p_{\mathrm{FDR}}$ \\
\midrule
0 & $4.535$ & $7.323e-05$ & $\mathbf{4.394e-04}$ & $1.953e-04$ \\
0.001 & $4.687$ & $4.668e-05$ & $\mathbf{3.267e-04}$ & $1.867e-04$ \\
0.01 & $-0.385$ & $0.7022$ & $1.0000$ & $0.8330$ \\
0.05 & $-0.806$ & $0.4260$ & $1.0000$ & $0.6816$ \\
0.1 & $-0.213$ & $0.8330$ & $1.0000$ & $0.8330$ \\
0.25 & $-0.314$ & $0.7553$ & $1.0000$ & $0.8330$ \\
0.5 & $-0.964$ & $0.3424$ & $1.0000$ & $0.6816$ \\
1 & $-6.951$ & $7.953e-08$ & $\mathbf{6.362e-07}$ & $6.362e-07$ \\

\bottomrule
\end{tabular}
\end{table}

\subsection{3D Taylor-Green Vortex, with outliers}
\label{3dwithoutlierstests}

\begin{table}[H]
\centering
\label{tab:pinn_fem_descriptive}
\caption{PINN and FEM results with Vargha-Delaney A effect size.}
\begin{tabular}{c c c c}
\toprule
Noise & PINN & FEM & Vargha--Delaney $A$ \\
\midrule
0 & $0.032690 \pm 0.009598$ & $0.002516 \pm 0.002596$ & $1.000$ \\
0.001 & $0.039061 \pm 0.009850$ & $0.011303 \pm 0.049650$ & $0.967$ \\
0.01 & $0.036553 \pm 0.010552$ & $0.030073 \pm 0.082802$ & $0.896$ \\
0.05 & $0.036072 \pm 0.011541$ & $0.082922 \pm 0.381518$ & $0.923$ \\
0.1 & $0.036208 \pm 0.010680$ & $0.012680 \pm 0.049622$ & $0.956$ \\
0.25 & $0.038357 \pm 0.010149$ & $0.020584 \pm 0.068931$ & $0.933$ \\
0.5 & $0.040931 \pm 0.009031$ & $0.003665 \pm 0.005002$ & $1.000$ \\
1 & $0.051235 \pm 0.004187$ & $0.013193 \pm 0.049524$ & $0.967$ \\

\bottomrule
\end{tabular}
\end{table}

\begin{table}[H]
\centering
\caption{Brunner-Munzel tests and adjusted p-values for PINN and FEM results.}
\label{tab:pinn_fem_tests}
\begin{tabular}{c c c c c}
\toprule
Noise &
Brunner--Munzel $W$ &
$p$ &
$p_{\mathrm{Holm}}$ &
$p_{\mathrm{FDR}}$ \\
\midrule
0 & $\mathrm{N/A}$ & $\mathrm{N/A}$ & $\mathrm{N/A}$ & $\mathrm{N/A}$ \\
0.001 & $-14.000$ & $1.964e-14$ & $\mathbf{1.178e-13}$ & $5.891e-14$ \\
0.01 & $-7.108$ & $7.914e-08$ & $\mathbf{7.914e-08}$ & $7.914e-08$ \\
0.05 & $-9.105$ & $4.065e-10$ & $\mathbf{8.786e-10}$ & $4.878e-10$ \\
0.1 & $-13.339$ & $3.640e-14$ & $\mathbf{1.456e-13}$ & $7.281e-14$ \\
0.25 & $-9.355$ & $2.929e-10$ & $\mathbf{8.786e-10}$ & $4.393e-10$ \\
0.5 & $\mathrm{N/A}$ & $\mathrm{N/A}$ & $\mathrm{N/A}$ & $\mathrm{N/A}$ \\
1 & $-14.000$ & $1.964e-14$ & $\mathbf{1.178e-13}$ & $5.891e-14$ \\

\bottomrule
\end{tabular}
\end{table}

\begin{table}[H]
\centering
\caption{FEM and PINN/FEM results with Vargha-Delaney A effect size.}
\label{tab:pinn_fem_descriptive}
\begin{tabular}{c c c c}
\toprule
Noise & FEM & PINN/FEM & Vargha--Delaney $A$ \\
\midrule
0 & $0.002516 \pm 0.002596$ & $0.006938 \pm 0.006173$ & $0.050$ \\
0.001 & $0.011303 \pm 0.049650$ & $0.012048 \pm 0.009203$ & $0.061$ \\
0.01 & $0.030073 \pm 0.082802$ & $0.009024 \pm 0.007570$ & $0.194$ \\
0.05 & $0.082922 \pm 0.381518$ & $0.011518 \pm 0.010064$ & $0.171$ \\
0.1 & $0.012680 \pm 0.049622$ & $0.010777 \pm 0.008486$ & $0.123$ \\
0.25 & $0.020584 \pm 0.068931$ & $0.013890 \pm 0.022805$ & $0.138$ \\
0.5 & $0.003665 \pm 0.005002$ & $0.012939 \pm 0.009445$ & $0.103$ \\
1 & $0.013193 \pm 0.049524$ & $0.010349 \pm 0.009772$ & $0.247$ \\

\bottomrule
\end{tabular}
\end{table}

\begin{table}[H]
\centering
\caption{Brunner-Munzel tests and adjusted p-values for PINN and PINN/FEM results.}
\label{tab:pinn_fem_tests}
\begin{tabular}{c c c c c}
\toprule
Noise &
Brunner--Munzel $W$ &
$p$ &
$p_{\mathrm{Holm}}$ &
$p_{\mathrm{FDR}}$ \\
\midrule
0 & $12.687$ & $1.423e-13$ & $\mathbf{1.138e-12}$ & $1.138e-12$ \\
0.001 & $11.501$ & $1.300e-12$ & $\mathbf{9.097e-12}$ & $5.198e-12$ \\
0.01 & $4.533$ & $4.887e-05$ & $\mathbf{9.774e-05}$ & $5.585e-05$ \\
0.05 & $5.359$ & $6.640e-06$ & $\mathbf{1.992e-05}$ & $8.853e-06$ \\
0.1 & $7.482$ & $1.137e-08$ & $\mathbf{5.683e-08}$ & $2.273e-08$ \\
0.25 & $6.107$ & $1.825e-07$ & $\mathbf{7.300e-07}$ & $2.920e-07$ \\
0.5 & $8.755$ & $3.877e-12$ & $\mathbf{2.326e-11}$ & $1.034e-11$ \\
1 & $3.963$ & $2.105e-04$ & $\mathbf{2.105e-04}$ & $2.105e-04$ \\

\bottomrule
\end{tabular}
\end{table}

\begin{table}[H]
\centering
\caption{PINN and PINN/FEM results with Vargha-Delaney A effect size.}
\label{tab:pinn_fem_descriptive}
\begin{tabular}{c c c c}
\toprule
Noise & PINN & PINN/FEM & Vargha--Delaney $A$ \\
\midrule
0 & $0.032690 \pm 0.009598$ & $0.006938 \pm 0.006173$ & $0.982$ \\
0.001 & $0.039061 \pm 0.009850$ & $0.012048 \pm 0.009203$ & $0.969$ \\
0.01 & $0.036553 \pm 0.010552$ & $0.009024 \pm 0.007570$ & $0.964$ \\
0.05 & $0.036072 \pm 0.011541$ & $0.011518 \pm 0.010064$ & $0.940$ \\
0.1 & $0.036208 \pm 0.010680$ & $0.010777 \pm 0.008486$ & $0.939$ \\
0.25 & $0.038357 \pm 0.010149$ & $0.013890 \pm 0.022805$ & $0.967$ \\
0.5 & $0.040931 \pm 0.009031$ & $0.012939 \pm 0.009445$ & $0.986$ \\
1 & $0.051235 \pm 0.004187$ & $0.010349 \pm 0.009772$ & $1.000$ \\

\bottomrule
\end{tabular}
\end{table}

\begin{table}[H]
\centering
\caption{Brunner-Munzel tests and adjusted p-values for PINN and PINN/FEM results.}
\label{tab:pinn_fem_tests}
\begin{tabular}{c c c c c}
\toprule
Noise &
Brunner--Munzel $W$ &
$p$ &
$p_{\mathrm{Holm}}$ &
$p_{\mathrm{FDR}}$ \\
\midrule
0 & $-40.513$ & $3.069e-44$ & $\mathbf{2.149e-43}$ & $2.149e-43$ \\
0.001 & $-24.209$ & $1.402e-29$ & $\mathbf{7.012e-29}$ & $3.272e-29$ \\
0.01 & $-24.061$ & $2.796e-31$ & $\mathbf{1.678e-30}$ & $9.787e-31$ \\
0.05 & $-15.114$ & $1.070e-21$ & $\mathbf{3.209e-21}$ & $1.498e-21$ \\
0.1 & $-14.911$ & $2.022e-19$ & $\mathbf{4.044e-19}$ & $2.359e-19$ \\
0.25 & $-14.000$ & $1.964e-14$ & $\mathbf{1.964e-14}$ & $1.964e-14$ \\
0.5 & $-32.882$ & $5.131e-26$ & $\mathbf{2.053e-25}$ & $8.980e-26$ \\
1 & $\mathrm{N/A}$ & $\mathrm{N/A}$ & $\mathrm{N/A}$ & $\mathrm{N/A}$ \\

\bottomrule
\end{tabular}
\end{table}

\subsection{3D Taylor-Green Vortex, without outliers}
\label{3dstat_test2}

\begin{table}[H]
\centering
\caption{PINN and FEM results with Vargha-Delaney A effect size.}
\label{tab:pinn_fem_descriptive}
\begin{tabular}{c c c c}
\toprule
Noise & PINN & FEM & Vargha--Delaney $A$ \\
\midrule
0 & $0.032690 \pm 0.009598$ & $0.001923 \pm 0.000019$ & $1.000$ \\
0.001 & $0.039061 \pm 0.009850$ & $0.001920 \pm 0.000001$ & $1.000$ \\
0.01 & $0.036553 \pm 0.010552$ & $0.001924 \pm 0.000018$ & $1.000$ \\
0.05 & $0.036072 \pm 0.011541$ & $0.001919 \pm 0.000001$ & $1.000$ \\
0.1 & $0.036208 \pm 0.010680$ & $0.001922 \pm 0.000006$ & $1.000$ \\
0.25 & $0.037936 \pm 0.010059$ & $0.001935 \pm 0.000022$ & $1.000$ \\
0.5 & $0.040479 \pm 0.008837$ & $0.001996 \pm 0.000111$ & $1.000$ \\
1 & $0.051235 \pm 0.004187$ & $0.002609 \pm 0.001327$ & $1.000$ \\

\bottomrule
\end{tabular}
\end{table}

\begin{table}[H]
\centering
\caption{Brunner-Munzel tests and adjusted p-values for PINN and FEM results.}
\label{tab:pinn_fem_tests}
\begin{tabular}{c c c c c}
\toprule
Noise &
Brunner--Munzel $W$ &
$p$ &
$p_{\mathrm{Holm}}$ &
$p_{\mathrm{FDR}}$ \\
\midrule
0 & $\mathrm{N/A}$ & $\mathrm{N/A}$ & $\mathrm{N/A}$ & $\mathrm{N/A}$ \\
0.001 & $\mathrm{N/A}$ & $\mathrm{N/A}$ & $\mathrm{N/A}$ & $\mathrm{N/A}$ \\
0.01 & $\mathrm{N/A}$ & $\mathrm{N/A}$ & $\mathrm{N/A}$ & $\mathrm{N/A}$ \\
0.05 & $\mathrm{N/A}$ & $\mathrm{N/A}$ & $\mathrm{N/A}$ & $\mathrm{N/A}$ \\
0.1 & $\mathrm{N/A}$ & $\mathrm{N/A}$ & $\mathrm{N/A}$ & $\mathrm{N/A}$ \\
0.25 & $\mathrm{N/A}$ & $\mathrm{N/A}$ & $\mathrm{N/A}$ & $\mathrm{N/A}$ \\
0.5 & $\mathrm{N/A}$ & $\mathrm{N/A}$ & $\mathrm{N/A}$ & $\mathrm{N/A}$ \\
1 & $\mathrm{N/A}$ & $\mathrm{N/A}$ & $\mathrm{N/A}$ & $\mathrm{N/A}$ \\

\bottomrule
\end{tabular}
\end{table}

\begin{table}[H]
\centering
\caption{FEM and PINN/FEM results with Vargha-Delaney A effect size.}
\label{tab:pinn_fem_descriptive}
\begin{tabular}{c c c c}
\toprule
Noise & FEM & PINN/FEM & Vargha--Delaney $A$ \\
\midrule
0 & $0.001923 \pm 0.000019$ & $0.006938 \pm 0.006173$ & $0.000$ \\
0.001 & $0.001920 \pm 0.000001$ & $0.012048 \pm 0.009203$ & $0.000$ \\
0.01 & $0.001924 \pm 0.000018$ & $0.009024 \pm 0.007570$ & $0.036$ \\
0.05 & $0.001919 \pm 0.000001$ & $0.011518 \pm 0.010064$ & $0.000$ \\
0.1 & $0.001922 \pm 0.000006$ & $0.010777 \pm 0.008486$ & $0.000$ \\
0.25 & $0.001935 \pm 0.000022$ & $0.013890 \pm 0.022805$ & $0.033$ \\
0.5 & $0.001996 \pm 0.000111$ & $0.012939 \pm 0.009445$ & $0.041$ \\
1 & $0.002609 \pm 0.001327$ & $0.010349 \pm 0.009772$ & $0.155$ \\

\bottomrule
\end{tabular}
\end{table}

\begin{table}[H]
\centering
\caption{Brunner-Munzel tests and adjusted p-values for FEM and PINN/FEM results.}
\label{tab:pinn_fem_tests}
\begin{tabular}{c c c c c}
\toprule
Noise &
Brunner--Munzel $W$ &
$p$ &
$p_{\mathrm{Holm}}$ &
$p_{\mathrm{FDR}}$ \\
\midrule
0 & $\mathrm{N/A}$ & $\mathrm{N/A}$ & $\mathrm{N/A}$ & $\mathrm{N/A}$ \\
0.001 & $\mathrm{N/A}$ & $\mathrm{N/A}$ & $\mathrm{N/A}$ & $\mathrm{N/A}$ \\
0.01 & $13.868$ & $1.880e-14$ & $\mathbf{7.518e-14}$ & $3.927e-14$ \\
0.05 & $\mathrm{N/A}$ & $\mathrm{N/A}$ & $\mathrm{N/A}$ & $\mathrm{N/A}$ \\
0.1 & $\mathrm{N/A}$ & $\mathrm{N/A}$ & $\mathrm{N/A}$ & $\mathrm{N/A}$ \\
0.25 & $14.000$ & $1.964e-14$ & $\mathbf{7.518e-14}$ & $3.927e-14$ \\
0.5 & $13.473$ & $4.124e-14$ & $\mathbf{8.247e-14}$ & $5.498e-14$ \\
1 & $6.546$ & $2.526e-08$ & $\mathbf{2.526e-08}$ & $2.526e-08$ \\

\bottomrule
\end{tabular}
\end{table}

\begin{table}[H]
\centering
\caption{PINN and PINN/FEM results with Vargha-Delaney A effect size.}
\label{tab:pinn_fem_descriptive}
\begin{tabular}{c c c c}
\toprule
Noise & PINN & PINN/FEM & Vargha--Delaney $A$ \\
\midrule
0 & $0.032690 \pm 0.009598$ & $0.006938 \pm 0.006173$ & $0.982$ \\
0.001 & $0.039061 \pm 0.009850$ & $0.012048 \pm 0.009203$ & $0.969$ \\
0.01 & $0.036553 \pm 0.010552$ & $0.009024 \pm 0.007570$ & $0.964$ \\
0.05 & $0.036072 \pm 0.011541$ & $0.011518 \pm 0.010064$ & $0.940$ \\
0.1 & $0.036208 \pm 0.010680$ & $0.010777 \pm 0.008486$ & $0.939$ \\
0.25 & $0.037936 \pm 0.010059$ & $0.013890 \pm 0.022805$ & $0.967$ \\
0.5 & $0.040479 \pm 0.008837$ & $0.012939 \pm 0.009445$ & $0.985$ \\
1 & $0.051235 \pm 0.004187$ & $0.010349 \pm 0.009772$ & $1.000$ \\

\bottomrule
\end{tabular}
\end{table}

\begin{table}[H]
\centering
\caption{Brunner-Munzel tests and adjusted p-values for PINN and PINN/FEM results.}
\label{tab:pinn_fem_tests}
\begin{tabular}{c c c c c}
\toprule
Noise &
Brunner--Munzel $W$ &
$p$ &
$p_{\mathrm{Holm}}$ &
$p_{\mathrm{FDR}}$ \\
\midrule
0 & $-40.513$ & $3.069e-44$ & $\mathbf{2.149e-43}$ & $2.149e-43$ \\
0.001 & $-24.209$ & $1.402e-29$ & $\mathbf{7.012e-29}$ & $3.272e-29$ \\
0.01 & $-24.061$ & $2.796e-31$ & $\mathbf{1.678e-30}$ & $9.787e-31$ \\
0.05 & $-15.114$ & $1.070e-21$ & $\mathbf{3.209e-21}$ & $1.498e-21$ \\
0.1 & $-14.911$ & $2.022e-19$ & $\mathbf{4.044e-19}$ & $2.359e-19$ \\
0.25 & $-14.000$ & $1.964e-14$ & $\mathbf{1.964e-14}$ & $1.964e-14$ \\
0.5 & $-31.771$ & $1.608e-25$ & $\mathbf{6.434e-25}$ & $2.815e-25$ \\
1 & $\mathrm{N/A}$ & $\mathrm{N/A}$ & $\mathrm{N/A}$ & $\mathrm{N/A}$ \\

\bottomrule
\end{tabular}
\end{table}

\section{Parameter results}
\label{paramres}

In this section we report the mean and median bias of each model. The bias is defined as the deviation between the parameter predicted by each model and the ground truth parameter. Bootstrap 95\% confidence intervals \cite{bootstrap} are also reported for each bias value as well. We characterize the distribution of the bias by reporting IQR, skewness and minimum/maximum values. To determine whether the bias deviates from 0 in a statistically significant manner, we used a two-sided Wilcoxon signed-rank test \cite{wilcox} as the distributions of the biases are substantially skewed, in particular at high noise levels. The resulting $p$-values are reported both in the original form and adjusted using the Holm family-wise error rate correction \cite{holmsture} as well as the Benjamini-Hochberg false discovery rate correction \cite{benjamini}. The Holm-based $p$ values below 0.05\% are shown in bold. Note that for the deterministic FEM results for 1D Burgers' equation with $\sigma=0$, skewness is undefined and the Wilcoxon test is uninformative. In these cases, we report the results as $\mathrm{N/A}$.

\subsection{1D Burgers' equation}
\label{1dparametertest}

Ground truth value for unknown parameter in this problem is $0.01/\pi$. 

\begin{table}[H]
\centering
\label{tab:parameter_bias_descriptive}
\caption{Bias statistics for 1D Burgers' equation tests.}
\begin{tabular}{c c c c c}
\toprule
Noise & Model &
Mean Bias &
Median Bias &
95\% Bootstrap CI for Bias \\
\midrule
0 & PINN & $-0.000098$ & $-0.000115$ &$[-0.000115,\ -0.000081]$ \\
0.5 & PINN & $-0.000342$ & $-0.000256$ &$[-0.000670,\ 0.000002]$ \\
1 & PINN & $-0.000394$ & $-0.001424$ &$[-0.001181,\ 0.000473]$ \\
2 & PINN & $-0.000536$ & $-0.002193$ &$[-0.001672,\ 0.000756]$ \\
3 & PINN & $-0.000644$ & $-0.002362$ &$[-0.002166,\ 0.001495]$ \\
5 & PINN & $0.000348$ & $-0.002549$ &$[-0.002431,\ 0.004670]$ \\
7 & PINN & $-0.001204$ & $-0.002519$ &$[-0.002468,\ 0.000703]$ \\
10 & PINN & $0.360167$ & $-0.001840$ &$[0.030164,\ 0.854303]$ \\
25 & PINN & $25.445983$ & $6.099766$ &$[9.407605,\ 49.872131]$ \\
0 & FEM & $0.000131$ & $0.000131$ &$[0.000131,\ 0.000131]$ \\
0.5 & FEM & $0.000425$ & $0.000489$ &$[0.000112,\ 0.000760]$ \\
1 & FEM & $0.000276$ & $-0.000314$ &$[-0.000482,\ 0.001151]$ \\
2 & FEM & $0.000763$ & $-0.001043$ &$[-0.000637,\ 0.002663]$ \\
3 & FEM & $0.006658$ & $0.002247$ &$[0.002790,\ 0.010983]$ \\
5 & FEM & $0.150866$ & $0.002336$ &$[0.009093,\ 0.359999]$ \\
7 & FEM & $0.371069$ & $0.007848$ &$[0.020457,\ 0.811692]$ \\
10 & FEM & $0.487935$ & $0.008563$ &$[0.075662,\ 1.016190]$ \\
25 & FEM & $2.002900$ & $0.254178$ &$[1.200200,\ 2.825378]$ \\

\bottomrule
\end{tabular}
\end{table}

\begin{table}[H]
\centering
\label{tab:parameter_distribution}
\caption{Characteristics of distribution for 1D Burgers' equation test errors.}
\begin{tabular}{c c c c c c c}
\toprule
Noise & Model &
IQR &
Skewness &
Minimum &
Maximum \\
\midrule
0 & PINN & $0.000050$ & $0.928$ & $-0.000188$ & $0.000038$ \\
0.5 & PINN & $0.001440$ & $0.589$ & $-0.001841$ & $0.002329$ \\
1 & PINN & $0.002127$ & $1.410$ & $-0.002722$ & $0.005951$ \\
2 & PINN & $0.002275$ & $2.044$ & $-0.002882$ & $0.010696$ \\
3 & PINN & $0.000790$ & $3.815$ & $-0.002874$ & $0.023615$ \\
5 & PINN & $0.000558$ & $4.101$ & $-0.002897$ & $0.048962$ \\
7 & PINN & $0.000783$ & $4.262$ & $-0.003113$ & $0.021295$ \\
10 & PINN & $0.028501$ & $4.200$ & $-0.003022$ & $6.106257$ \\
25 & PINN & $20.202263$ & $4.443$ & $-0.003014$ & $322.680601$ \\
0 & FEM & $0.000000$ & $\mathrm{N/A}$ & $0.000131$ & $0.000131$ \\
0.5 & FEM & $0.001205$ & $0.827$ & $-0.001199$ & $0.003356$ \\
1 & FEM & $0.002442$ & $1.824$ & $-0.002789$ & $0.008702$ \\
2 & FEM & $0.003205$ & $3.426$ & $-0.002492$ & $0.021940$ \\
3 & FEM & $0.011753$ & $1.847$ & $-0.003101$ & $0.043822$ \\
5 & FEM & $0.029121$ & $3.654$ & $-0.003147$ & $2.155088$ \\
7 & FEM & $0.046579$ & $3.314$ & $-0.003162$ & $4.996817$ \\
10 & FEM & $0.136785$ & $3.025$ & $-0.003150$ & $4.996817$ \\
25 & FEM & $4.980715$ & $0.463$ & $-0.003185$ & $4.996817$ \\

\bottomrule
\end{tabular}
\end{table}

\begin{table}[H]
\centering
\label{tab:parameter_bias_tests}
\caption{Wilcoxon signed-rank test statistics for 1D Burgers' equation.}
\begin{tabular}{c c c c c c c}
\toprule
Noise & Model &
$W$ &
$p$ &
$p_{\mathrm{Holm}}$ &
$p_{\mathrm{FDR}}$ \\
\midrule
0 & PINN & $5.0$ & $1.863e-08$ & $\mathbf{3.166e-07}$ & \textbf{$1.583e-07$} \\
0.5 & PINN & $138.0$ & $0.0523$ & $0.2613$ & $0.0683$ \\
1 & PINN & $170.0$ & $0.2054$ & $0.7923$ & $0.2328$ \\
2 & PINN & $126.0$ & $0.0277$ & $0.1664$ & \textbf{$3.930e-02$} \\
3 & PINN & $96.0$ & $0.0040$ & $\mathbf{4.032e-02}$ & \textbf{$8.568e-03$} \\
5 & PINN & $59.0$ & $1.529e-04$ & $\mathbf{2.140e-03}$ & \textbf{$6.412e-04$} \\
7 & PINN & $61.0$ & $1.886e-04$ & $\mathbf{2.452e-03}$ & \textbf{$6.412e-04$} \\
10 & PINN & $169.0$ & $0.1981$ & $0.7923$ & $0.2328$ \\
25 & PINN & $10.0$ & $8.009e-08$ & $\mathbf{1.201e-06}$ & \textbf{$4.539e-07$} \\
0 & FEM & $\mathrm{N/A}$ & $\mathrm{N/A}$ & $\mathrm{N/A}$ & $\mathrm{N/A}$ \\
0.5 & FEM & $116.0$ & $0.0155$ & $0.1082$ & \textbf{$2.389e-02$} \\
1 & FEM & $223.0$ & $0.8553$ & $1.0000$ & $0.8553$ \\
2 & FEM & $215.0$ & $0.7303$ & $1.0000$ & $0.7760$ \\
3 & FEM & $110.0$ & $0.0106$ & $0.0894$ & \textbf{$1.802e-02$} \\
5 & FEM & $109.0$ & $0.0099$ & $0.0894$ & \textbf{$1.802e-02$} \\
7 & FEM & $71.0$ & $5.055e-04$ & $\mathbf{6.066e-03}$ & \textbf{$1.432e-03$} \\
10 & FEM & $73.0$ & $6.084e-04$ & $\mathbf{6.692e-03}$ & \textbf{$1.478e-03$} \\
25 & FEM & $5.0$ & $1.863e-08$ & $\mathbf{3.166e-07}$ & \textbf{$1.583e-07$} \\

\bottomrule
\end{tabular}
\end{table}

\subsection{1D Burgers' equation with two parameters}

Ground truth value for viscosity parameter in this problem is $0.01/\pi$. Ground truth value for advection parameter in this problem is $0.55$. 

\begin{table}[H]
\centering
\label{tab:parameter_bias_descriptive}
\caption{Bias statistics for viscosity prediction in 1D Burgers' equation tests.}
\begin{tabular}{c c c c c}
\toprule
Noise & Model &
Mean Bias &
Median Bias &
95\% Bootstrap CI for Bias \\
\midrule
0 & PINN & $0.000193$ & $0.000198$ &$[0.000181,\ 0.000206]$ \\
0.5 & PINN & $-0.000583$ & $-0.000531$ &$[-0.000947,\ -0.000226]$ \\
1 & PINN & $-0.001792$ & $-0.002553$ &$[-0.002328,\ -0.001170]$ \\
2 & PINN & $-0.002821$ & $-0.003034$ &$[-0.003021,\ -0.002507]$ \\
3 & PINN & $-0.002983$ & $-0.003060$ &$[-0.003058,\ -0.002897]$ \\
5 & PINN & $-0.000553$ & $-0.003103$ &$[-0.003084,\ 0.004446]$ \\
7 & PINN & $0.061586$ & $-0.003117$ &$[-0.003069,\ 0.186921]$ \\
10 & PINN & $0.256946$ & $-0.002925$ &$[0.009165,\ 0.617358]$ \\
25 & PINN & $9.494277$ & $1.600279$ &$[3.282049,\ 17.611974]$ \\
0 & FEM & $0.000152$ & $0.000152$ &$[0.000152,\ 0.000152]$ \\
0.5 & FEM & $0.000162$ & $-0.000221$ &$[-0.000148,\ 0.000493]$ \\
1 & FEM & $0.000824$ & $0.000817$ &$[0.000287,\ 0.001370]$ \\
2 & FEM & $-0.000152$ & $-0.001336$ &$[-0.001145,\ 0.000951]$ \\
3 & FEM & $0.001838$ & $0.000419$ &$[0.000374,\ 0.003564]$ \\
5 & FEM & $0.102238$ & $-0.000375$ &$[0.002358,\ 0.297202]$ \\
7 & FEM & $0.562331$ & $-0.000480$ &$[0.077626,\ 1.136288]$ \\
10 & FEM & $1.182931$ & $0.002623$ &$[0.511359,\ 1.932753]$ \\
25 & FEM & $1.268733$ & $0.005378$ &$[0.566791,\ 2.052807]$ \\

\bottomrule
\end{tabular}
\end{table}

\begin{table}[H]
\centering
\label{tab:parameter_distribution}
\caption{Characteristics of distribution for viscosity prediction in 1D Burgers' equation.}
\begin{tabular}{c c c c c c c}
\toprule
Noise & Model &
IQR &
Skewness &
Minimum &
Maximum \\
\midrule
0 & PINN & $0.000060$ & $-0.116$ & $0.000134$ & $0.000247$ \\
0.5 & PINN & $0.001390$ & $-0.360$ & $-0.002745$ & $0.001270$ \\
1 & PINN & $0.001450$ & $2.036$ & $-0.003182$ & $0.003860$ \\
2 & PINN & $0.000392$ & $4.604$ & $-0.003183$ & $0.001037$ \\
3 & PINN & $0.000287$ & $1.605$ & $-0.003183$ & $-0.002260$ \\
5 & PINN & $0.000266$ & $5.475$ & $-0.003183$ & $0.071309$ \\
7 & PINN & $0.000263$ & $5.469$ & $-0.003183$ & $1.856309$ \\
10 & PINN & $0.005934$ & $3.645$ & $-0.003183$ & $3.807769$ \\
25 & PINN & $5.715438$ & $3.112$ & $-0.003183$ & $94.327661$ \\
0 & FEM & $0.000000$ & $\mathrm{N/A}$ & $0.000152$ & $0.000152$ \\
0.5 & FEM & $0.001013$ & $0.987$ & $-0.001184$ & $0.002327$ \\
1 & FEM & $0.002302$ & $0.289$ & $-0.001783$ & $0.003989$ \\
2 & FEM & $0.004060$ & $1.225$ & $-0.002987$ & $0.008324$ \\
3 & FEM & $0.005226$ & $1.853$ & $-0.002894$ & $0.018063$ \\
5 & FEM & $0.015898$ & $5.470$ & $-0.003346$ & $2.883208$ \\
7 & FEM & $0.014215$ & $2.555$ & $-0.003350$ & $4.996817$ \\
10 & FEM & $2.289449$ & $1.209$ & $-0.003343$ & $4.996817$ \\
25 & FEM & $1.494229$ & $1.246$ & $-0.003355$ & $4.996817$ \\

\bottomrule
\end{tabular}
\end{table}

\begin{table}[H]
\centering
\label{tab:parameter_bias_tests}
\caption{Wilcoxon signed-rank tests for viscosity prediction in 1D Burgers' equation.}
\begin{tabular}{c c c c c c c}
\toprule
Noise & Model &
$W$ &
$p$ &
$p_{\mathrm{Holm}}$ &
$p_{\mathrm{FDR}}$ \\
\midrule
0 & PINN & $0.0$ & $1.863e-09$ & $\mathbf{3.166e-08}$ & \textbf{$1.583e-08$} \\
0.5 & PINN & $103.0$ & $0.0066$ & $0.0598$ & \textbf{$1.254e-02$} \\
1 & PINN & $45.0$ & $3.050e-05$ & $\mathbf{3.659e-04}$ & \textbf{$8.640e-05$} \\
2 & PINN & $1.0$ & $3.725e-09$ & $\mathbf{5.588e-08}$ & \textbf{$2.111e-08$} \\
3 & PINN & $0.0$ & $1.863e-09$ & $\mathbf{3.166e-08}$ & \textbf{$1.583e-08$} \\
5 & PINN & $30.0$ & $3.790e-06$ & $\mathbf{4.928e-05}$ & \textbf{$1.289e-05$} \\
7 & PINN & $87.0$ & $0.0020$ & $\mathbf{2.020e-02}$ & \textbf{$4.293e-03$} \\
10 & PINN & $199.0$ & $0.5028$ & $1.0000$ & $0.5342$ \\
25 & PINN & $27.0$ & $2.349e-06$ & $\mathbf{3.288e-05}$ & \textbf{$9.982e-06$} \\
0 & FEM & $\mathrm{N/A}$ & $\mathrm{N/A}$ & $\mathrm{N/A}$ & $\mathrm{N/A}$ \\
0.5 & FEM & $219.0$ & $0.7922$ & $1.0000$ & $0.7922$ \\
1 & FEM & $107.0$ & $0.0087$ & $0.0696$ & \textbf{$1.480e-02$} \\
2 & FEM & $191.0$ & $0.4045$ & $1.0000$ & $0.4584$ \\
3 & FEM & $155.0$ & $0.1142$ & $0.6851$ & $0.1618$ \\
5 & FEM & $178.0$ & $0.2710$ & $1.0000$ & $0.3291$ \\
7 & FEM & $173.0$ & $0.2286$ & $1.0000$ & $0.2989$ \\
10 & FEM & $116.0$ & $0.0155$ & $0.1082$ & \textbf{$2.389e-02$} \\
25 & FEM & $85.0$ & $0.0017$ & $\mathbf{1.890e-02}$ & \textbf{$4.174e-03$} \\

\bottomrule
\end{tabular}
\end{table}

\begin{table}[H]
\centering
\label{tab:parameter_bias_descriptive}
\caption{Bias statistics for advection prediction in 1D Burgers' equation tests.}
\begin{tabular}{c c c c c}
\toprule
Noise & Model &
Mean Bias &
Median Bias &
95\% Bootstrap CI for Bias \\
\midrule
0 & PINN & $0.002167$ & $0.002157$ &$[0.002081,\ 0.002256]$ \\
0.5 & PINN & $-0.005446$ & $0.001737$ &$[-0.014797,\ 0.003965]$ \\
1 & PINN & $-0.038576$ & $-0.040225$ &$[-0.059253,\ -0.018518]$ \\
2 & PINN & $-0.080095$ & $-0.059642$ &$[-0.115137,\ -0.046315]$ \\
3 & PINN & $-0.111381$ & $-0.106573$ &$[-0.173821,\ -0.051607]$ \\
5 & PINN & $-0.217104$ & $-0.199926$ &$[-0.277550,\ -0.157503]$ \\
7 & PINN & $-0.231970$ & $-0.256692$ &$[-0.314987,\ -0.148053]$ \\
10 & PINN & $0.083602$ & $-0.322993$ &$[-0.299589,\ 0.727331]$ \\
25 & PINN & $0.742806$ & $-0.085778$ &$[0.031044,\ 1.713769]$ \\
0 & FEM & $0.001291$ & $0.001291$ &$[0.001291,\ 0.001291]$ \\
0.5 & FEM & $-0.004102$ & $-0.006705$ &$[-0.012382,\ 0.004651]$ \\
1 & FEM & $0.009455$ & $0.001445$ &$[-0.005047,\ 0.025132]$ \\
2 & FEM & $-0.002922$ & $-0.006106$ &$[-0.034036,\ 0.028089]$ \\
3 & FEM & $-0.023035$ & $-0.038868$ &$[-0.068635,\ 0.022375]$ \\
5 & FEM & $0.027904$ & $-0.047465$ &$[-0.056421,\ 0.114937]$ \\
7 & FEM & $-0.017300$ & $-0.059742$ &$[-0.109609,\ 0.075631]$ \\
10 & FEM & $0.007344$ & $0.007008$ &$[-0.134890,\ 0.144440]$ \\
25 & FEM & $-0.046517$ & $-0.007033$ &$[-0.192782,\ 0.101437]$ \\

\bottomrule
\end{tabular}
\end{table}

\begin{table}[H]
\centering
\label{tab:parameter_distribution}
\caption{Distribution characteristics for advection prediction in 1D Burgers' equation.}
\begin{tabular}{c c c c c c c}
\toprule
Noise & Model &
IQR &
Skewness &
Minimum &
Maximum \\
\midrule
0 & PINN & $0.000439$ & $0.036$ & $0.001703$ & $0.002567$ \\
0.5 & PINN & $0.038467$ & $0.064$ & $-0.052701$ & $0.057314$ \\
1 & PINN & $0.071762$ & $-0.135$ & $-0.153387$ & $0.079692$ \\
2 & PINN & $0.115984$ & $-0.403$ & $-0.258359$ & $0.068890$ \\
3 & PINN & $0.132001$ & $-0.376$ & $-0.540691$ & $0.177078$ \\
5 & PINN & $0.190839$ & $-0.605$ & $-0.550000$ & $0.050067$ \\
7 & PINN & $0.413568$ & $0.201$ & $-0.550000$ & $0.203899$ \\
10 & PINN & $0.454597$ & $5.029$ & $-0.550000$ & $8.383439$ \\
25 & PINN & $1.174251$ & $3.109$ & $-0.550000$ & $10.835713$ \\
0 & FEM & $0.000000$ & $\mathrm{N/A}$ & $0.001291$ & $0.001291$ \\
0.5 & FEM & $0.020729$ & $0.996$ & $-0.049999$ & $0.063621$ \\
1 & FEM & $0.063369$ & $0.543$ & $-0.051569$ & $0.105989$ \\
2 & FEM & $0.111149$ & $0.183$ & $-0.158631$ & $0.171450$ \\
3 & FEM & $0.170158$ & $-0.113$ & $-0.349669$ & $0.268005$ \\
5 & FEM & $0.365696$ & $0.558$ & $-0.337237$ & $0.450000$ \\
7 & FEM & $0.280093$ & $-0.057$ & $-0.550000$ & $0.450000$ \\
10 & FEM & $0.809269$ & $-0.266$ & $-0.550000$ & $0.450000$ \\
25 & FEM & $0.991587$ & $-0.040$ & $-0.550000$ & $0.450000$ \\

\bottomrule
\end{tabular}
\end{table}

\begin{table}[H]
\centering
\label{tab:parameter_bias_tests}
\caption{Wilcoxon signed-rank tests for advection prediction in 1D Burgers' equation.}
\begin{tabular}{c c c c c c c}
\toprule
Noise & Model &
$W$ &
$p$ &
$p_{\mathrm{Holm}}$ &
$p_{\mathrm{FDR}}$ \\
\midrule
0 & PINN & $0.0$ & $1.863e-09$ & $\mathbf{3.166e-08}$ & \textbf{$3.166e-08$} \\
0.5 & PINN & $184.0$ & $0.3285$ & $1.0000$ & $0.5584$ \\
1 & PINN & $83.0$ & $0.0015$ & $\mathbf{1.895e-02}$ & \textbf{$4.955e-03$} \\
2 & PINN & $56.0$ & $1.106e-04$ & $\mathbf{1.548e-03}$ & \textbf{$4.700e-04$} \\
3 & PINN & $94.0$ & $0.0035$ & $\mathbf{4.170e-02}$ & \textbf{$9.846e-03$} \\
5 & PINN & $8.0$ & $4.657e-08$ & $\mathbf{7.451e-07}$ & \textbf{$3.958e-07$} \\
7 & PINN & $45.0$ & $3.050e-05$ & $\mathbf{4.574e-04}$ & \textbf{$1.728e-04$} \\
10 & PINN & $106.0$ & $0.0081$ & $0.0896$ & \textbf{$1.978e-02$} \\
25 & PINN & $209.0$ & $0.6408$ & $1.0000$ & $0.7781$ \\
0 & FEM & $\mathrm{N/A}$ & $\mathrm{N/A}$ & $\mathrm{N/A}$ & $\mathrm{N/A}$ \\
0.5 & FEM & $157.0$ & $0.1241$ & $1.0000$ & $0.2638$ \\
1 & FEM & $191.0$ & $0.4045$ & $1.0000$ & $0.5730$ \\
2 & FEM & $223.0$ & $0.8553$ & $1.0000$ & $0.8553$ \\
3 & FEM & $189.0$ & $0.3818$ & $1.0000$ & $0.5730$ \\
5 & FEM & $220.0$ & $0.8078$ & $1.0000$ & $0.8553$ \\
7 & FEM & $205.0$ & $0.5838$ & $1.0000$ & $0.7634$ \\
10 & FEM & $221.0$ & $0.8236$ & $1.0000$ & $0.8553$ \\
25 & FEM & $177.0$ & $0.2621$ & $1.0000$ & $0.4951$ \\

\bottomrule
\end{tabular}
\end{table}

\subsection{2D Taylor-Green Vortex}

Ground truth value for unknown parameter in this problem is $0.1$.

\begin{table}[H]
\centering
\label{tab:parameter_bias_descriptive}
\caption{Bias statistics for viscosity prediction in 2D Taylor-Green Vortex tests.}
\begin{tabular}{c c c c c}
\toprule
Noise & Model &
Mean Bias &
Median Bias &
95\% Bootstrap CI for Bias \\
\midrule
0 & PINN & $-0.000558$ & $0.000026$ &$[-0.001053,\ -0.000148]$ \\
0.001 & PINN & $-0.000524$ & $-0.000053$ &$[-0.001027,\ -0.000145]$ \\
0.01 & PINN & $-0.000184$ & $-0.000039$ &$[-0.000496,\ 0.000041]$ \\
0.05 & PINN & $-0.000889$ & $-0.000776$ &$[-0.001333,\ -0.000462]$ \\
0.1 & PINN & $-0.002075$ & $-0.002093$ &$[-0.002659,\ -0.001515]$ \\
0.25 & PINN & $-0.005316$ & $-0.005385$ &$[-0.006187,\ -0.004425]$ \\
0.5 & PINN & $-0.008795$ & $-0.007857$ &$[-0.009940,\ -0.007670]$ \\
1 & PINN & $-0.009789$ & $-0.009772$ &$[-0.012159,\ -0.007637]$ \\
0 & FEM & $0.000104$ & $0.000104$ &$[0.000104,\ 0.000104]$ \\
0.001 & FEM & $0.000104$ & $0.000104$ &$[0.000103,\ 0.000104]$ \\
0.01 & FEM & $0.000099$ & $0.000099$ &$[0.000092,\ 0.000106]$ \\
0.05 & FEM & $0.000084$ & $0.000065$ &$[0.000041,\ 0.000126]$ \\
0.1 & FEM & $0.000170$ & $0.000208$ &$[0.000101,\ 0.000237]$ \\
0.25 & FEM & $0.000117$ & $0.000077$ &$[-0.000069,\ 0.000304]$ \\
0.5 & FEM & $0.000087$ & $0.000007$ &$[-0.000284,\ 0.000469]$ \\
1 & FEM & $0.000120$ & $-0.000161$ &$[-0.000419,\ 0.000667]$ \\

\bottomrule
\end{tabular}
\end{table}

\begin{table}[H]
\centering
\caption{Distribution characteristics for viscosity prediction in 2D Taylor-Green vortex.}
\label{tab:parameter_distribution}
\begin{tabular}{c c c c c c c}
\toprule
Noise & Model &
IQR &
Skewness &
Minimum &
Maximum \\
\midrule
0 & PINN & $0.000214$ & $-2.142$ & $-0.004590$ & $0.000106$ \\
0.001 & PINN & $0.000161$ & $-2.803$ & $-0.005437$ & $0.000245$ \\
0.01 & PINN & $0.000244$ & $-3.614$ & $-0.003713$ & $0.000770$ \\
0.05 & PINN & $0.001946$ & $-0.077$ & $-0.002847$ & $0.001184$ \\
0.1 & PINN & $0.002040$ & $-0.125$ & $-0.005230$ & $0.001228$ \\
0.25 & PINN & $0.002774$ & $0.181$ & $-0.010830$ & $0.000586$ \\
0.5 & PINN & $0.004669$ & $-0.353$ & $-0.015567$ & $-0.003149$ \\
1 & PINN & $0.006672$ & $-1.402$ & $-0.028845$ & $-0.000376$ \\
0 & FEM & $0.000000$ & $-0.187$ & $0.000103$ & $0.000104$ \\
0.001 & FEM & $0.000002$ & $0.816$ & $0.000101$ & $0.000108$ \\
0.01 & FEM & $0.000018$ & $-0.331$ & $0.000044$ & $0.000142$ \\
0.05 & FEM & $0.000162$ & $0.006$ & $-0.000164$ & $0.000283$ \\
0.1 & FEM & $0.000229$ & $-0.413$ & $-0.000343$ & $0.000570$ \\
0.25 & FEM & $0.000998$ & $0.056$ & $-0.000906$ & $0.000950$ \\
0.5 & FEM & $0.001626$ & $0.389$ & $-0.001635$ & $0.002386$ \\
1 & FEM & $0.002415$ & $0.416$ & $-0.002274$ & $0.002966$ \\

\bottomrule
\end{tabular}
\end{table}

\begin{table}[H]
\centering
\label{tab:parameter_bias_tests}
\caption{Wilcoxon signed-rank tests for viscosity prediction in 2D Taylor-Green Vortex.}
\begin{tabular}{c c c c c c c}
\toprule
Noise & Model &
$W$ &
$p$ &
$p_{\mathrm{Holm}}$ &
$p_{\mathrm{FDR}}$ \\
\midrule
0 & PINN & $189.0$ & $0.3818$ & $1.0000$ & $0.4699$ \\
0.001 & PINN & $112.0$ & $0.0120$ & $0.0723$ & \textbf{$1.752e-02$} \\
0.01 & PINN & $200.0$ & $0.5158$ & $1.0000$ & $0.5895$ \\
0.05 & PINN & $79.0$ & $0.0010$ & $\mathbf{7.267e-03}$ & \textbf{$1.661e-03$} \\
0.1 & PINN & $16.0$ & $3.148e-07$ & $\mathbf{3.148e-06}$ & \textbf{$7.195e-07$} \\
0.25 & PINN & $1.0$ & $3.725e-09$ & $\mathbf{4.098e-08}$ & \textbf{$9.934e-09$} \\
0.5 & PINN & $0.0$ & $1.863e-09$ & $\mathbf{2.980e-08}$ & \textbf{$5.960e-09$} \\
1 & PINN & $0.0$ & $1.863e-09$ & $\mathbf{2.980e-08}$ & \textbf{$5.960e-09$} \\
0 & FEM & $0.0$ & $1.863e-09$ & $\mathbf{2.980e-08}$ & \textbf{$5.960e-09$} \\
0.001 & FEM & $0.0$ & $1.863e-09$ & $\mathbf{2.980e-08}$ & \textbf{$5.960e-09$} \\
0.01 & FEM & $0.0$ & $1.863e-09$ & $\mathbf{2.980e-08}$ & \textbf{$5.960e-09$} \\
0.05 & FEM & $72.0$ & $5.548e-04$ & $\mathbf{4.439e-03}$ & \textbf{$9.864e-04$} \\
0.1 & FEM & $51.0$ & $6.287e-05$ & $\mathbf{5.658e-04}$ & \textbf{$1.257e-04$} \\
0.25 & FEM & $184.0$ & $0.3285$ & $1.0000$ & $0.4380$ \\
0.5 & FEM & $217.0$ & $0.7611$ & $1.0000$ & $0.8118$ \\
1 & FEM & $224.0$ & $0.8712$ & $1.0000$ & $0.8712$ \\

\bottomrule
\end{tabular}
\end{table}

\subsection{3D Taylor-Green Vortex, with outliers}

Ground truth value for unknown parameter in this problem is $0.01$. 

\begin{table}[H]
\centering
\caption{Bias statistics for viscosity prediction in 3D Taylor-Green Vortex tests.}
\label{tab:parameter_bias_descriptive}
\begin{tabular}{c c c c c}
\toprule
Noise & Model &
Mean Bias &
Median Bias &
95\% Bootstrap CI for Bias \\
\midrule
0 & PINN & $0.000635$ & $0.001489$ &$[-0.000784,\ 0.001951]$ \\
0.001 & PINN & $-0.001079$ & $-0.000067$ &$[-0.002878,\ 0.000755]$ \\
0.01 & PINN & $0.000541$ & $0.000964$ &$[-0.001189,\ 0.002187]$ \\
0.05 & PINN & $-0.001477$ & $-0.000084$ &$[-0.003103,\ 0.000115]$ \\
0.1 & PINN & $-0.001392$ & $-0.000025$ &$[-0.003050,\ 0.000192]$ \\
0.25 & PINN & $0.003627$ & $0.001124$ &$[-0.002388,\ 0.013921]$ \\
0.5 & PINN & $-0.000971$ & $-0.001744$ &$[-0.003294,\ 0.001756]$ \\
1 & PINN & $-0.002776$ & $-0.002551$ &$[-0.003954,\ -0.001583]$ \\
0 & FEM & $-0.000041$ & $0.000294$ &$[-0.000603,\ 0.000295]$ \\
0.001 & FEM & $0.166339$ & $0.000294$ &$[-0.000220,\ 0.499160]$ \\
0.01 & FEM & $0.493919$ & $0.000293$ &$[-0.000174,\ 0.996085]$ \\
0.05 & FEM & $0.165154$ & $0.000289$ &$[-0.001825,\ 0.498047]$ \\
0.1 & FEM & $0.165684$ & $0.000313$ &$[-0.001089,\ 0.498493]$ \\
0.25 & FEM & $0.329266$ & $0.000243$ &$[-0.000306,\ 0.824930]$ \\
0.5 & FEM & $-0.000425$ & $0.000278$ &$[-0.001250,\ 0.000238]$ \\
1 & FEM & $0.165420$ & $0.000112$ &$[-0.001460,\ 0.498441]$ \\

\bottomrule
\end{tabular}
\end{table}

\begin{table}[H]
\centering
\caption{Distribution characteristics for viscosity prediction in 3D Taylor-Green vortex.}
\label{tab:parameter_distribution}
\begin{tabular}{c c c c c c c}
\toprule
Noise & Model &
IQR &
Skewness &
Minimum &
Maximum \\
\midrule
0 & PINN & $0.004957$ & $-0.759$ & $-0.006858$ & $0.006629$ \\
0.001 & PINN & $0.009734$ & $0.065$ & $-0.006858$ & $0.008105$ \\
0.01 & PINN & $0.005261$ & $-0.349$ & $-0.006858$ & $0.009968$ \\
0.05 & PINN & $0.009177$ & $-0.185$ & $-0.006858$ & $0.004507$ \\
0.1 & PINN & $0.009413$ & $-0.173$ & $-0.006858$ & $0.005126$ \\
0.25 & PINN & $0.009551$ & $5.233$ & $-0.006858$ & $0.141775$ \\
0.5 & PINN & $0.009039$ & $1.981$ & $-0.006858$ & $0.026158$ \\
1 & PINN & $0.006602$ & $0.210$ & $-0.006858$ & $0.004680$ \\
0 & FEM & $0.000000$ & $-4.316$ & $-0.006475$ & $0.000316$ \\
0.001 & FEM & $0.000001$ & $5.477$ & $-0.003751$ & $4.988961$ \\
0.01 & FEM & $0.000008$ & $2.810$ & $-0.006858$ & $4.987737$ \\
0.05 & FEM & $0.000054$ & $5.477$ & $-0.006858$ & $4.985034$ \\
0.1 & FEM & $0.000093$ & $5.477$ & $-0.006858$ & $4.985623$ \\
0.25 & FEM & $0.000173$ & $3.661$ & $-0.005090$ & $4.989280$ \\
0.5 & FEM & $0.000429$ & $-2.445$ & $-0.006858$ & $0.001019$ \\
1 & FEM & $0.001526$ & $5.477$ & $-0.006858$ & $4.988669$ \\

\bottomrule
\end{tabular}
\end{table}

\begin{table}[H]
\centering
\label{tab:parameter_bias_tests}
\caption{Wilcoxon signed-rank tests for viscosity prediction in 3D Taylor-Green Vortex.}
\begin{tabular}{c c c c c c c}
\toprule
Noise & Model &
$W$ &
$p$ &
$p_{\mathrm{Holm}}$ &
$p_{\mathrm{FDR}}$ \\
\midrule
0 & PINN & $173.0$ & $0.2286$ & $1.0000$ & $0.2813$ \\
0.001 & PINN & $167.0$ & $0.1840$ & $1.0000$ & $0.2453$ \\
0.01 & PINN & $192.0$ & $0.4161$ & $1.0000$ & $0.4756$ \\
0.05 & PINN & $160.0$ & $0.1403$ & $1.0000$ & $0.2209$ \\
0.1 & PINN & $162.0$ & $0.1519$ & $1.0000$ & $0.2209$ \\
0.25 & PINN & $205.0$ & $0.5838$ & $1.0000$ & $0.5838$ \\
0.5 & PINN & $155.0$ & $0.1142$ & $1.0000$ & $0.2209$ \\
1 & PINN & $65.0$ & $2.833e-04$ & $\mathbf{3.682e-03}$ & \textbf{$1.133e-03$} \\
0 & FEM & $59.0$ & $1.529e-04$ & $\mathbf{2.140e-03}$ & \textbf{$8.153e-04$} \\
0.001 & FEM & $57.0$ & $1.233e-04$ & $\mathbf{1.850e-03}$ & \textbf{$8.153e-04$} \\
0.01 & FEM & $78.0$ & $9.518e-04$ & $\mathbf{1.142e-02}$ & \textbf{$3.046e-03$} \\
0.05 & FEM & $159.0$ & $0.1347$ & $1.0000$ & $0.2209$ \\
0.1 & FEM & $135.0$ & $0.0449$ & $0.4940$ & $0.1198$ \\
0.25 & FEM & $56.0$ & $1.106e-04$ & $\mathbf{1.769e-03}$ & \textbf{$8.153e-04$} \\
0.5 & FEM & $144.0$ & $0.0699$ & $0.6989$ & $0.1598$ \\
1 & FEM & $204.0$ & $0.5699$ & $1.0000$ & $0.5838$ \\

\bottomrule
\end{tabular}
\end{table}

\subsection{3D Taylor-Green Vortex, without outliers}

Ground truth value for unknown parameter in this problem is $0.01$. 

\begin{table}[H]
\centering
\label{tab:parameter_bias_descriptive}
\caption{Bias statistics for viscosity prediction in 3D Taylor-Green Vortex tests.}
\begin{tabular}{c c c c c}
\toprule
Noise & Model &
Mean Bias &
Median Bias &
95\% Bootstrap CI for Bias \\
\midrule
0 & PINN & $0.000635$ & $0.001489$ &$[-0.000784,\ 0.001951]$ \\
0.001 & PINN & $-0.001079$ & $-0.000067$ &$[-0.002878,\ 0.000755]$ \\
0.01 & PINN & $0.000541$ & $0.000964$ &$[-0.001189,\ 0.002187]$ \\
0.05 & PINN & $-0.001477$ & $-0.000084$ &$[-0.003103,\ 0.000115]$ \\
0.1 & PINN & $-0.001392$ & $-0.000025$ &$[-0.003050,\ 0.000192]$ \\
0.25 & PINN & $-0.001137$ & $0.001096$ &$[-0.002790,\ 0.000444]$ \\
0.5 & PINN & $-0.001906$ & $-0.002592$ &$[-0.003642,\ -0.000099]$ \\
1 & PINN & $-0.002776$ & $-0.002551$ &$[-0.003955,\ -0.001550]$ \\
0 & FEM & $0.000294$ & $0.000294$ &$[0.000294,\ 0.000294]$ \\
0.001 & FEM & $0.000294$ & $0.000294$ &$[0.000293,\ 0.000294]$ \\
0.01 & FEM & $0.000294$ & $0.000294$ &$[0.000292,\ 0.000296]$ \\
0.05 & FEM & $0.000298$ & $0.000298$ &$[0.000289,\ 0.000307]$ \\
0.1 & FEM & $0.000312$ & $0.000329$ &$[0.000291,\ 0.000331]$ \\
0.25 & FEM & $0.000246$ & $0.000243$ &$[0.000198,\ 0.000294]$ \\
0.5 & FEM & $0.000350$ & $0.000328$ &$[0.000217,\ 0.000474]$ \\
1 & FEM & $-0.000206$ & $0.000133$ &$[-0.000704,\ 0.000229]$ \\

\bottomrule
\end{tabular}
\end{table}

\begin{table}[H]
\centering
\label{tab:parameter_distribution}
\caption{Distribution characteristics for viscosity prediction in 3D Taylor-Green vortex.}
\begin{tabular}{c c c c c c c}
\toprule
Noise & Model &
IQR &
Skewness &
Minimum &
Maximum \\
\midrule
0 & PINN & $0.004957$ & $-0.759$ & $-0.006858$ & $0.006629$ \\
0.001 & PINN & $0.009734$ & $0.065$ & $-0.006858$ & $0.008105$ \\
0.01 & PINN & $0.005261$ & $-0.349$ & $-0.006858$ & $0.009968$ \\
0.05 & PINN & $0.009177$ & $-0.185$ & $-0.006858$ & $0.004507$ \\
0.1 & PINN & $0.009413$ & $-0.173$ & $-0.006858$ & $0.005126$ \\
0.25 & PINN & $0.009237$ & $-0.319$ & $-0.006858$ & $0.005572$ \\
0.5 & PINN & $0.009002$ & $0.411$ & $-0.006858$ & $0.009137$ \\
1 & PINN & $0.006602$ & $0.210$ & $-0.006858$ & $0.004680$ \\
0 & FEM & $0.000000$ & $0.658$ & $0.000294$ & $0.000294$ \\
0.001 & FEM & $0.000001$ & $-0.474$ & $0.000292$ & $0.000295$ \\
0.01 & FEM & $0.000005$ & $0.298$ & $0.000283$ & $0.000307$ \\
0.05 & FEM & $0.000031$ & $-0.264$ & $0.000255$ & $0.000334$ \\
0.1 & FEM & $0.000074$ & $-0.200$ & $0.000216$ & $0.000423$ \\
0.25 & FEM & $0.000105$ & $0.141$ & $-0.000024$ & $0.000554$ \\
0.5 & FEM & $0.000356$ & $-0.336$ & $-0.000417$ & $0.001019$ \\
1 & FEM & $0.000897$ & $-1.577$ & $-0.003265$ & $0.001259$ \\

\bottomrule
\end{tabular}
\end{table}

\begin{table}[H]
\centering
\label{tab:parameter_bias_tests}
\caption{Wilcoxon signed-rank tests for viscosity prediction in 3D Taylor-Green Vortex.}
\begin{tabular}{c c c c c c c}
\toprule
Noise & Model &
$W$ &
$p$ &
$p_{\mathrm{Holm}}$ &
$p_{\mathrm{FDR}}$ \\
\midrule
0 & PINN & $173.0$ & $0.2286$ & $0.9820$ & $0.2813$ \\
0.001 & PINN & $167.0$ & $0.1840$ & $0.9820$ & $0.2453$ \\
0.01 & PINN & $192.0$ & $0.4161$ & $1.0000$ & $0.4439$ \\
0.05 & PINN & $160.0$ & $0.1403$ & $0.9820$ & $0.2209$ \\
0.1 & PINN & $162.0$ & $0.1519$ & $0.9820$ & $0.2209$ \\
0.25 & PINN & $175.0$ & $0.3692$ & $1.0000$ & $0.4220$ \\
0.5 & PINN & $125.0$ & $0.0455$ & $0.3642$ & $0.0809$ \\
1 & PINN & $65.0$ & $2.833e-04$ & $\mathbf{2.549e-03}$ & \textbf{$5.665e-04$} \\
0 & FEM & $0.0$ & $9.537e-07$ & $\mathbf{1.049e-05}$ & \textbf{$2.543e-06$} \\
0.001 & FEM & $0.0$ & $5.960e-08$ & $\mathbf{9.537e-07}$ & \textbf{$4.768e-07$} \\
0.01 & FEM & $0.0$ & $4.768e-07$ & $\mathbf{5.722e-06}$ & \textbf{$1.526e-06$} \\
0.05 & FEM & $0.0$ & $2.384e-07$ & $\mathbf{3.099e-06}$ & \textbf{$9.537e-07$} \\
0.1 & FEM & $0.0$ & $1.192e-07$ & $\mathbf{1.669e-06}$ & \textbf{$6.358e-07$} \\
0.25 & FEM & $1.0$ & $5.960e-08$ & $\mathbf{9.537e-07}$ & \textbf{$4.768e-07$} \\
0.5 & FEM & $30.0$ & $6.032e-05$ & $\mathbf{6.032e-04}$ & \textbf{$1.379e-04$} \\
1 & FEM & $151.0$ & $0.7712$ & $1.0000$ & $0.7712$ \\

\bottomrule
\end{tabular}
\end{table}

\section{1D Burgers' equation with two unknown parameters}
\label{twoparamet}

We also tested 1D Burgers' equation with two parameters.

\begin{align}
\frac{\partial \mathbf{u}(\bm x,t)}{\partial t} 
+ a \cdot \mathbf{u}(\bm x,t) \cdot \nabla \mathbf{u}(\bm x,t) 
&= \nu \nabla^2 \mathbf{u}(\bm x,t), \notag \\
\mathbf{x} &\in \Omega \subset \mathbb{R}^n, 
\quad t \in [0, \infty), 
\quad \mathbf{u} \in \mathbb{R}^n.
\end{align}

\noindent
The parameter added was an advection parameter, $a$, which scales the advection. These tests may give insight in how the behavior of the models is affected when identifying multiple parameters. The ground truth data was created from an analytical solution using the Cole-Hopf transformation. To be consistent with the scaling, the equation was reformulated in terms of the rescaled variable $v=au$. After applying the standard Cole-Hopf transform to $v$, the physical velocity field was recovered using $u=v/a$. The ground truth solutions for the unknown parameters were $a=0.55$ and $\nu=0.01/\pi$. 

The ranking between the models were very similar as with one parameter, however overall all the models performed slightly worse. Compared to the Navier-Stokes experiments, the standard 1D Burgers' equation tests showed less of the biased parameter learning and exploding physics loss that we saw in the more complex problems. However, in the 1D Burgers' equation tests with two unknown parameters, these phenomena were more pronounced.

\begin{figure}[!ht!]
\centering
\includegraphics[width=0.49\textwidth]{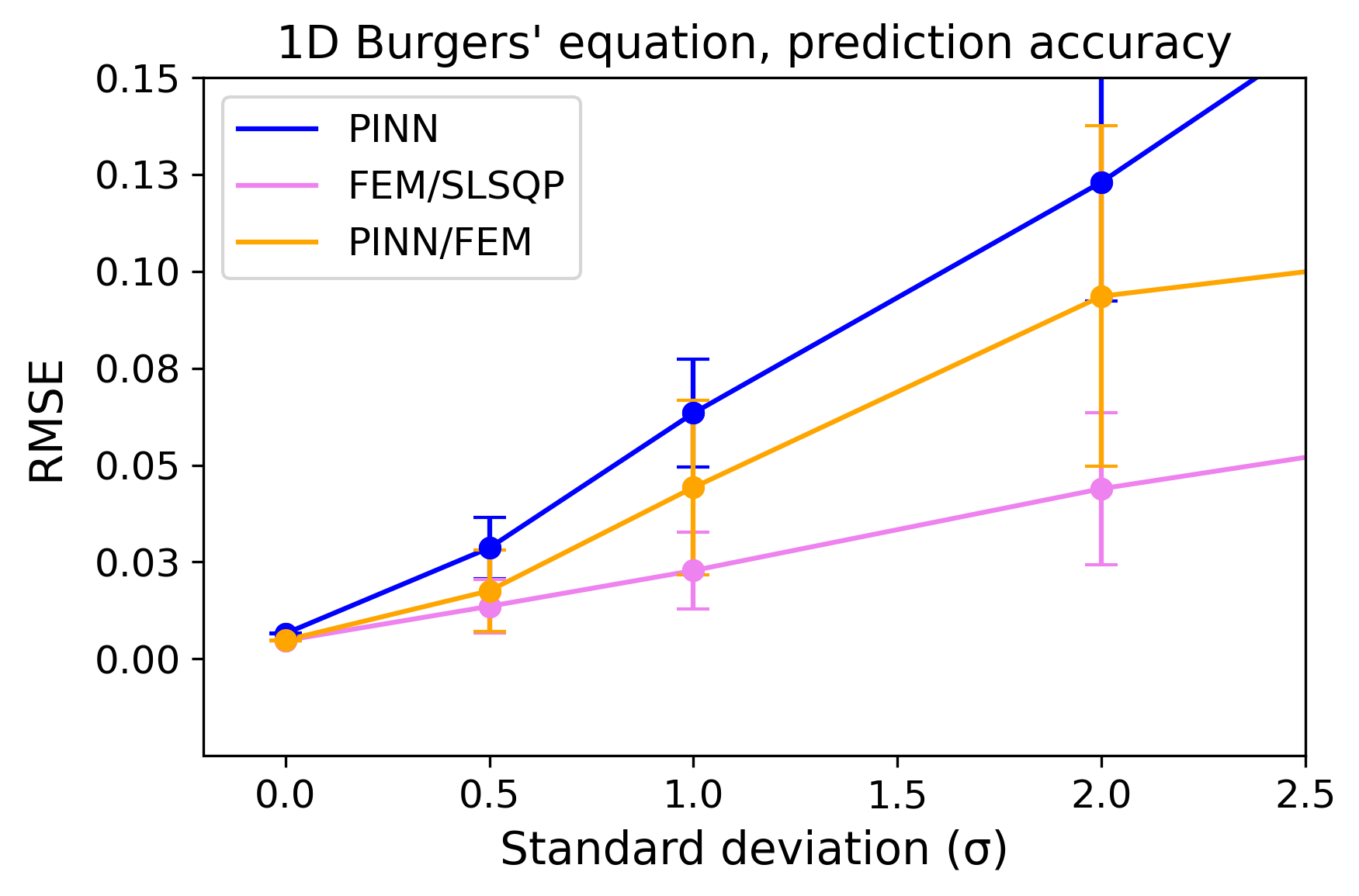}
\includegraphics[width=0.49\textwidth]{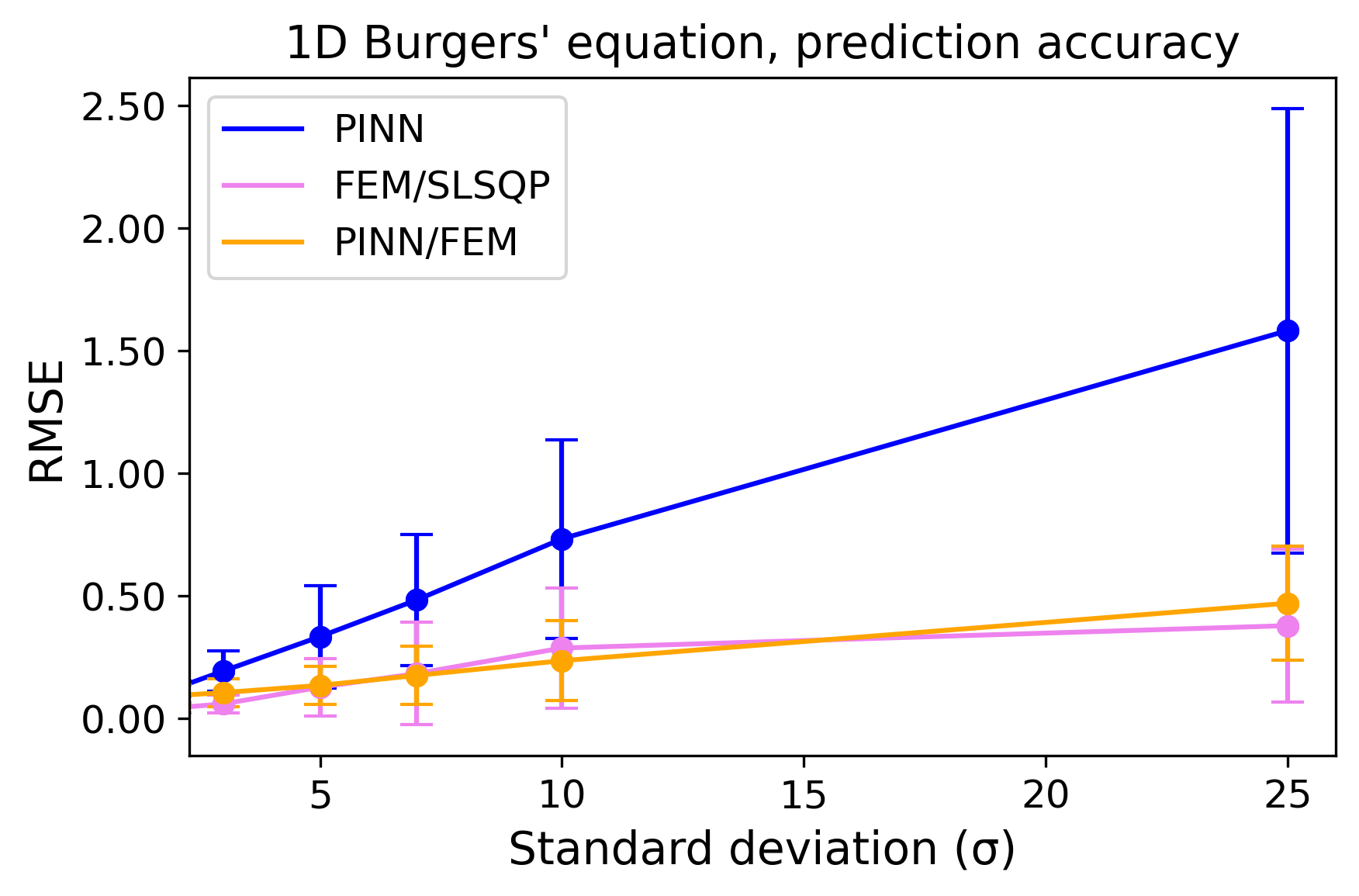}
\includegraphics[width=0.49\textwidth]{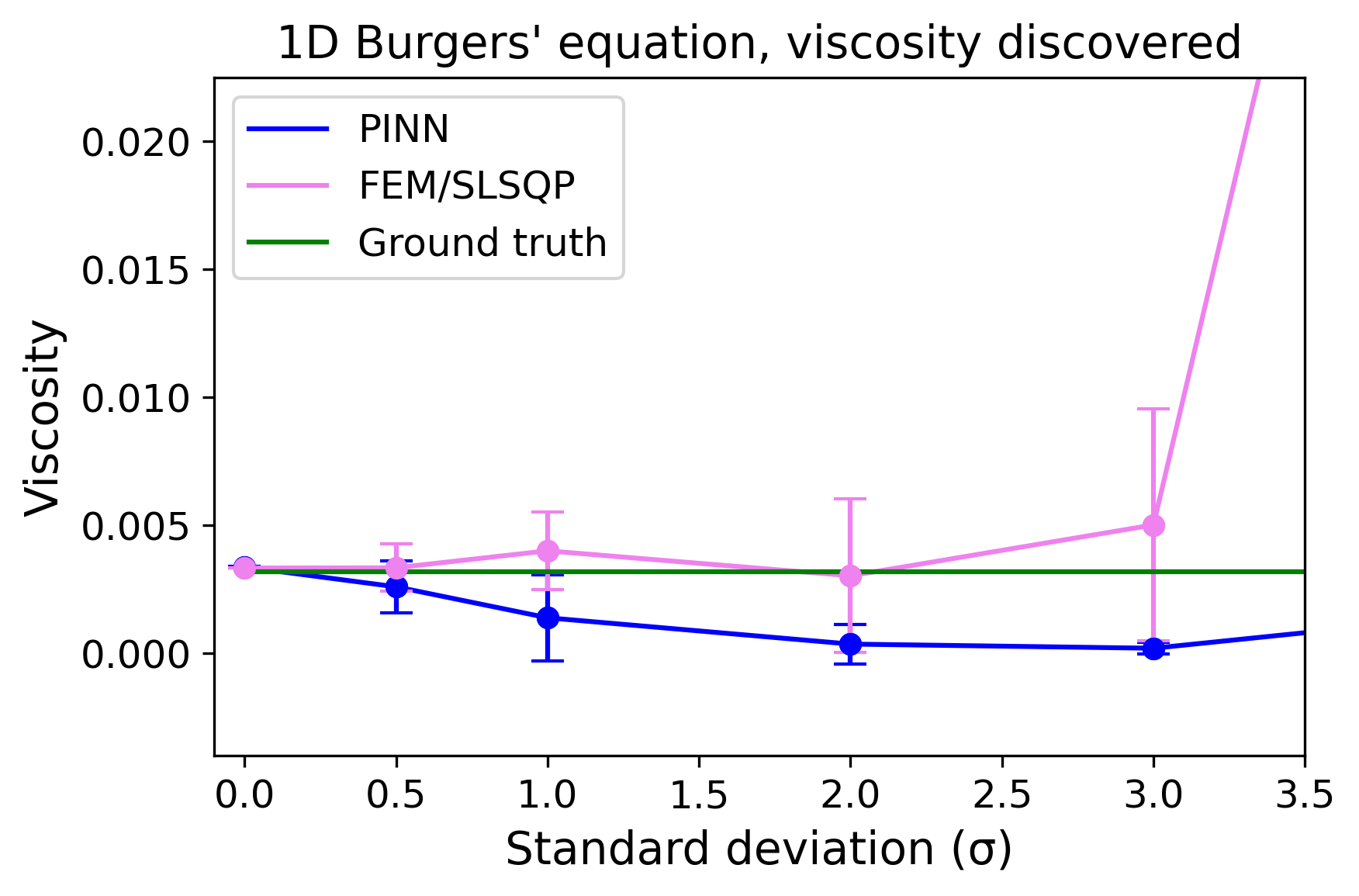}
\includegraphics[width=0.49\textwidth]{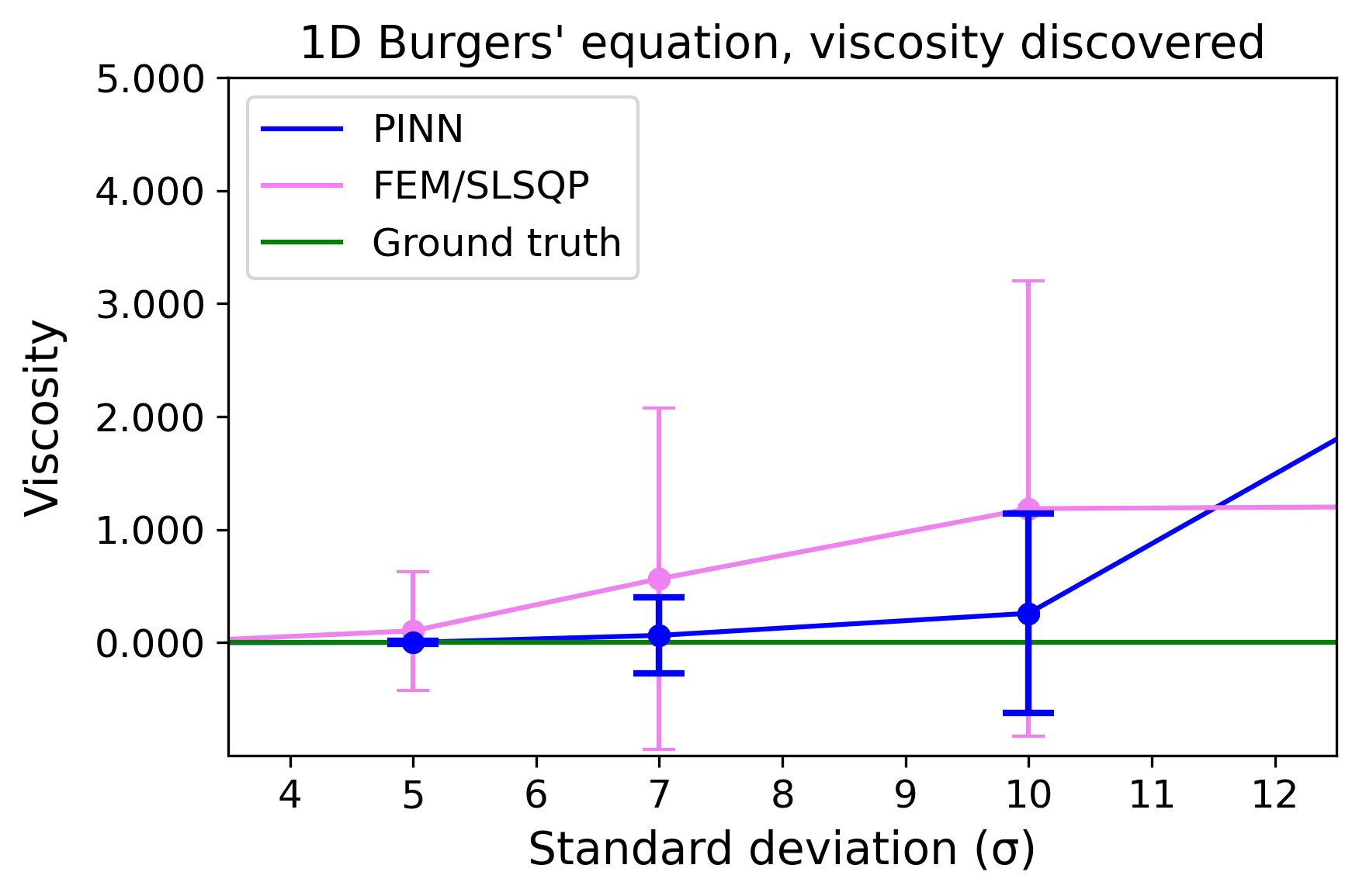}
\includegraphics[width=0.49\textwidth]{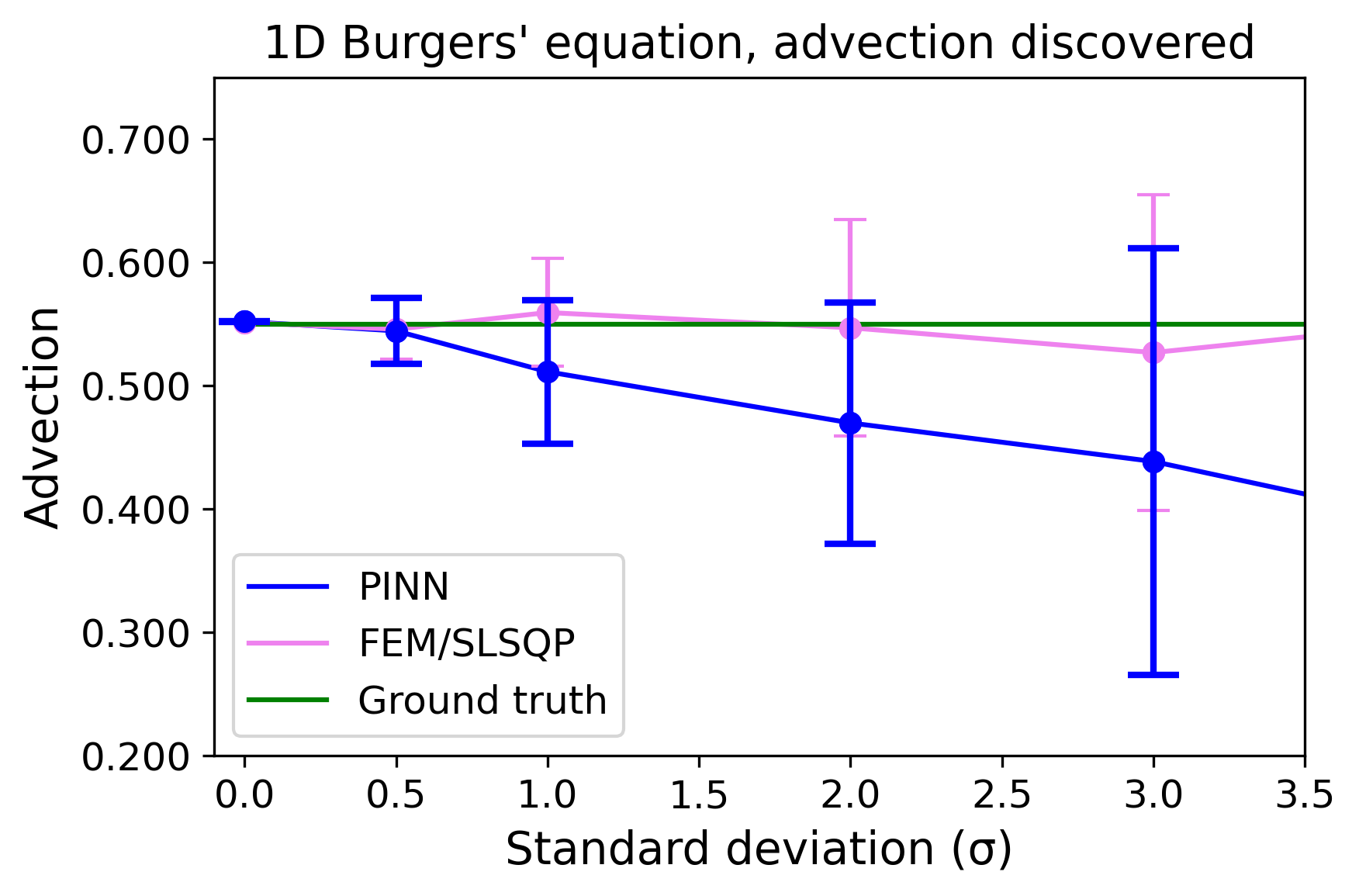}
\includegraphics[width=0.49\textwidth]{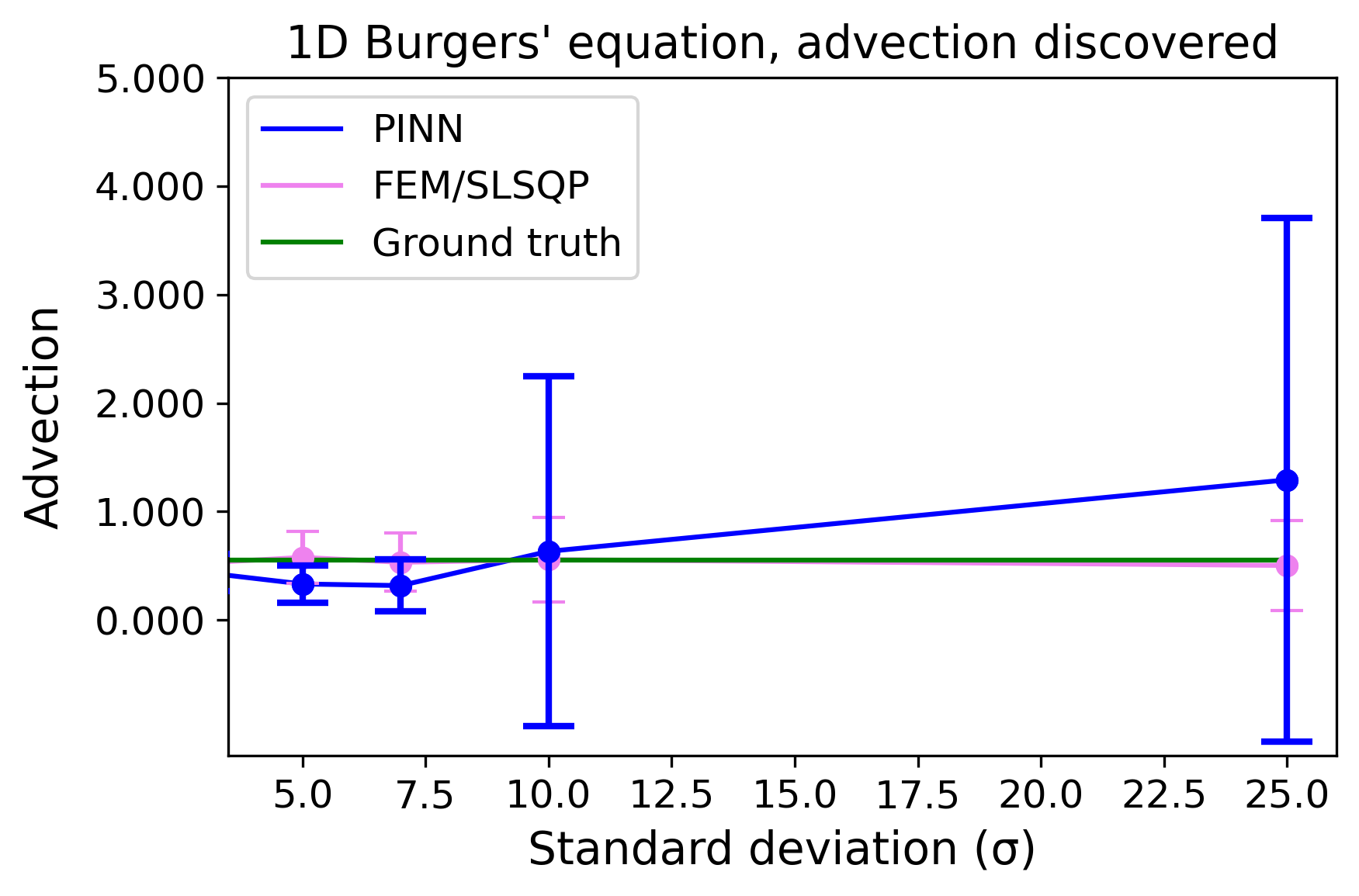}
\caption{Results for 1D Burgers' equation with two unknown parameters on test set based on 30 runs of each model. Standard deviation ($\sigma$) is the noise added to the training data. Top left: Prediction accuracies, $\sigma \in [0, 2]$. Top right: Prediction accuracies, $\sigma \in [3, 25]$. Center left: Viscosity identification accuracies, $\sigma \in [0, 3]$. Center right: Viscosity identification accuracies, $\sigma \in [5, 10]$. Bottom left: Advection identification accuracies, $\sigma \in [0, 3]$. Bottom right: Advection identification accuracies, $\sigma \in [5, 25]$} \label{fig:burgersapp}
\end{figure}

\clearpage

\section{Predictions graphs in MSE}
\label{msegraphs}

\subsection{1D Burgers' equation}

\begin{figure}[H]
\centering
\includegraphics[width=0.49\textwidth]{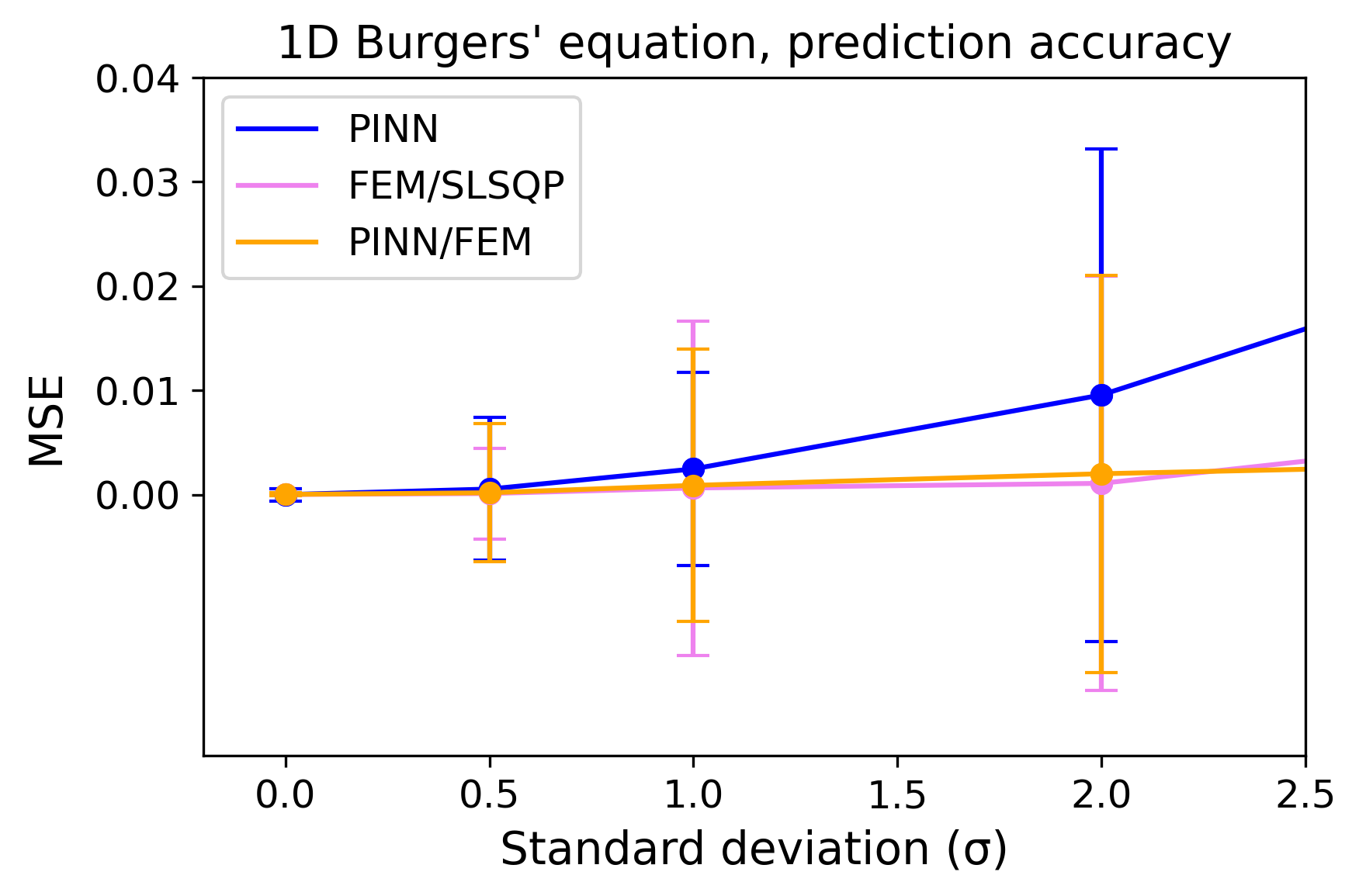}
\includegraphics[width=0.49\textwidth]{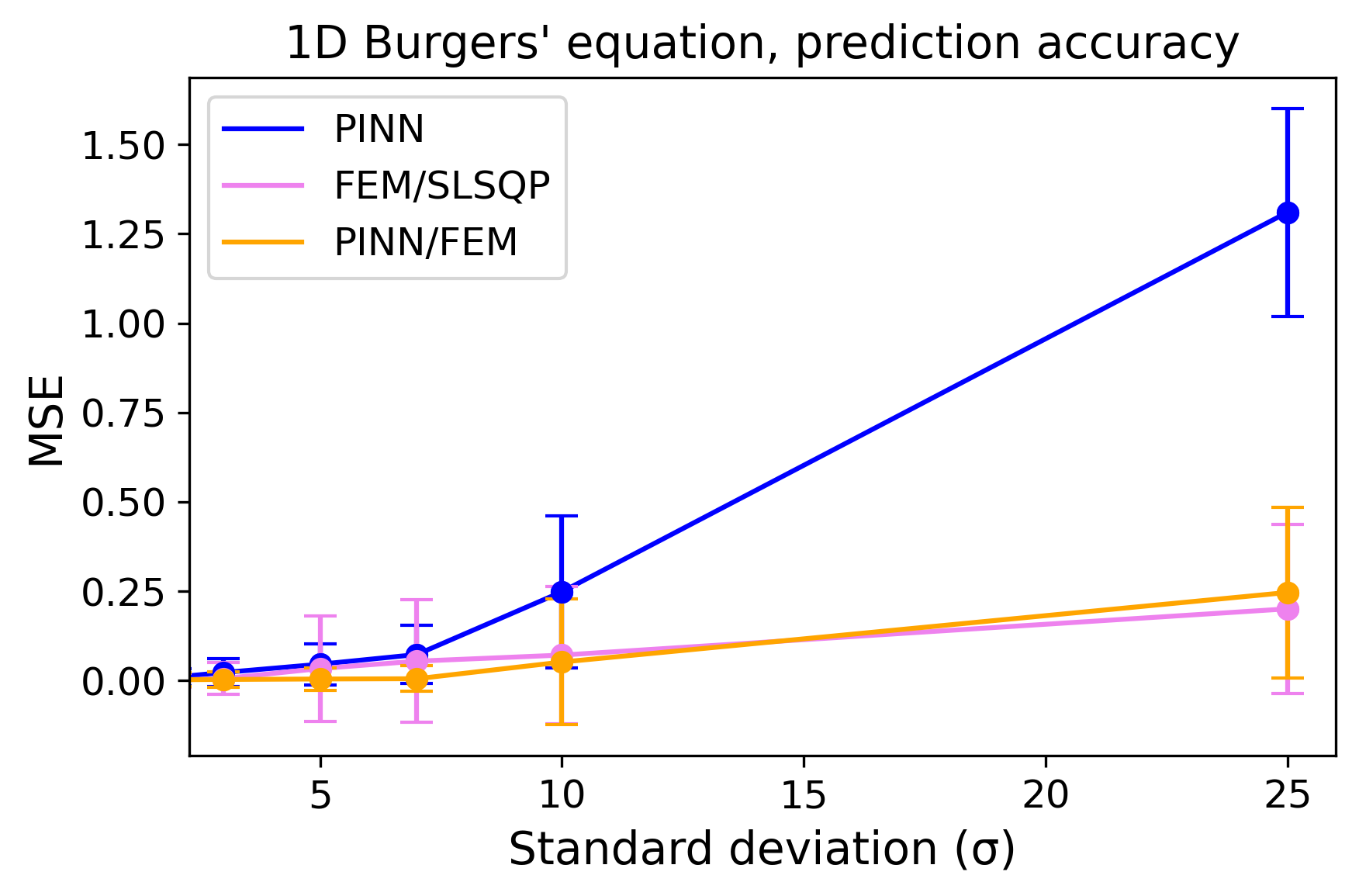}
\caption{Results for 1D Burgers' equation on test set based on 30 runs of each model. Left: Prediction accuracies, $\sigma \in [0, 2]$. Right: Prediction accuracies, $\sigma \in [3, 25]$.}
\end{figure}

\subsection{1D Burgers' equation with two parameters}

\begin{figure}[H]
\centering
\includegraphics[width=0.49\textwidth]{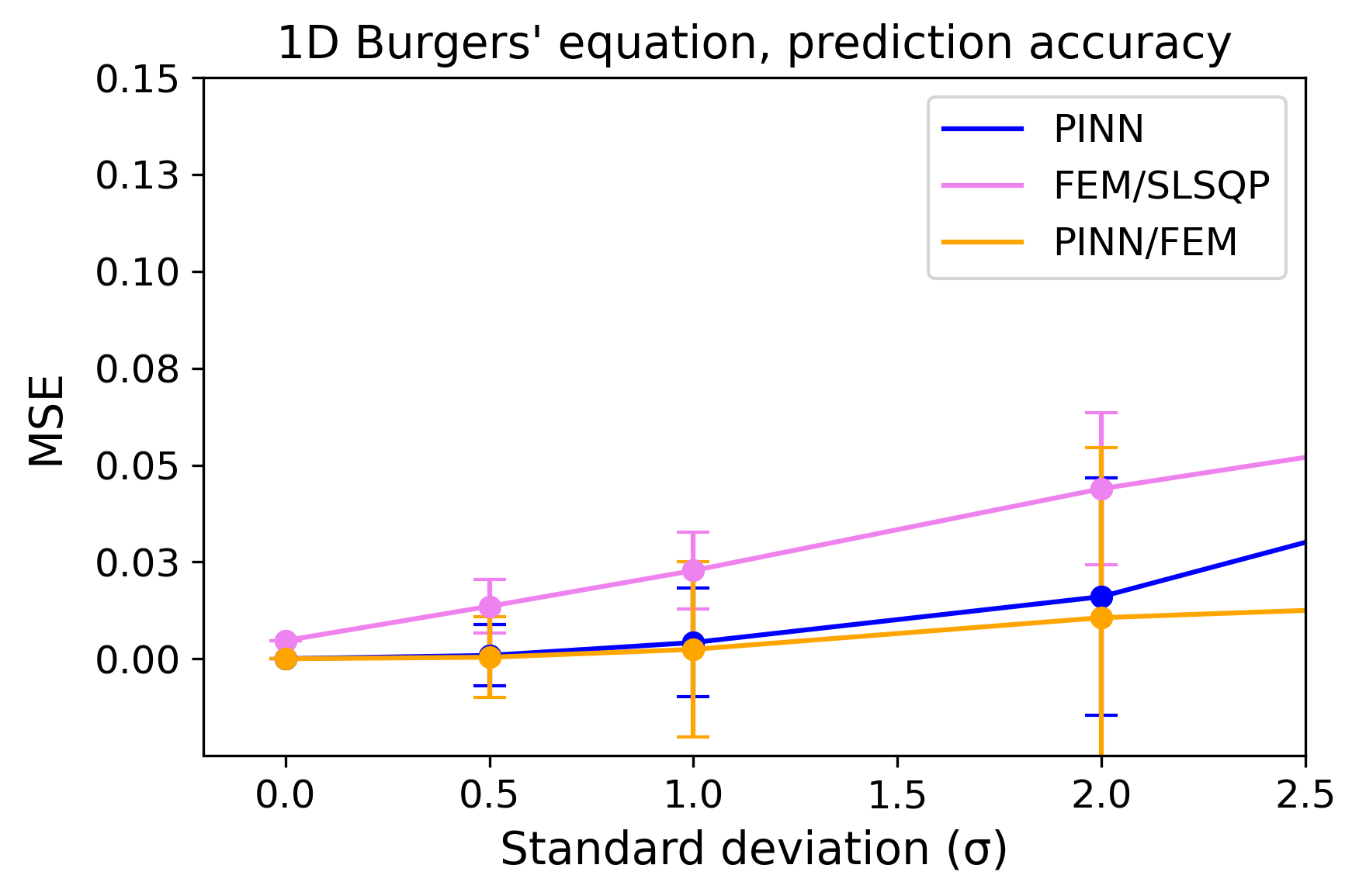}
\includegraphics[width=0.49\textwidth]{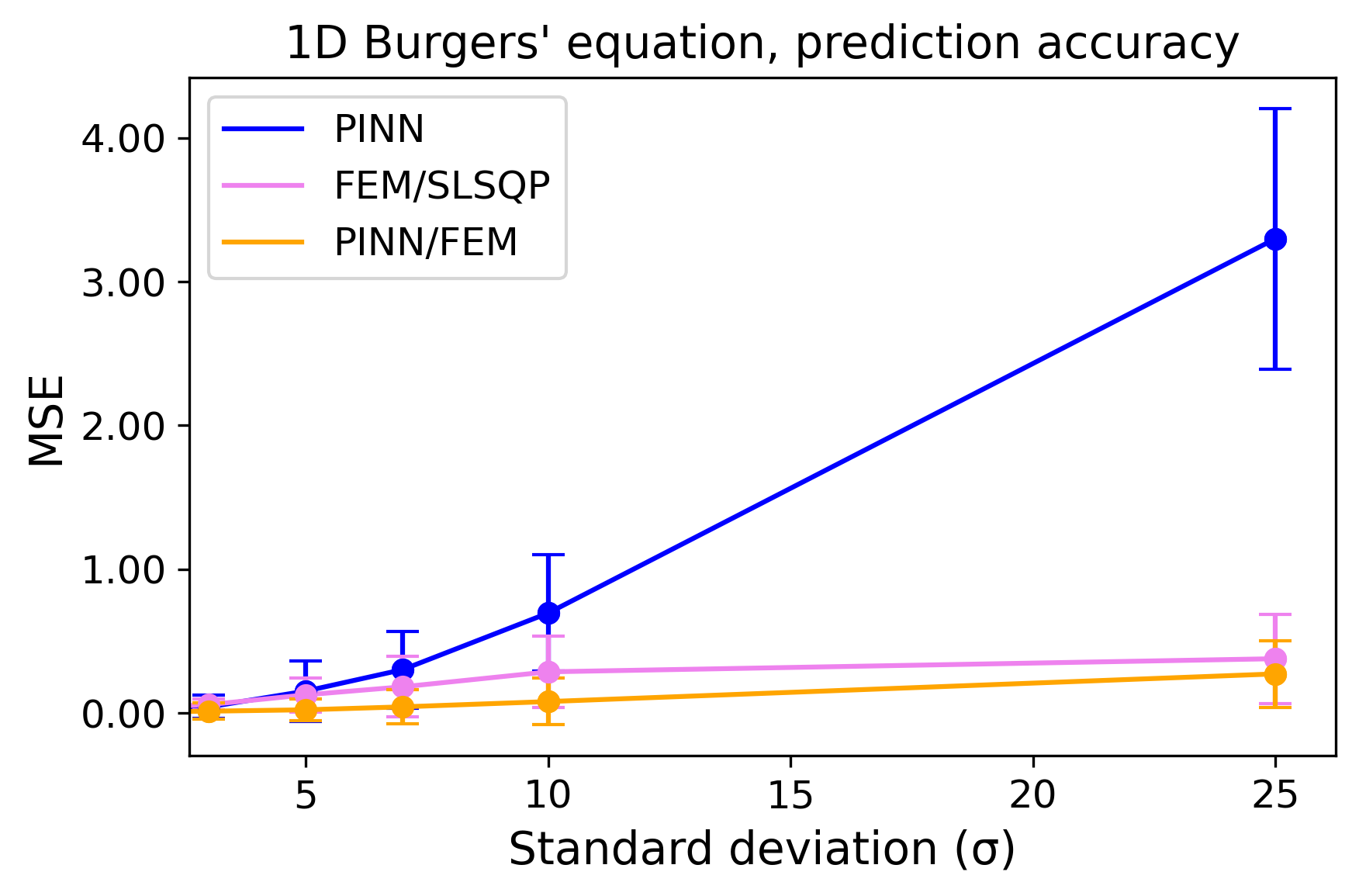}
\caption{Results for 1D Burgers' equation with two parameters on test set based on 30 runs of each model. Left: Prediction accuracies, $\sigma \in [0, 2]$. Right: Prediction accuracies, $\sigma \in [3, 25]$.}
\end{figure}

\subsection{2D Taylor-Green Vortex}

\begin{figure}[H]
\centering
\includegraphics[width=0.49\textwidth]{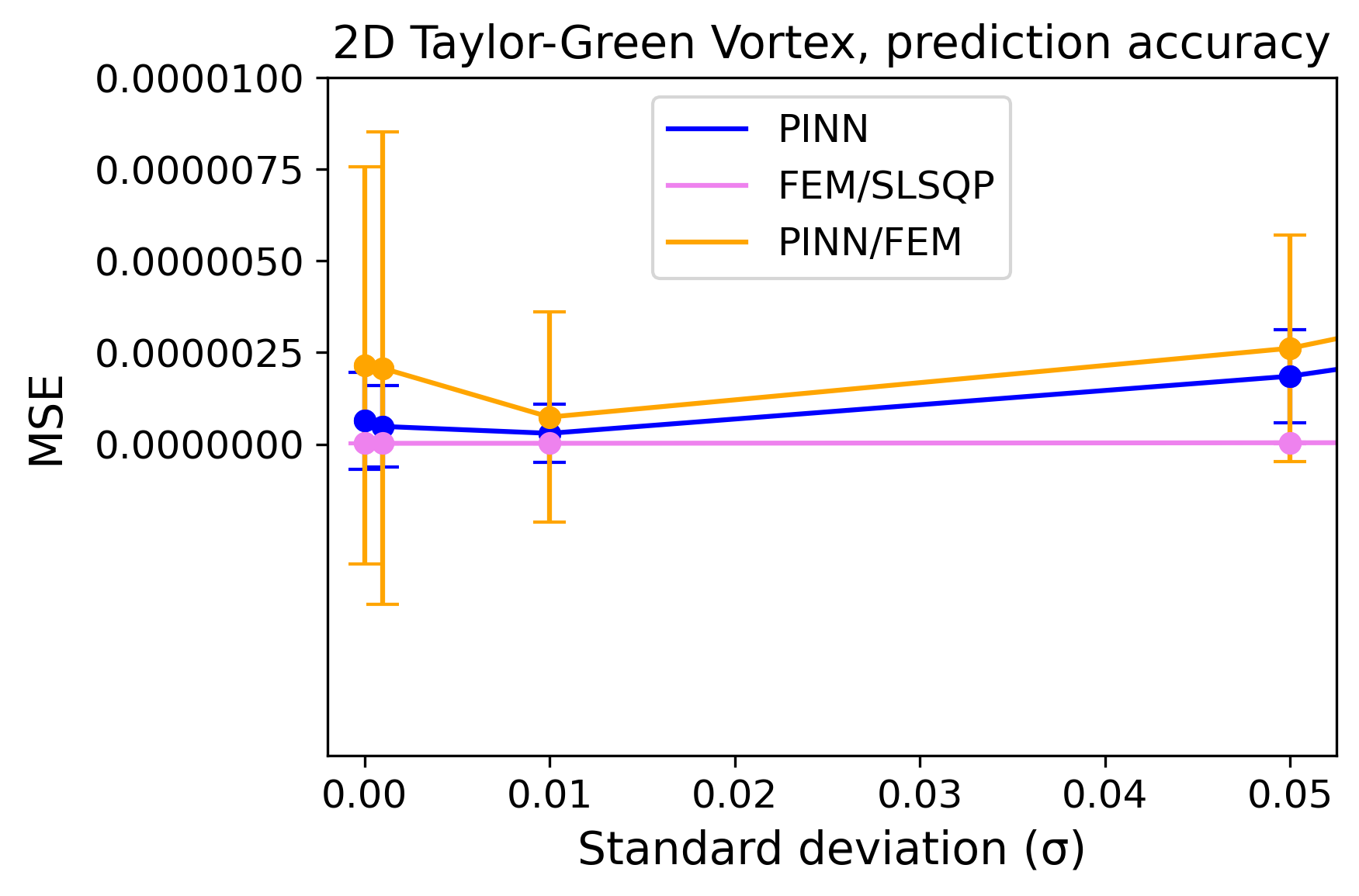}
\includegraphics[width=0.49\textwidth]{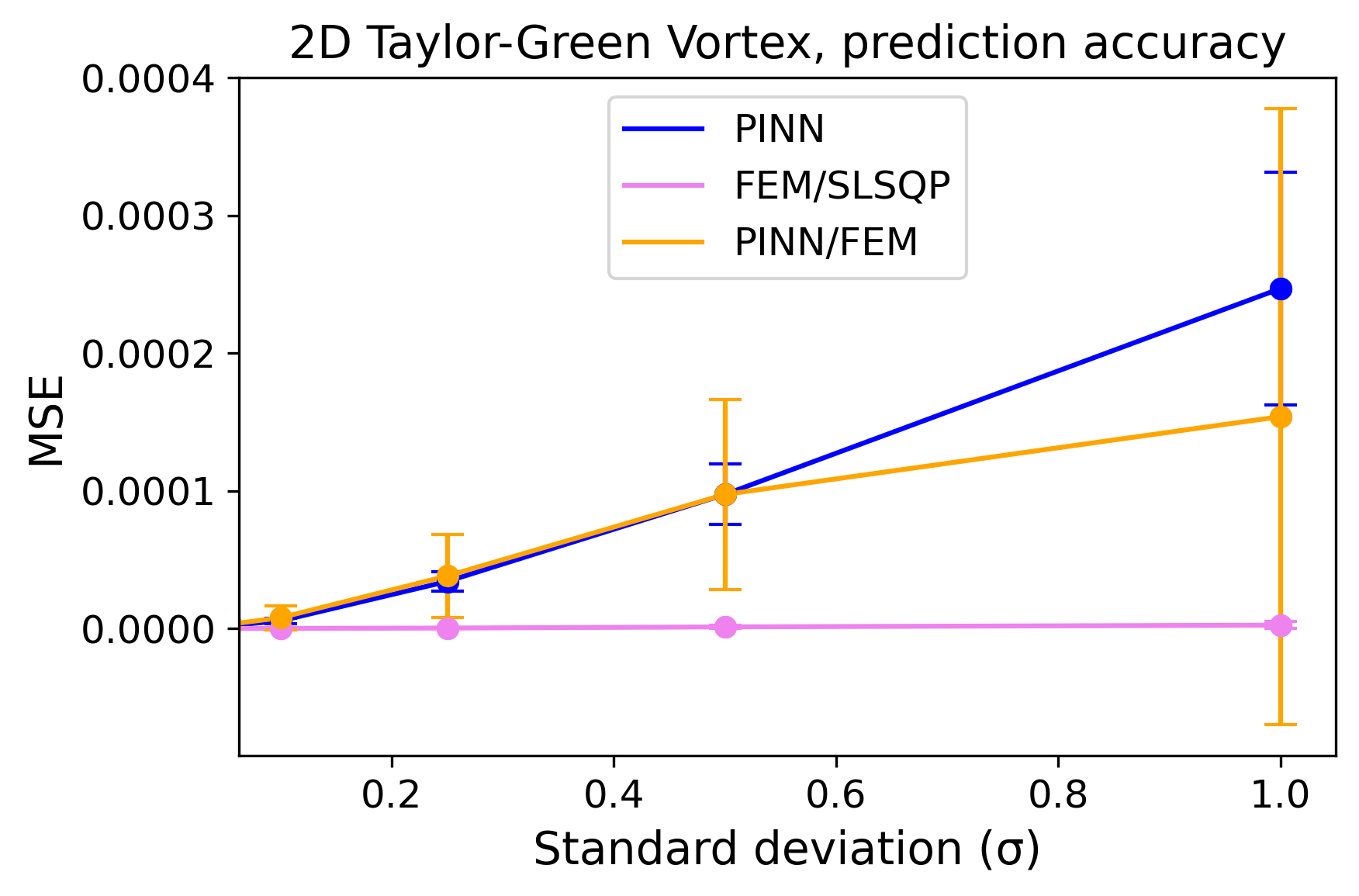}
\caption{Results for 2D Taylor-Green Vortex on test set based on 30 runs of each model. Left: Prediction accuracies, $\sigma \in [0, 0.05]$. Right: Prediction accuracies, $\sigma \in [0.05, 1]$.}
\end{figure}

\subsection{3D Taylor-Green Vortex}

\begin{figure}[H]
\centering
\includegraphics[width=0.49\textwidth]{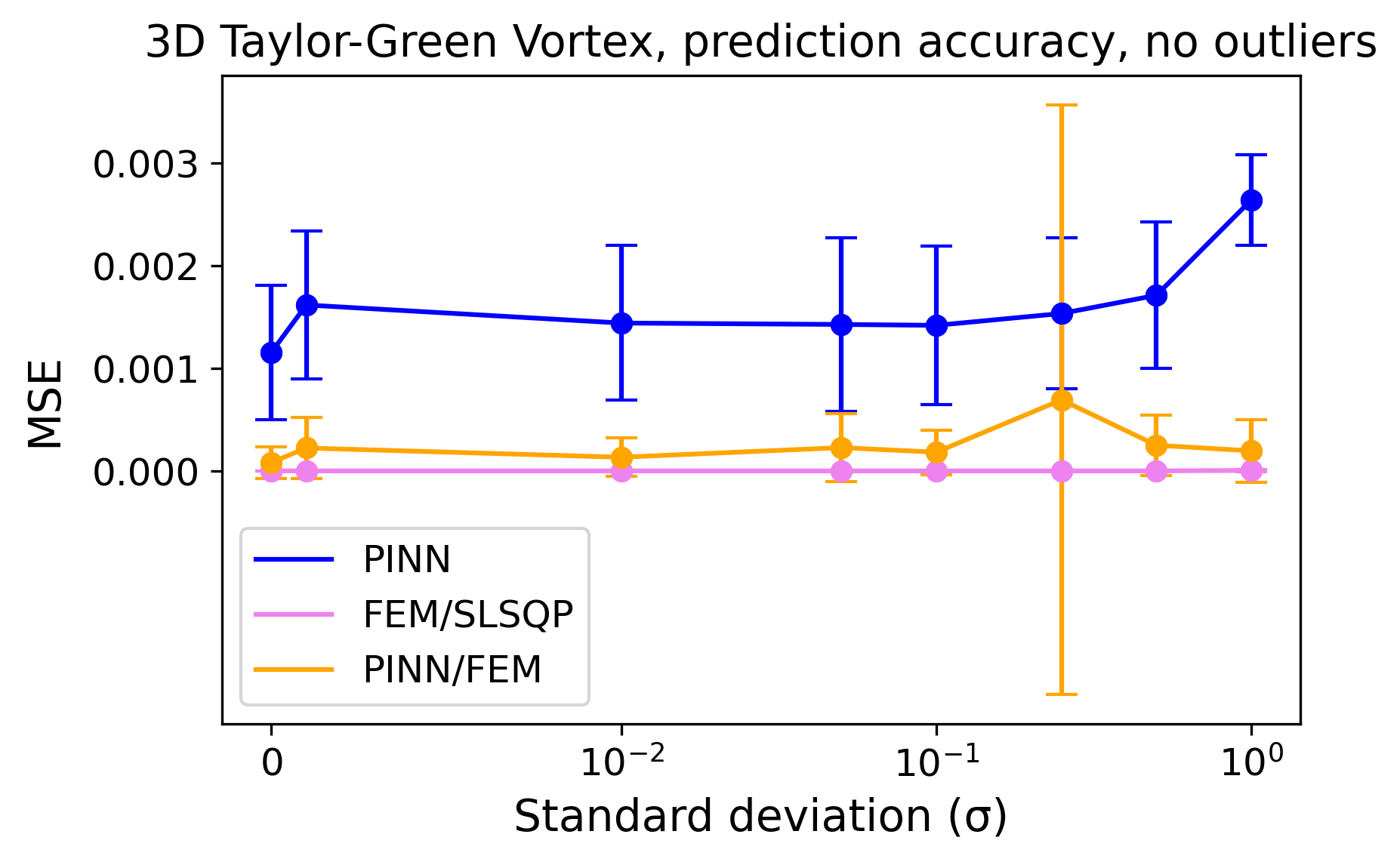}
\includegraphics[width=0.49\textwidth]{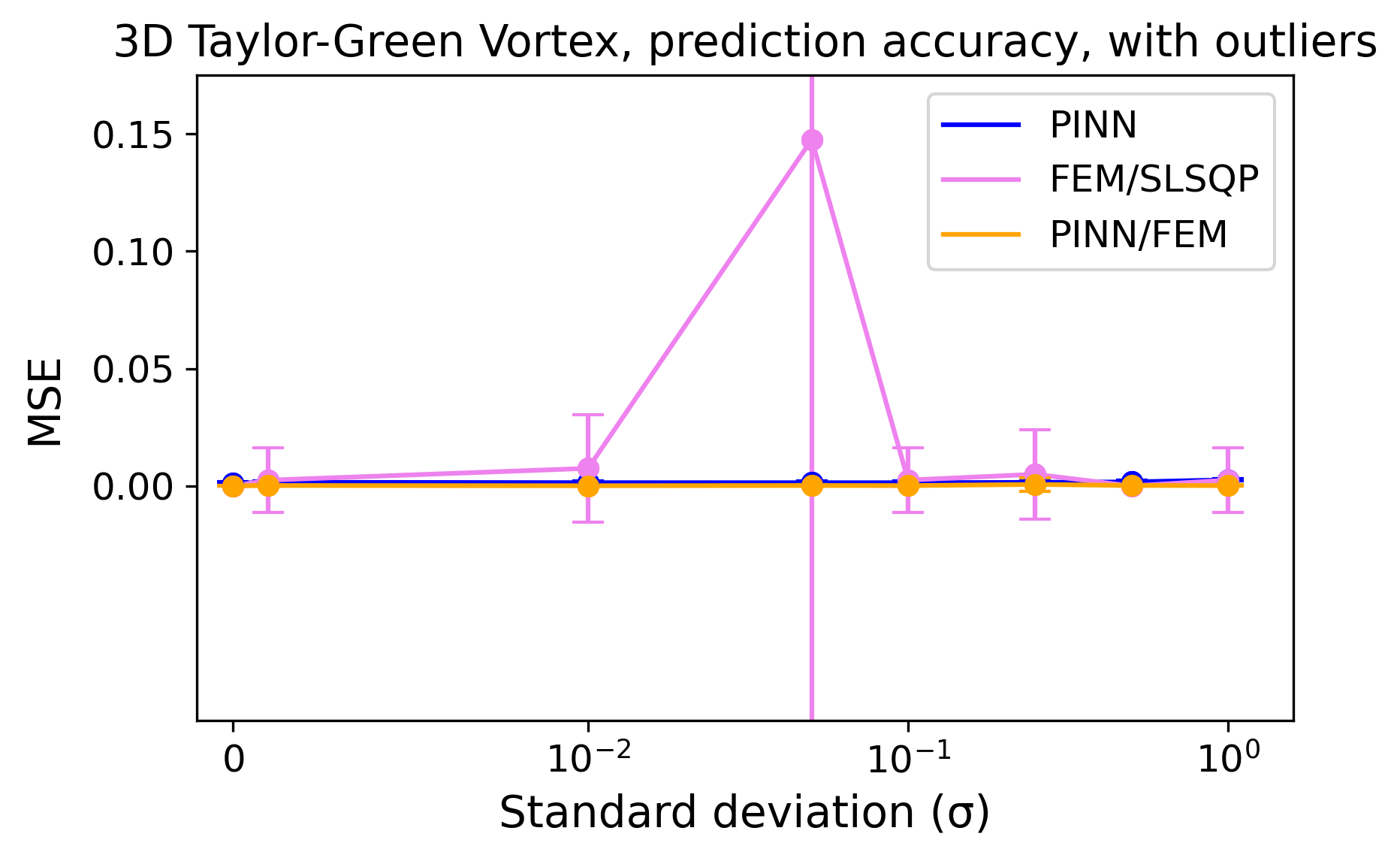}
\caption{Results for 3D Taylor-Green Vortex on test set based on 30 runs of each model. Left: Prediction accuracies with outliers excluded. Right: Prediction accuracies with outliers included.}
\end{figure}

\section{Noise levels tested}
\label{noiselevels}

\subsection{1D Burgers' equation}
\label{noiselevels1d}

\begin{table}[H]
\begin{tabular}{c c c}
\hline
$\sigma$ & SNR & SNR (dB) \\
\hline
0    & $\infty$ & --- \\
0.5  & 1.5048   & 1.77 \\
1    & 0.3762   & -4.25 \\
2    & 0.0940   & -10.27 \\
3    & 0.0418   & -13.79 \\
5    & 0.0150   & -18.23 \\
7    & 0.0077   & -21.15 \\
10   & 0.0038   & -24.25 \\
25   & 0.0006   & -32.20 \\
\hline
\end{tabular}
\label{tab:snr_sigma}
\end{table}

\subsection{1D Burgers' equation with two parameters}

\begin{table}[H]
\begin{tabular}{c c c}
\hline
$\sigma$ & SNR & SNR (dB) \\
\hline
0    & $\infty$ & --- \\
0.5  & 1.8476   & 2.67 \\
1    & 0.4619   & -3.35 \\
2    & 0.1155   & -9.38 \\
3    & 0.0513   & -12.90 \\
5    &  0.0185   & -17.33 \\
7    & 0.0094   & -20.26 \\
10   & 0.0046   & -23.35 \\
25   & 0.0007   & -31.31 \\
\hline
\end{tabular}
\end{table}

\subsection{2D Taylor-Green Vortex}
\label{noiselevels2d}

\begin{table}[H]
\begin{tabular}{c c c}
\hline
$\sigma$ & SNR & SNR (dB) \\
\hline
0    & $\infty$ & --- \\
0.001 & 114885.2596 & 50.60 \\
0.01 & 1148.8526 & 30.60 \\
0.05 & 45.9541 & 16.62 \\
0.1 & 11.4885 & 10.60 \\
0.25 & 1.8382 & 2.64 \\
0.5 & 0.4595 & -3.38 \\
1  & 0.1149 & -9.40 \\
\hline
\end{tabular}
\label{tab:snr_sigma}
\end{table}

\subsection{3D Taylor-Green Vortex}
\label{noiselevels3d}

\begin{table}[H]
\begin{tabular}{c c c}
\hline
$\sigma$ & SNR & SNR (dB) \\
\hline
0    & $\infty$ & --- \\
0.001 & 69943.5822 & 48.45 \\
0.01 & 699.4358 & 28.45 \\
0.05 & 27.9774 & 14.47 \\
0.1 & 6.9944 & 8.45 \\
0.25 & 1.1191 & 0.49 \\
0.5 & 0.2798 & -5.53 \\
1  & 0.0699 & -11.55 \\
\hline
\end{tabular}
\label{tab:snr_sigma}
\end{table}

\section{Additional graph}
\label{addgraph}

\begin{figure}[H]
    \centering
    \includegraphics[width=1\linewidth]{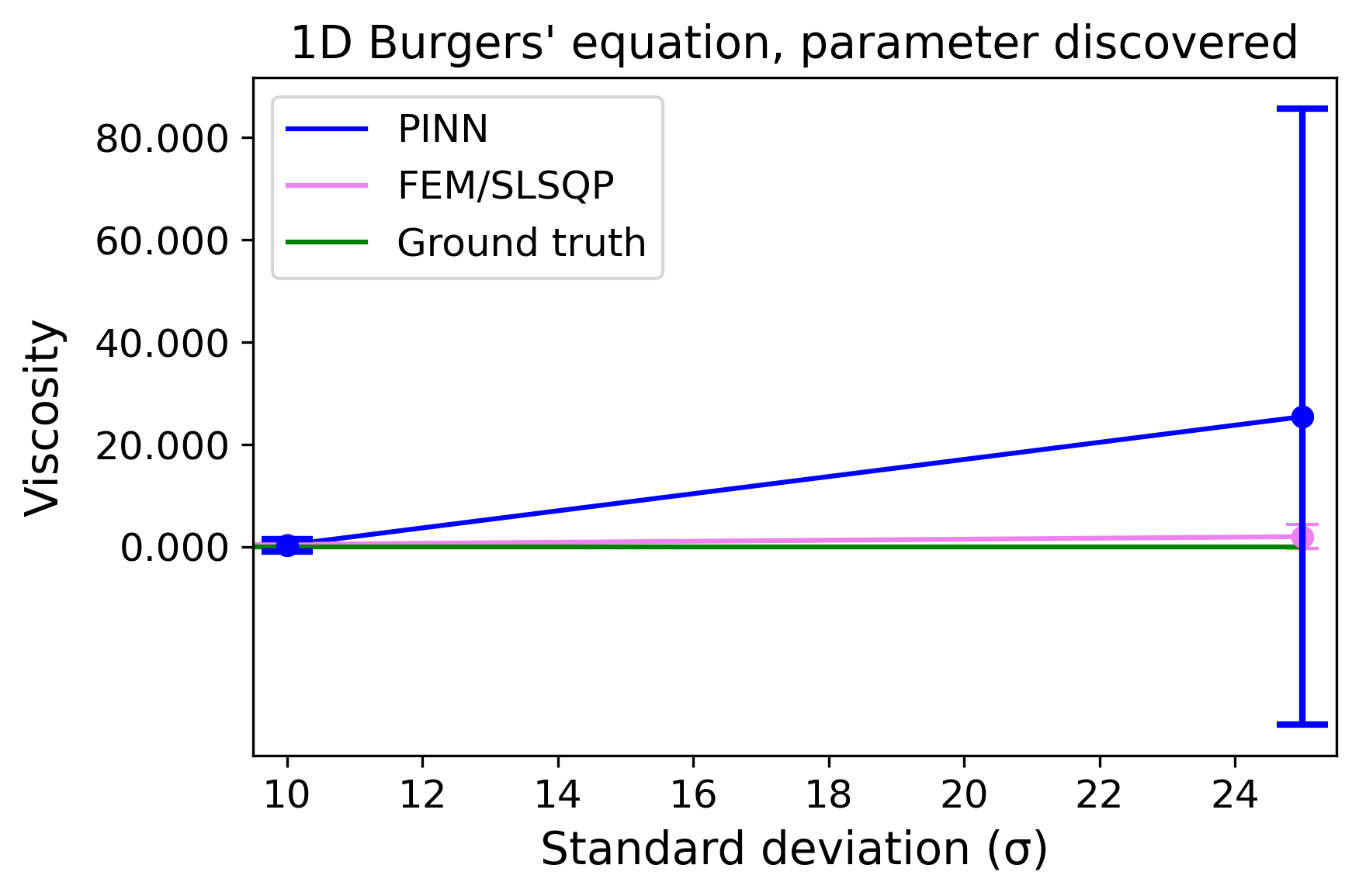}
\end{figure}

\end{document}